\documentclass[prd,twocolumn,aps,superscriptaddress,nofootinbib]{revtex4-2}
\usepackage{amsfonts}
\usepackage{mathtools}
\usepackage[svgnames,table]{xcolor} \usepackage[T1]{fontenc}
\usepackage{txfonts}
\usepackage{graphicx}
\usepackage{hyperref}
\hypersetup{
        colorlinks=true,
        citecolor=SteelBlue,
        filecolor=LimeGreen,
        linkcolor=SlateBlue,
        urlcolor=MediumPurple
}
\usepackage[caption=false]{subfig}
\usepackage{amsmath}
\usepackage[]{overpic}

\newcommand{\be}{\begin{equation}}
\newcommand{\ee}{\end{equation}}

\graphicspath{{Figures/}}

\usepackage[normalem]{ulem}
\newcommand{\equ}[1]{\begin{equation}#1\end{equation}}

\newcommand{\ttrans}{t_{\rm trans}}
\newcommand{\chimin}{\chi^2_{\rm min}}

\begin{document}
\preprint{APS/123-QED}
\title{Angular emission patterns of remnant black holes}
\author{Xiang Li}
\affiliation{Burke Institute for Theoretical Physics, California Institute of Technology, Pasadena, CA 91125, USA}
\author{Ling Sun}
\affiliation{LIGO Laboratory, California Institute of Technology, Pasadena, CA 91125, USA}
\affiliation{OzGrav-ANU, Centre for Gravitational Astrophysics, College of Science, The Australian National University, ACT 2601, Australia}
\author{Rico Ka Lok Lo}
\affiliation{LIGO Laboratory, California Institute of Technology, Pasadena, CA 91125, USA}
\author{Ethan Payne}
\affiliation{LIGO Laboratory, California Institute of Technology, Pasadena, CA 91125, USA}
\affiliation{OzGrav-ANU, Centre for Gravitational Astrophysics, College of Science, The Australian National University, ACT 2601, Australia}
\affiliation{School of Physics and Astronomy, Monash University, Clayton, VIC 3800, Australia}
\affiliation{OzGrav: The ARC Centre of Excellence for Gravitational Wave Discovery, Clayton, VIC 3800, Australia}
\author{Yanbei Chen}
\affiliation{Burke Institute for Theoretical Physics, California Institute of Technology, Pasadena, CA 91125, USA}
\date{\today}

\begin{abstract}
    The gravitational radiation from the ringdown of a binary black hole merger is described by the solution of the Teukolsky equation, which predicts both the temporal dependence and the angular distribution of the emission. Many studies have explored the temporal feature of the ringdown wave through black hole spectroscopy. In this work, we further study the spatial distribution, by introducing a global fitting procedure over both temporal and spatial dependences, to propose a more complete test of General Relativity. We show that spin-weighted spheroidal harmonics are the better representation of the ringdown angular emission patterns compared to spin-weighted spherical harmonics. The differences are distinguishable in numerical relativity waveforms. We also study the correlation between progenitor binary properties and the excitation of quasinormal modes, including higher-order angular modes, overtones, prograde and retrograde modes. Specifically, we show that the excitation of retrograde modes is dominant when the remnant spin is anti-aligned with the binary orbital angular momentum. This study seeks to provide an analytical strategy and inspire the future development of ringdown test using real gravitational wave events.
\end{abstract}

\maketitle

\section{Introduction}

The gravitational waves emitted at the final stage of a binary black hole (BBH) merger -- the ringdown stage, consist of a series of quasinormal modes (QNMs)~\cite{Leaver1985Ana,Kokkotas1999Quasinormal,Nollert1999Quasinormal,Berti2009Quasinormal,Isi2021Analyzing}. QNMs are solutions to the homogeneous Teukolsky equation~\cite{Teukolsky1972Rotating,Teukolsky1973Perturbations,Press1973Perturbations,Cook2014Gravitational}, i.e. the linearized Einstein's equations in the background of a Kerr black hole
\cite{Price1972NonsphericalI,Price1972NonsphericalII,Cunningham1978Radiation,Cunningham1979Radiation,Teukolsky2015Kerr}. 
The foundation for doing so follows models that describe stellar collapses~\cite{Price1972NonsphericalI,Price1972NonsphericalII,Cunningham1978Radiation,Cunningham1979Radiation} -- the strong-field region ``falls down'' toward the future horizon of the final black hole, revealing a spacetime region in which perturbations satisfy the homogeneous Teukolsky equation with ingoing condition near the horizon, and outgoing condition near infinity. 

The homogeneous Teukolsky equation not only predicts the {\it temporal} dependence of the ringdown waves, in terms of their {\it complex} spectra, but also their {\it spatial} distributions, in terms of angular emission patterns. 
There have been many studies on black hole ringdown spectroscopy involving multiple angular frequencies by modeling it as the superposition of exponentially damped sinusoids~\cite{Damour2014New,LSC2016TGRGW150914,Del2017Analytic,Brito2018Black-hole,Carullo2019Observational,Cook2020Aspects}, or using other methods of frequency extraction~\cite{Berti2007Mining,Nakano2019Comparison}. Recently, it has been shown that the inclusion of overtones~\cite{Berti2006Quasinormal,Berti2009Quasinormal} can improve the fitting of numerical relativity (NR) waveforms and lead to better estimation of ringdown model parameters~\cite{Giesler2019Black,Forteza2020Spectroscopy,Cook2020Aspects}, because of the better characterization of the post-merger signal from an earlier time. 

Many phenomenological fitting studies based on NR waveforms have been done~\cite{London2014Modeling,Forteza2017Hierarchical,London2020Modeling}, while most previous works only focused on the ringdown temporal properties. Our study further includes spatial dependence on different models, explicitly. 
Specifically, when the final spin is not aligned with the initial orbital angular momentum, the spatial properties for retrograde excitations~\cite{Taracchini2014Small,Hughes2019Learning,Apte2019Exciting1,Lim2019Exciting2} have not been carefully studied. 
As the temporal-spatial consistency check of ringdown emission can provide a more complete test of General Relativity~\cite{Cardoso2016Testing,Isi2019Testing,LVC2019Tests,LVC2020Tests}, exploring such a problem defines the theme of this paper.

In gravitational wave observations, the prospective searches for ringdown waveforms~\cite{Berti2007Matched,Maselli2017Observing} would enable the test of the spatial-temporal features. With the rapidly increasing number of binary coalescences observed~\cite{LIGO2016O1,LVC2019GWTC-1,LVC2020O2,LVC2021GWTC-2} by ground-based detectors like Advanced LIGO~\cite{ALIGO2015} and Advanced Virgo~\cite{AVirgo2014}, events with detectable higher-order modes~\cite{Healy2013Template,Pekowsky2013Impact,Mills2021Measuring} are observed, e.g., GW190412~\cite{Abbott2020GW190412} and GW190814~\cite{GW190814AJ}. The detectability of higher-order modes not only impacts the parameter estimation~\cite{Brito2018Black-hole,Cotesta2018Enriching,Carullo2019Observational,LVC2020Tests, Payne2019, lange2018rapid}, but can be used to study angular emission as well. Currently, in the ringdown stage, a high signal-to-noise ratio (SNR) is difficult to achieve due to the lack of post-merger cycles and the degraded detector sensitivity at high frequency range. However, the sensitivity of the proposed next-generation detectors, including Einstein Telescope~\cite{ET2010,ETSC2020}, Cosmic Explorer~\cite{reitze2019CE,CE2020low-frequency}, and NEMO~\cite{NEMO2020}, will be significantly improved~\cite{Divyajyoti2021Detectability}, especially at the high frequencies, opening more possibilities in the BBH post-merger studies~\cite{Forteza2020Spectroscopy}. 
Although a single event could provide limited information about angular dependence, combining multiple events and extracting angular-dependent features will become possible with the expected large number of events in the future. That calls for strategic studies of temporal-spatial emission patterns before more events with high ringdown SNR are observed.

In this paper, we study ringdown gravitational waves and show that the spin-weighted spheroidal harmonics are essential in the faithful representation of the \emph{temporal-spatial} ringdown emission pattern~\cite{Berti2006Eigenvalues}, further confirming that the Teukolsky equation describes the ringdown dynamics. We fit the NR simulated strain data of merging binaries with different parameters provided in the Simulating eXtreme Spacetimes (SXS) Collaboration catalog~\cite{Boyle2019SXS}, without adding simulated noise. 
The exclusion of nonlinear gravitational wave memory effects in SXS waveforms~\cite{Mitman2021Adding} provides an ideal test bed for the linear perturbation theory. We infer the final black hole spin and mass via parameter estimation~\cite{Tichy2008Final,Rezzolla2008Final}, as in the usual temporal-only fitting. Our results show that when the spatial distribution is considered in addition to the spectrum, more information could be extracted from the ringdown stage, which would benefit the determination of the progenitor BBH properties and their formation channels~\cite{Rodriguez2016Illuminating,Gerosa2018Spin}. 
Along this procedure, we describe various features of the temporal-spatial emission pattern; in particular, we illustrate the physical meaning of the two complex amplitudes of each QNM. 
We also study the cases where retrograde modes~\cite{Taracchini2014Small,Khanna2017Black,Hughes2019Learning,Apte2019Exciting1,Lim2019Exciting2} are excited when the initial spin of the primary progenitor black hole is large and anti-aligned with the orbital angular momentum~\cite{Berti2008Cosmological,Rodriguez2016Illuminating}, and thus the remnant spin is left anti-aligned with the binary orbital angular momentum.

The structure of this paper is organized as follows. In Sec.~\ref{sec:model}, we review the QNM decomposition model under the spin-weighted spheroidal harmonic basis and the spin-weighted spherical harmonic basis, respectively, and describe the temporal-spatial fitting strategy used in this paper. 
In Sec.~\ref{sec:nonspinning}, we describe the fitting results for waveforms of a benchmark binary and several nonspinning binaries with different mass ratios, and discuss the distinguishability of the two decomposition models and the contributions of higher-order modes and overtones. 
Then, in Sec.~\ref{sec:retrograde}, we investigate the excitation of retrograde modes when the remnant spin is anti-aligned with the binary orbital angular momentum. 
Finally, in Sec.~\ref{sec:conclusion}, we summarize and further discuss our results. Details about conventions, fitting strategy, and results are given in the appendices.

\section{\label{sec:model}Ringdown Models}

In this work, the discussion is restricted to non-precessing binaries, for which we establish two coordinate systems: the ``orbital frame'' adapted to the binary orbital angular momentum, and the ``final spin frame'' adapted to the remnant spin angular momentum.  In the final spin frame, we decompose the temporal and angular distribution of gravitational waves into a sum over QNMs. This analytical decomposition is {\it fitted} to the waveforms from NR simulations. 
In Sec.~\ref{sec:frames}, we introduce the orbital and final spin frames, describe the QNM decomposition in the final spin frame, and discuss the excitations of the modes analytically. In Sec.~\ref{sec:fitting_method}, we present the fitting strategy in the orbital frame.

\subsection{\label{sec:frames}Frames and QNM models}

\begin{figure}
	\begin{center}
		\includegraphics[width=0.63\columnwidth]{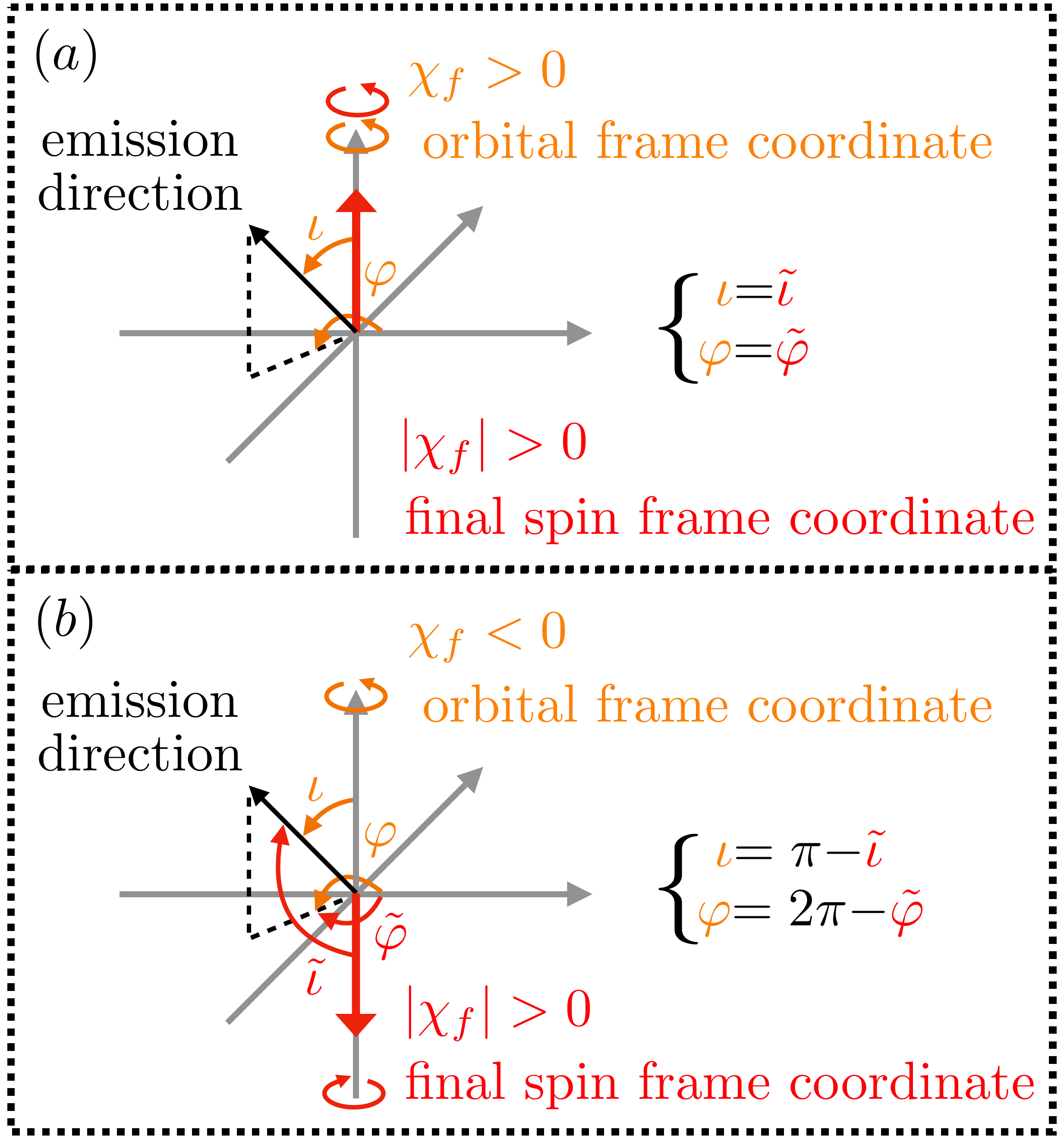}
		\caption{\label{fig:aliant_coord} 
		Convention of coordinate continuation from positive to negative values of the final spin $\chi_f$. (a) Spin aligned case ($\chi_f>0$): when the final spin is aligned with the orbital angular momentum, the final spin frame coincides with the orbital frame, with $\iota=\tilde{\iota}$ and $\varphi=\tilde{\varphi}$. (b) Spin anti-aligned case ($\chi_f<0$): when the final spin is anti-aligned with the orbital angular momentum, we have $\iota=\pi-\tilde{\iota}$ and $\varphi=2\pi-\tilde{\varphi}$.}
	\end{center}
\end{figure}

In this subsection, we define the coordinate frames used to describe outgoing waves near infinity, carry out QNM decomposition in two types of bases, describe the temporal and spatial dependences of the QNMs, and discuss the excitations of prograde and retrograde modes. 

\subsubsection{Coordinate frames}

For the majority of BBH events detected so far, the inspiral stage often contributes most of the SNR~\cite{LVC2019GWTC-1,LVC2021GWTC-2,LVC2019Tests,LVC2020Tests}. The most natural coordinate system to describe non-precessing binaries has its $\hat z$ axis aligned with the direction of the orbital angular momentum. 
We refer to this as the \emph{orbital frame} and use $(\iota,\varphi)$ to label the polar (inclination) and azimuthal angles, respectively. In particular, waveforms of non-preceesing binaries from the SXS catalog adopt the orbital frame. 
On the other hand, QNM decomposition is most easily performed by taking the $\hat z$ axis along the spin direction of the remnant black hole. We shall refer to this as the \emph{final spin frame} and use $(\tilde{\iota}, \tilde{\varphi})$ for its polar (inclination) and azimuthal angles. 

The transformation between the orbital frame and the final spin frame in two specific cases are illustrated in Fig.~\ref{fig:aliant_coord}: (a) When the final spin of the remnant black hole is aligned with the orbital angular momentum (spin aligned case)~\cite{London2020Modeling}, the two frames coincide, i.e. $\iota=\tilde{\iota},\ \varphi=\tilde{\varphi}$. (b)  When the final spin is anti-aligned with the orbital angular momentum (spin anti-aligned case), we have $\iota=\pi-\tilde{\iota}$ and $\varphi=2\pi-\tilde{\varphi}$.
With such coordinate transformation, the parameter space of $\chi_f$ is constructed to be continuous across zero, thus linking the cases of aligned and anti-aligned remnant spins. 
For more general cases of spin misaligned with the orbital angular momentum, the orbital angular momentum precesses in time, and is neither aligned nor anti-aligned with the final spin -- this is left for future studies.



For consistency with the SXS data structure and most of the literature, we adopt the orbital frame when describing the final remnant spin $\chi_f$, with $\chi_f<0$ corresponding to the spin anti-aligned case, and use $|\chi_f|$ to denote the spin magnitude. For simplicity, we use the symbol $\vec{\Omega}\equiv (\iota,\varphi)$ to represent coordinates in the orbital frame.

\subsubsection{QNM decomposition models}

Let us now perform QNM decomposition of outgoing gravitational waves near infinity in the final spin frame. 
From a start time $t_0$, the complex-valued ringdown waveform can be written as~\cite{Teukolsky1973Perturbations,Press1973Perturbations,Leaver1985Ana,Berti2006Gravitational,Isi2021Analyzing}:
\begin{align}
\label{eq:S_model}
& h^{(S)}(\tilde{\iota}, \tilde{\varphi},t) \nonumber\\
=&(h_{+}-i\,h_{\times})^{(S)}(\tilde{\iota}, \tilde{\varphi},t) \nonumber\\ 
=&\frac{M_f}{r}\sum_{l=2}^{l_{\rm max}}\sum_{m=-l}^{m=l} \sum_{n=0}^{n_{\rm max}}\large[B^{(S+)}_{lmn} e^{-i\omega_{lmn}(t-t_0)} {}_{-2}S_{lmn}(\gamma_{lmn},\tilde{\iota}, \tilde{\varphi})+ \nonumber\\
&\qquad \qquad \qquad \;\; B^{(S-)}_{lmn} e^{i\omega^*_{lmn}(t-t_0)} {}_{-2}S^*_{lmn}(\gamma_{lmn},\pi-\tilde{\iota}, \tilde{\varphi}) \large],
\end{align}
where $h_{+}$ and $h_{\times}$ are the plus and cross polarization components, respectively, $r$ is the distance from the source binary to the detector on Earth. 
The QNMs summed over here are labeled by three integers: the angular indices $(l,m)$ with $l=2,3,...$ and $|m|\leq l$, and the overtone index $n=0,1,...$. Here we have only carried out the summations up to finite maximal values of the angular quantum number $l_{\rm max}$ and the overtone number $n_{\rm max}$. 
%
%
The real and imaginary parts of each $\omega_{lmn}$ correspond to the (angular) frequency and decay rate of the QNM; the entire spectrum $\{\omega_{lmn}\}$ is exclusively determined by the remnant black hole's mass $M_f$ and dimensionless spin $\chi_f$ -- the only two parameters that characterize a stationary uncharged black hole, according to the no-hair theorem~\cite{Israel1967Event,hawking1972black,Carter1971Axisymmetric,Gurlebeck2015No-Hair,Carullo2018Empirical,Isi2019Testing,Forteza2020Spectroscopy}. 
The angular functions ${}_{-2}S_{lmn}$ in Eq.~\eqref{eq:S_model} are the spin-weighted {\it spheroidal} harmonics, with dimensionless {\it spheroidicity parameter} $\gamma_{lmn}$ given by~\cite{Press1973Perturbations}:    
\begin{equation}\label{eq:spheroidicity}
    \gamma_{lmn}=\chi_f M_f\omega_{lmn}.
\end{equation} 
The dependence of the angular function on $\gamma_{lmn}$ can be attributed to the deformation of bounded photon orbits due to a Kerr black hole's spin. 


In practice, on the other hand, the ringdown waveform is often {\it approximated} by an expansion of decaying sinusoids with angular dependence given by spin-weighted {\it spherical} harmonics $\{{}_{-2}Y_{lm}\}$:
\begin{align}
\label{eq:Y_model}
& h^{(Y)}(\tilde{\iota}, \tilde{\varphi},t)\nonumber\\ 
=&(h_{+}-i\,h_{\times})^{(Y)}(\tilde{\iota}, \tilde{\varphi},t) \nonumber \\
=&\frac{M_f}{r}\sum_{l=2}^{l_{\rm max}}\sum_{m=-l}^{m=l} \sum_{n=0}^{n_{\rm max}}\large[B^{(Y+)}_{lmn} e^{-i\omega_{lmn}(t-t_0)} {}_{-2}Y_{lm}(\tilde{\iota}, \tilde{\varphi})+ \nonumber \\
&\qquad \qquad \qquad \; \; B^{(Y-)}_{lmn} e^{i\omega^*_{lmn}(t-t_0)} {}_{-2}Y^*_{lm}(\pi-\tilde{\iota}, \tilde{\varphi}) \large].
\end{align}
For simplicity, we refer to the decompositions in Eqs.\,\eqref{eq:S_model} and~\eqref{eq:Y_model} as $S$ model and $Y$ model, respectively. The mode mixing between the two bases depends on the spheroidicity $\gamma_{lmn}$~\cite{Press1973Perturbations,Berti2006Gravitational,berti2014mixing,London2014Modeling}:

\begin{align}
\label{eq:SY_mixing}
{}_{-2}S_{lmn}(\gamma_{lmn},\tilde{\iota}, \tilde{\varphi})={}_{-2}Y_{lm}(\tilde{\iota}, \tilde{\varphi}) +\gamma_{lmn}\sum_{l \neq l^{\prime}} & c_{l' l m}\  {}_{-2}Y_{l'm}(\tilde{\iota}, \tilde{\varphi})\nonumber \\
&+\mathcal{O}(\gamma_{lmn})^{2},
\end{align}
where $c_{l'lm}$ are the mixing coefficients between spin-weighted spheroidal harmonics and spin-weighted spherical harmonics with different $l'$ and $l$ but the same $m$ index. To test the angular emission patterns of remnant black holes, we compare the two models and check which one is more consistent with the QNM expansions. 

It is worth pointing out that the conventions for writing the QNM expansion in Eq.~\eqref{eq:S_model} are not all consistent in the existing literature, e.g., Refs~\cite{Berti2006Gravitational,London2014Modeling,Isi2021Analyzing}. We will briefly comment on them in Sec.~\ref{subsec:QNM_comments} and summarize in Table~\ref{tab:QNM_convention}.
For the reference of the readers, the notation and terminology specifically defined in this paper are listed in Appendix~\ref{app:notation}.

\subsubsection{Temporal and spatial dependences of the modes}

\begin{figure}[t]
	\begin{center}
		\includegraphics[width=\columnwidth]{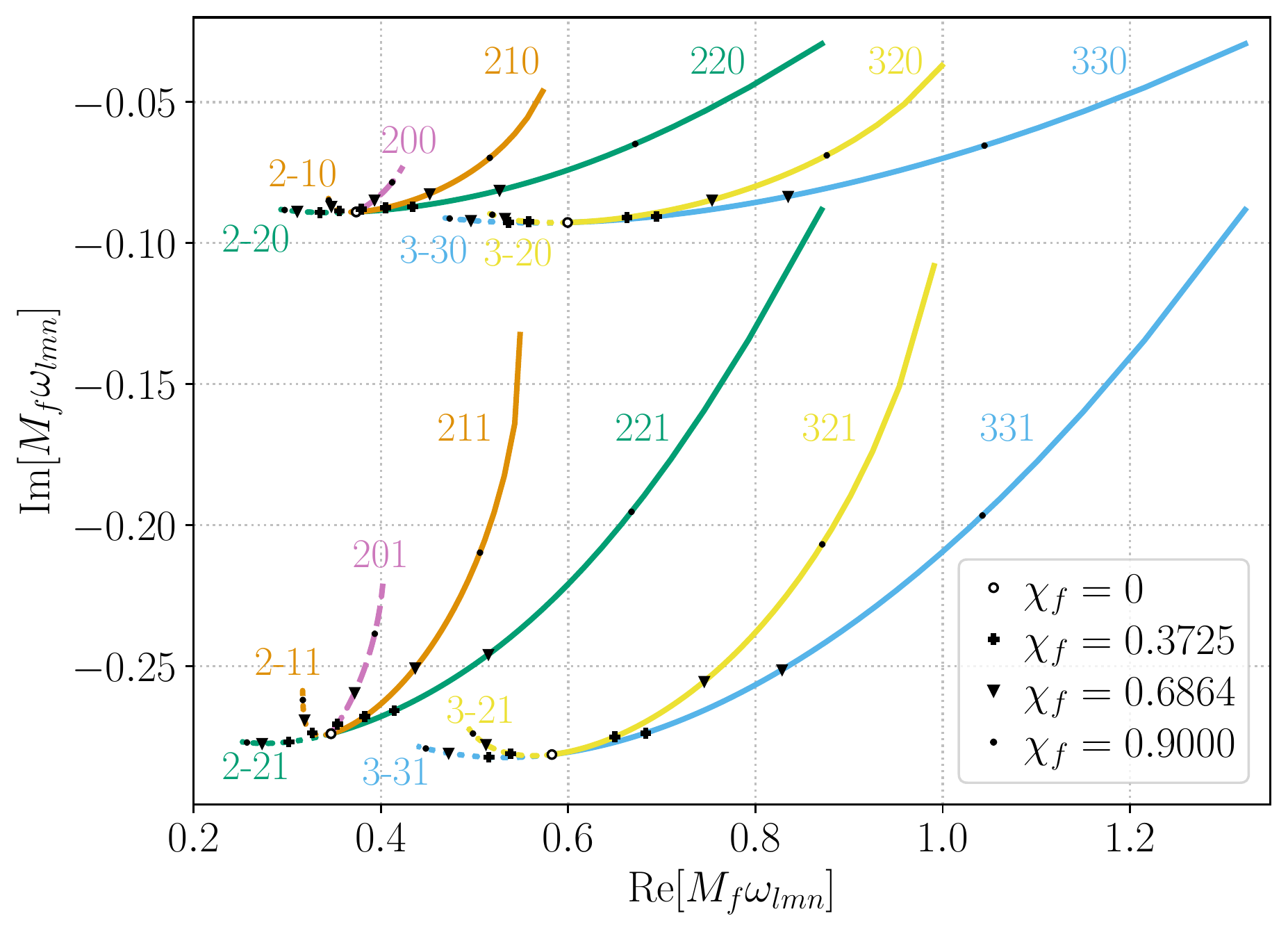}
		\caption{\label{fig:freq_branches} Examples of complex QMN frequencies for $\chi_f\in [0,0.99]$ (similar to Fig.~6 in Ref.~\cite{Berti2006Gravitational}). The horizontal axis represents the frequency (the real part) and the vertical axis represents the decay rate (the imaginary part). The solid, dashed, and dotted curves stand for prograde modes ($m>0$), $m=0$ modes and retrograde modes ($m<0$), respectively. Each curve is labeled by its $lmn$ indices, starting with the frequency value of $\chi_f=0$ and ending at the frequency value of $\chi_f=0.99$. When \mbox{$\chi_f=0$}, QNM frequencies with the same $l$ index coincide, as indicated by the black circles. We mark the frequencies corresponding to \mbox{$\chi_f=0.3725$} (N9; see waveform label in Table~\ref{tab:SXS_N}), \mbox{$\chi_f=0.6864$} (N1) and \mbox{$\chi_f=0.9$} (the example in Fig.~\ref{fig:Bpm_afpm}) by plus, inverted triangle and dot markers, respectively. For the modes with the same $lm$ indices, the ones with overtone $n=1$ (lower curves) have larger decay rates than the ones with $n=0$ (upper curves).}
	\end{center}
\end{figure}

The physical meanings of the parameters and terms in Eqs.\,\eqref{eq:S_model}\,and\,\eqref{eq:Y_model} are explained as follows:

\noindent {\it Time dependence.---} Each $\omega_{lmn}$ represents one complex frequency of the QNMs with index $lmn$, with the imaginary part being the decay rate. Among the QNMs, the $m>0$ modes are {\it prograde}, while the $m<0$ terms are {\it retrograde}. 
In this paper, we obtain $\omega_{lmn}$ numerically using the \texttt{qnm} python package~\cite{Stein:2019mop}. As we vary $\chi_f$, the trajectories of a selected set of $\omega_{lmn}$ appear as branches in the complex plane in Fig.~\ref{fig:freq_branches}. 
Real parts of prograde-mode frequencies (solid curves) increase with the spin magnitude $|\chi_f|$, while those of the retrograde-mode frequencies (dotted curves) decrease with $|\chi_f|$. Moreover, the decay rate increases with increasing overtone number $n$, as shown in the upper ($n=0$) and lower ($n=1$) parts of Fig.~\ref{fig:freq_branches}. 
Higher overtones play more important roles in the earlier stage of the ringdown~\cite{Berti2006Quasinormal,Berti2009Quasinormal,Giesler2019Black,Forteza2020Spectroscopy}.

\noindent {\it Spatial dependence.---} For prograde (retrograde) modes with indices $m>0$ ($m<0$), the $\{B^{(S+)}_{lmn}{}_{-2}S_{lmn}(\gamma_{lmn},\tilde{\iota}, \tilde{\varphi})\}$, $\{B^{(Y+)}_{lmn}{}_{-2}Y_{lm}(\tilde{\iota}, \tilde{\varphi})\}$ terms describe the emission mainly towards the north (south) hemisphere, while the $\{B^{(S-)}_{lmn}{}_{-2}S^*_{lmn}(\gamma_{lmn},\pi-\tilde{\iota}, \tilde{\varphi})\}$, $\{B^{(Y-)}_{lmn}{}_{-2}Y^*_{lm}(\pi-\tilde{\iota}, \tilde{\varphi})\}$ ones describe the emission mainly towards the south (north) hemisphere. The coefficients $B^{(S\pm/Y\pm)}_{lmn}$ are the corresponding excitations of the $lmn$ mode and are governed by the merging dynamics of the progenitor binary. The temporal and spatial features of different terms are summarized in Table~\ref{tab:QNM_convention}.

\begin{table*}[]
\begin{tabular}{l|c|c|c|c}
\hline
\hline
 & [a] & [b] & [c] & [d] \\
\hline 
Temporal-spatial profile & $ e^{-i\omega_{220} t} {} _{-2}S_{220} (\gamma_{220},\tilde\iota,\tilde\varphi)$ &
$ e^{i\omega_{220}^* t} {}_{-2}S_{220}^* (\gamma_{220},\pi-\tilde\iota,\tilde\varphi)$ &
$ e^{-i\omega_{2-20} t} {} _{-2}S_{2-20} (\gamma_{2-20},\tilde\iota,\tilde\varphi)$ & 
$  e^{i\omega^*_{2-20} t} {}_{-2}S_{2-20}^* (\gamma_{2-20},\pi-\tilde\iota,\tilde\varphi)$\\
Amplitude in Eq.~\eqref{eq:S_model} & $B_{220}^{(+)}$  & $B_{220}^{(-)} $
& $B_{2-20}^{(+)} $ & $B_{2-20}^{(-)}$ \\
Right/Left-handed (R/L) & R & L & R& L \\
Prograde/Retrograde & Prograde & Prograde & Retrograde & Retrograde \\
\begin{tabular}{l}
Emission direction\\
(in final spin frame)
\end{tabular}
& North & South & South & North \\
\hline
\begin{tabular}{l}
Angular eigenfunction\\
(standard form)
\end{tabular}
&
${}_{-2}S_{220} (\gamma_{220},\tilde\iota,\tilde\varphi)$ & 
${}_{-2}S_{2-20} (-\gamma_{220}^*,\tilde\iota,\tilde\varphi)$ &
${}_{-2}S_{2-20} (\gamma_{2-20},\tilde\iota,\tilde\varphi)$ &
${}_{-2}S_{220} (-\gamma_{2-20}^*,\tilde\iota,\tilde\varphi)$ \\ 
Amplitude in \cite{Berti2006Gravitational}\footnotemark[1] (2006) & $\mathcal{A}_{220}$ & $\mathcal{A}'_{220}$ (mirror mode of [a]) & $\mathcal{A}_{2-20}$ &  $\mathcal{A}'_{2-20}$ (mirror mode of [c])\\
Terminology in \cite{London2014Modeling} (2014) & Regular mode & Regular mode & Mirror mode of [b] & Mirror mode of [a] \\ 
Amplitude in \cite{Hughes2019Learning,Lim2019Exciting2} (2019) & $\mathcal{A}_{220}$ &  $\mathcal{A}_{220}'$ & $\mathcal{A}_{2-20}$ & $\mathcal{A}'_{2-20}$\\ 
Amplitude in \cite{LVC2020Tests} (2021) & $\mathcal{A}_{220}$ & $\mathcal{A}_{2-20}$  & (not included) & (not included) \\
Amplitude in \cite{Isi2021Analyzing} (2021) & $\mathcal{C}_{[1]220}$ & $\mathcal{C}_{[1]2-20}$ & $\mathcal{C}_{[-1]220}$ & $\mathcal{C}_{[-1]2-20}$ \\
Amplitude in \cite{Finch2021modelling} (2021) & $C_{220}$ & $C'_{2-20}$ (mirror mode of [c]) & $C_{2-20}$ & $C'_{220}$ (mirror mode of [a]) \\
Amplitude in \cite{Dhani2020Importance,Dhani2021Overtones} (2021) & $\mathcal{C}_{220}$ & $\mathcal{C}_{2-20}$ & $\mathcal{C}'_{2-20}$ (mirror mode of [b])  & $\mathcal{C}'_{220}$ (mirror mode of [a]) \\ 
\hline
\hline
\end{tabular}
\footnotetext[1]{Ref.~\cite{Berti2006Gravitational} denotes GW strain tensor as $h_{+}+i\,h_{\times}$, thus we include a sign change in angular frequency when comparing the convention in \cite{Berti2006Gravitational} with other papers.} 
\caption{\label{tab:QNM_convention}Conventions and notations for writing the QNM expansion, taking $(l,m,n)=(2,\pm 2, 0)$ for example.}
\end{table*}

\subsubsection{Excitations of prograde and retrograde modes}
\label{subsec:pro_retro}

\begin{figure}
\begin{center}
    \includegraphics[width=\columnwidth]{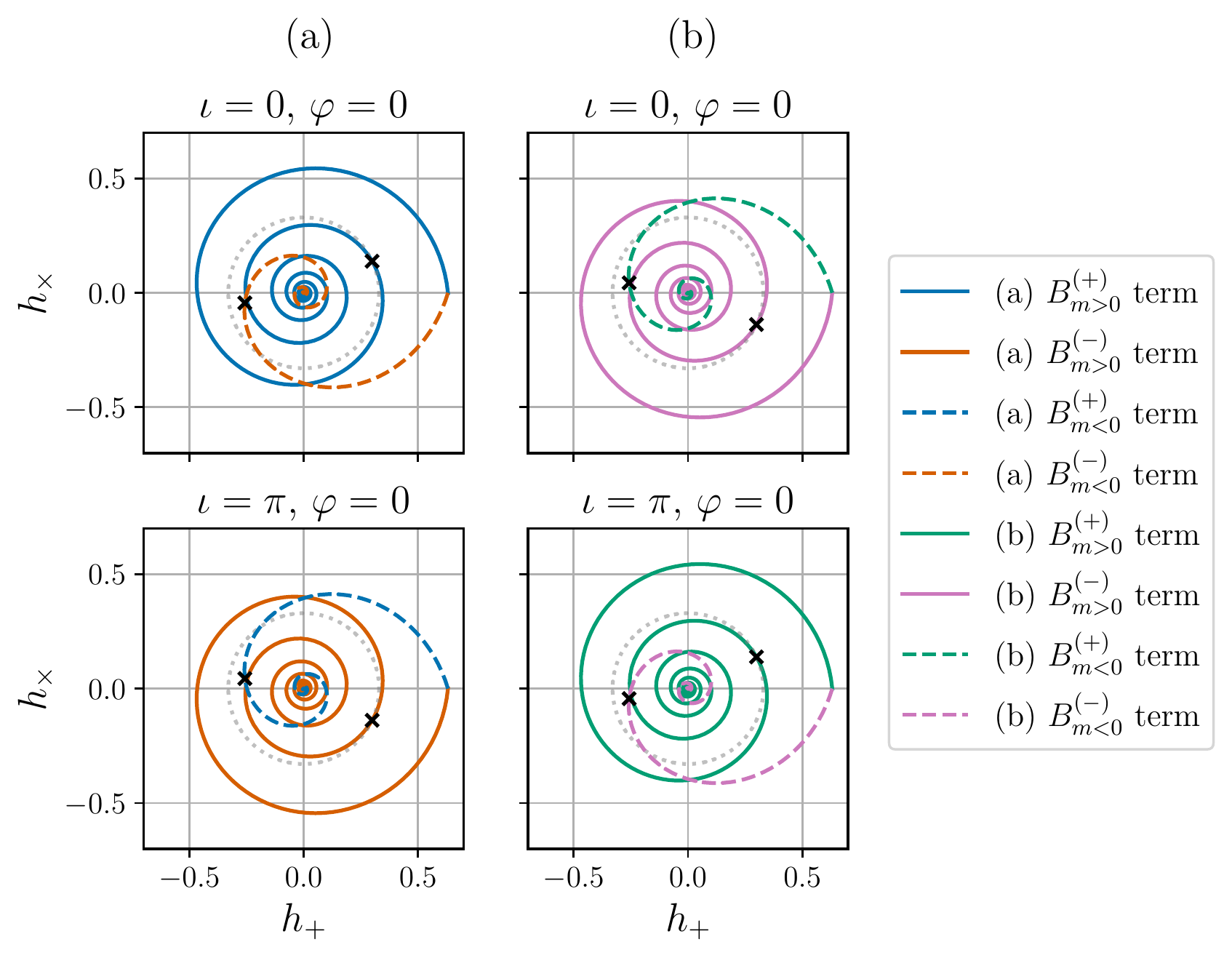}
    \caption{\label{fig:Bpm_afpm} 
    Polarization contents ($h_{+}(t),h_{\times}(t)$) of the template ringdown waveform indexed with $(l,m)=(2,2),(2,-2)$, observed from the north pole ($\iota=0,\varphi=0$) and south pole ($\iota=\pi,\varphi=0$) in the orbital frame for (a) spin aligned ($\chi_f=0.9$) and (b) spin anti-aligned ($\chi_f=-0.9$) cases. Each parametric curve starts at $t=0$ from ($h_{+},h_{\times}$)=(0.63,0) and evolve as a spiral over $t\in [0,100 M_f]$. The ``$\times$'' marker indicates the strain at $t=10 M_f$ and the gray dotted circle indicates the strain magnitude of prograde mode at that moment for reference. 
    The $B_{m>0}$ and $B_{m<0}$ terms represent the prograde and retrograde QNMs excitation, respectively. For $B_{m>0}$ ($B_{m<0}$), the ``$(+)$'' superscript stands for the emission in the same (opposite) direction as the spin. 
    For (a) spin aligned case, the $B_{m>0}^{(+)}$ and $B_{m<0}^{(-)}$ terms represent the emission towards the north pole direction; for (b) spin anti-aligned case, the $B_{m>0}^{(-)}$ and $B_{m<0}^{(+)}$ terms represent the emission towards the north pole direction. In either case, the $B_{m>0}^{(+)}$ and $B_{m<0}^{(-)}$ terms represent the emission towards the remnant spin direction.} 
\end{center}
\end{figure}

Let us now discuss the features of prograde and retrograde modes, and how they should be excited in a merging binary. In Fig.~\ref{fig:Bpm_afpm}, we illustrate the polarization contents of each ringdown mode by plotting in 2D graphs the $(h_+(t),h_\times(t))$ observed from the north ($\iota=0,\varphi=0$) and south ($\iota=\pi,\varphi=0$) poles in the orbital frame, for $t\in [0, 100M_f]$. For illustration purpose, we choose $\chi_f=0.9$, $l=2$, and $m =\pm 2$ in the plot. We assign the same, arbitrarily chosen starting amplitude for all the modes, $(h_+(0),h_\times(0)) = (0.63,0)$. Since the modes oscillate and decay over time, each mode traces a trajectory that spirals toward the center as time passes.

Here we emphasize that plotting $h_+(t)$ along the $x$ direction and $h_\times(t)$ along the $y$ direction in the graph (instead of the opposite) illustrates the way that the polarization patterns rotate: if a binary along with its emission pattern rotates about the $(\iota,\varphi)$ emission direction by an angle, say $\zeta$, following right-hand rule, the complex strain value $h =h_+ -i h_\times$ will become $e^{-2i\zeta}h =e^{-2i\zeta} (h_+ -i h_\times)$ and the pattern of $(h_+(t),h_\times(t))$ in the plot will rotate by $2\zeta$ counterclockwise.


\noindent {\it Spin aligned case.---} In the left panels (a) of Fig.~\ref{fig:Bpm_afpm}, we study the spin aligned case, with $\chi_f > 0$. At the north pole ($\iota=0$, upper left panel), we observe that the $B_{m>0}^{(+)}$ (blue solid curve) and $B_{m<0}^{(-)}$ (orange dashed curve) terms correspond to the counterclockwise and clockwise trajectories in the $(h_+,h_\times)$ plane, respectively. It is important to note that the other two terms, with amplitudes $B_{m>0}^{(-)}$ and $B_{m<0}^{(+)}$, vanish at the north pole due to the properties of spin-weighted spheroidal harmonics. 
The counterclockwise (clockwise) trajectory corresponds to the positive (negative) value of the real part in the QNM frequency of the $B_{m>0}^{(+)}$ ($B_{m<0}^{(-)}$) term. In this case, the spin of the black hole is counterclockwise, therefore $B_{m>0}^{(+)}$ ($B_{m<0}^{(-)}$) corresponds to a prograde (retrograde) pattern of polarization rotation. One can also see that the prograde $B_{m>0}^{(+)}$ mode spirals decay ``slower'' toward the center than the retrograde $B_{m<0}^{(-)}$ mode (the blue solid curve reaches the gray dotted circle at $t=10 M_f$ while the orange dashed curve reaches it before $t=10 M_f$), because the prograde mode has a higher quality factor (the ratio between absolute values of real and imaginary parts of the QNM frequency). 

At the south pole ($\iota=\pi$, lower left panel), again according to properties of the spin-weighted spheroidal harmonics, we only observe the  $B_{m>0}^{(-)}$ (solid orange curve) and $B_{m<0}^{(+)}$ (dashed blue curve) terms, which represent the prograde and retrograde modes, respectively.  In comparison with the view from the north pole, the prograde (retrograde) mode now spirals clockwise (counterclockwise). This is consistent with the fact that the black hole rotates clockwise when viewed from the south pole. Mathematically, the signs of the real parts of the eigenfrequencies are flipped, while the imaginary parts remain unchanged.

\noindent {\it Spin anti-aligned case.---} In the right panels (b) of Fig.~\ref{fig:Bpm_afpm}, we study the spin anti-aligned case, with $\chi_f < 0$. We still follow an expansion in the final spin frame $(\tilde \iota,\tilde\varphi)$ using Eq.~\eqref{eq:S_model} with dimensionless spin equal to $|\chi_f|$, but now need to carry out the transformation $(\iota=\pi-\tilde\iota,\varphi = 2\pi-\tilde\varphi)$ to obtain the complex strain $h$ in the orbital frame. Because of this transformation, at the north pole ($\iota=0$, upper right panel), we observe the $B_{m>0}^{(-)}$ (solid magenta curve) and $B_{m<0}^{(+)}$ (dashed green curve) modes. They still correspond to prograde and retrograde modes, respectively, although now the prograde (retrograde) mode has frequency with negative real part and a spiral pattern that goes clockwise (counterclockwise). These are consistent with the fact that the spin direction of the black hole is clockwise viewed from the north pole. 

At the south pole ($\iota = \pi$, lower right panel), we observe the $B_{m>0}^{(+)}$ (solid green curve) and $B_{m<0}^{(-)}$ (dashed magenta curve) modes, which correspond to prograde and retrograde modes, respectively. The black hole spins counterclockwise when viewed from the south. 

It is worth pointing out that both the remnant spin direction and the excitation of prograde or retrograde mode are determined by the binary dynamics. They are related to each other because the polarization patterns of the inspiral wave would transition smoothly to the ringdown modes. During the inspiral stage, the orbital motion of the binary appears counterclockwise (clockwise) when viewed from the north (south) pole. When the progenitor black holes are nonspinning, the remnant spin is contributed solely from the orbital angular momentum and thus aligned with the orbital angular momentum. Therefore, we anticipate the prograde modes to be more strongly excited. On the other hand, when the initial spin is anti-aligned with the orbital angular momentum, there would be competition between the spin and orbital angular momentum during the merger. The left-over stronger contribution will determine the direction of remnant spin, as well as the excitation of prograde or retrograde modes. Specifically, when the contribution from the negative individual spin is larger, the remnant black hole is left with a spin anti-aligned with the orbital angular momentum, and we anticipate stronger excitations for retrograde modes. 

\subsubsection{QNM conventions}
\label{subsec:QNM_comments}

Before moving on to the following discussion, we briefly comment on the conventions for writing the QNM expansion in Eq.~\eqref{eq:S_model} or \eqref{eq:Y_model} in this paper and other literature. 
The QNM expansion consists of four parts corresponding to the combinations of two frequencies ($\omega_{lmn}$ and $-\omega^*_{lmn}$), as well as the prograde and retrograde modes. 

In our convention, we first add up the terms with conjugate frequencies explicitly, and then sum up prograde (the expressions with positive $m$'s) and retrograde modes (with negative $m$'s) over all the $lmn$ indices through the summation signs. 
The convention and summation order in Ref.~\cite{Isi2021Analyzing} is similar to ours. In Eq.~(1) of \cite{Isi2021Analyzing}, the $p=1$ terms represent the prograde modes (our $m>0$ terms), and $p=-1$ terms represent the retrograde modes (our $m<0$ terms). When the retrograde modes are not considered, the Eq.~(3) of Ref.~\cite{Isi2021Analyzing} is equivalent to our $m\geq 0$ terms in Eq.~\eqref{eq:S_model}, with $(-1)^l$ absorbed into the $B_{lmn}^{(S-)}$ coefficients to be determined. 

In an alternative convention, one can sum up the terms with the same sign of frequencies first. Specifically, the terminology of ``mirror mode'' was first mentioned in~\cite{Berti2006Gravitational} to describe the feature that half of the QNM frequencies are ``degenerate in modulus of the frequency and damping time,'' as shown in Fig.~6 of \cite{Berti2006Gravitational}. 
Later in Ref.~\cite{London2014Modeling}, the terminology ``mirror modes'' was used to refer to the retrograde modes with a different frequency sign from the corresponding prograde modes; while in Ref.~\cite{Finch2021modelling}, it was used to refer to all terms with negative frequencies. 
In recent work presented in~\cite{Dhani2020Importance,Dhani2021Overtones}, the same terminology ``mirror modes" is used to describe the retrograde modes that emit to the same direction as the considered prograde modes (Fig.~1 of Ref.~\cite{Dhani2020Importance}), while the remaining terms with the conjugate frequencies are taken into account by symmetry.
Since there is no consistent definition of mirror modes yet, we choose not to use such terminology and, instead, define and describe the prograde and retrograde modes, as well as the terms with conjugate frequencies explicitly.
We summarize the conventions and notations for writing the QNM expansion in some existing literature in Table~\ref{tab:QNM_convention}. More details are given in Appendix~\ref{app:QNM_convention}.

\subsubsection{Beyond linear ringdowns}

We finally point out that in addition to QNMs, the gravitational waveform after the merger phase also contains power-law tails~\cite{Ching1995Late} and gravitational wave memory~\cite{braginsky1987gravitational,Christodoulou1991Nonlinear,Thorne1991Gravitational}. Power-law tails arise from the long-range nature of the black hole's gravitational potential. 
This contribution decay with time following a power law and is generally believed to be negligible for binary black hole coalescence. 

Gravitational wave memory originally refers to the change in spacetime geometry at future null infinity before and after the passage of a transient gravitational wave. Linear memory refers to changes that can be related to the initial and final momentum distributions of the gravitational wave source~\cite{braginsky1987gravitational}, while non-linear memory is induced by the non-linearity of the Einstein's equations~\cite{Christodoulou1991Nonlinear}; it can be further interpreted as arising from the gravity effect caused by the energy and momentum of the gravitons~\cite{Thorne1991Gravitational}. For compact binaries, the memory waveform also refers to a non-oscillatory component of the total waveform that starts off at zero and gradually reaches the final value equal to the gravitational wave memory~\cite{Talbot2018Gravitational}. Memory waveform can be computed from those waveforms obtained from perturbation theory that do not account for the memory effect~\cite{Blanchet1992Hereditary,Favata2010gravitational}.  

Recent numerical simulations have also been able to decompose the full waveform at null infinity into a memory piece and a memory-free piece~\cite{Mitman2020Computation}. It is also shown that the previous SXS waveforms (including those approximated by the NR surrogate models) correspond to the memory-free piece. In this paper, we use the memory-free waveforms, and show that they can be decomposed into QNMs, in terms of both the temporal and spatial distributions. The memory waveform, on the other hand, {\it cannot} be decomposed into QNMs.

\subsection{\label{sec:fitting_method} Fit memory-free NR waveforms with QNM expansions}


We now develop a strategy to use QNM expansions~\eqref{eq:S_model} and~\eqref{eq:Y_model} to {\it fit} memory-free NR waveforms. In our strategy, we do not focus on one wave-emission direction, or one particular $(l,m)$ mode, but rather consider the joint temporal and spatial (angular) dependence of the ringdown gravitational waves. More specifically, for each binary, and its ringdown waveform starting from a particular time $t_0$, we find the optimal set of parameters $(M_f,\chi_f)$ and coefficients $\{B_{lmn}^{(S\pm/Y\pm)}\}$ with which the expansion in Eqs.~\eqref{eq:S_model} or \eqref{eq:Y_model} best describes the $h(\vec \Omega,t)$ obtained from numerical relativity. From this approach, we are able to determine whether the $S$ or the $Y$ model is the more faithful representation of the ringdown gravitational waves. 

\subsubsection{Target waveforms and templates}

In this paper, our (memory-free) {\it target waveform} $h(\vec \Omega,t)$ is obtained from the SXS catalog, which provides $h$ in terms of the expansion in spin-weighted spherical harmonics:
\begin{equation}
    h(\vec\Omega,t) = \sum_{lm} {}_{-2}Y_{lm}(\vec\Omega)h_{lm}\,(t).
\end{equation}
For each binary, we denote the time at which $\sum_{lm}|h_{lm}(t)|^2$ is maximum by $t_{\rm peak}$. We then use {\it templates} in the forms of QNM expansion, either  Eq.~\eqref{eq:S_model} or Eq.~\eqref{eq:Y_model}, to approximate the target waveform during $t \in [t_0,+\infty)$\footnote{Practically, for all value of $t_0$, we consider the waveform till $t_{\rm peak}+100M$ when it essentially damps to $0$.}. Here $t_0$ is a starting time not far from $t_{\rm peak}$, with an offset
\be\label{eq:time_offset}
t_{\rm offset}=t_0 - t_{\rm peak}.
\ee
In order to evaluate the quality of the fit, we first define a temporal-spatial inner product of the target waveform $h(\Vec{\Omega},t)$ and the template waveform $g(\Vec{\Omega},t)$ by doing a double integral over both time and angular coordinates:
\be\label{eq:2D_innerproduct}
\left\langle g \mid h\right\rangle = \int  d^2\vec\Omega \int_{t_0}^{+\infty}  dt \left[ g^*(\Vec{\Omega},t)h(\Vec{\Omega},t)\right].
\ee
We can then characterize the distance between $h(\Vec{\Omega},t)$ and $g(\Vec{\Omega},t)$ using
\be\label{eq:distance}
\chi^{2}[h,g]=\frac{\left\langle h-g \mid  h-g\right\rangle }{ \left\langle h \mid  h\right\rangle}.
\ee


For each binary configuration, the distance $\chi^2$ depends on the starting time $t_0$ of ringdown fit, the mass and dimensionless spin $(M_f,\chi_f)$ of the remnant black hole, the set of QNM modes summed over (including whether we use the $S$ or the $Y$ model), and their complex amplitudes $\{B_{lmn}^{(\pm)}\}$. The parameters $(M_f,\chi_f)$ are intrinsic and need to be optimized numerically, the mode amplitudes $\{B_{lmn}^{(\pm)}\}$ 
can be optimized analytically, while the starting time $t_0$ works as the hyperparameter that controls the fitting. 

\subsubsection{Strategy for minimizing $\chi^2$ over $\{B_{lmn}^{(\pm)}\}$}

As it turns out, minimization of $\chi^2$ over $\{B_{lmn}^{(\pm)}\}$ can be carried out analytically, because $\chi^2$ can be viewed as squared distance between $h$ and a linear subspace spanned by the spin-weighted harmonic basis by varying coefficients/coordinates $\{B_{lmn}^{(\pm)}\}$. 

Let us do this explicitly for the $S$ model, and the procedure for the $Y$ model can be obtained by simply switching $S$ to $Y$. Let us first define the following quantities for inner products: 
\begin{align}
    K_{lmn}^{(\sigma)} & = \int d^2\vec\Omega \int_{t_0}^{+\infty} dt\, g_{lmn}^{(\sigma)\,*}(\vec\Omega,t) h(\vec\Omega,t), \\
    G_{lmn,l'm'n'}^{(\sigma,\sigma')} & =\int d^2\vec\Omega \int_{t_0}^{+\infty} dt\, g_{lmn}^{(\sigma)\,*}(\vec\Omega,t) g_{l'm'n'}^{(\sigma')}(\vec\Omega,t).
\end{align}
Here $\sigma = \pm$, and we further define
\begin{align}
g_{lmn}^{(+)}(\vec\Omega,t) &= e^{-i\omega_{lmn}(t-t_0)}\,{}_{-2}S_{lmn}(\gamma_{lmn},\tilde\iota,\tilde\varphi),\\
g_{lmn}^{(-)}(\vec\Omega,t) &= e^{i\omega_{lmn}^*(t-t_0)}\,{}_{-2}S_{lmn}^*(\gamma_{lmn},\pi-\tilde\iota,\tilde\varphi).
\end{align}
Note that $(\tilde\iota,\tilde\varphi)$ need to be converted to $(\iota,\varphi)$ depending on whether we have spin aligned or spin anti-aligned binaries. We can now view $B_{lmn}^{(\sigma)}$ and $K_{lmn}^{(\sigma)}$ as the components of column vectors $\vec{B}$ and $\vec{K}$, and $G_{lmn,l'm'n'}^{(\sigma,\sigma')}$ as those of a Hermitian matrix $\mathbb{G}$. The inner product in Eq.\,\eqref{eq:2D_innerproduct} can thus be written as: 
\begin{equation}\label{eq:gh_gg_inner}
    \langle g | h\rangle = \vec{B}{}^\dagger \vec{K}\,,\quad \langle g |g\rangle =\vec{B}{}^\dagger \mathbb{G} \vec{B}.
\end{equation}
This leads to the least-squares value of distance
\begin{equation}\label{eq:chi2_ls}
    \chi^2_{\rm l-s} = 1- \frac{\vec{K}^\dagger \mathbb{G} \vec{K}}{\langle h | h\rangle},
\end{equation}
which is achieved when the coefficients satisfy
\begin{equation}\label{eq:coef_ls}
\vec{B} = \mathbb{G}^{-1}\vec{K}.
\end{equation}
We then search the 2D parameter space of $(M_f,\chi_f)$ to find the best estimates of $(M_{f,\rm est},\chi_{f,\rm est})$ that yield the optimal distance $\chi^{2}[h,g_{\rm opt}^{(S/Y)}]$ for $S$ and $Y$ models, respectively. Details of the numerical implementation can be found in Appendix~\ref{app:regression_algorithm}.



\begin{table}[t]
\caption{\label{tab:SXS_N}SXS BBH waveforms used in Sec.~\ref{sec:nonspinning}.}
\begin{ruledtabular}
\begin{tabular}{cccccccc}
Label & SXS ID/Lev\footnotemark[1] &$q_{\rm ref}$\footnotemark[2] &$(\vec{\chi}_{\rm ref,1})_z$\footnotemark[2] &  $(\vec{\chi}_{\rm ref,2})_z$\footnotemark[2] & $\chi_{\rm eff}$ & $(\vec \chi_f)_z$\footnotemark[2]${}^,$\footnotemark[3]\\
\hline
G0 & 0305/Lev6 & 1.221 & 0.3300 &  $-$0.4399 & $-$0.0166 & 0.6921 \\
N1 & 1154/Lev3 & 1.000 & 0.0000 & 0.0000 & 0.0000 & 0.6864 \\
N2 & 1143/Lev3 & 1.250 & $-$0.0001 & 0.0000 & $-$0.0001 & 0.6795 \\
N3 & 0593/Lev3 & 1.500 & 0.0000 & 0.0001 & 0.0001 & 0.6641 \\
N4 & 1354/Lev3 & 1.832 & $-$0.0002 & 0.0001 & $-$0.0001 & 0.6377 \\
N5 & 1166/Lev3 & 2.000 & 0.0000 & 0.0000 & 0.0000 & 0.6234 \\
N6 & 2265/Lev3 & 3.000 & 0.0000 & 0.0000 & 0.0000 & 0.5406 \\
N7 & 1906/Lev3 & 4.000 & 0.0001 & $-$0.0001 & 0.0000 & 0.4718 \\
N8 & 0187/Lev3 & 5.039 & 0.0000 & 0.0000 & 0.0000 & 0.4148 \\
N9 & 0181/Lev4 & 6.000 & 0.0000 & 0.0000 & 0.0000 & 0.3725 \\
\end{tabular}
\end{ruledtabular}
\footnotetext[1]{All SXS waveforms~\cite{Boyle2019SXS} used in this paper have the ID type ``BBH\_SKS'', and the levels listed are the maximum available ones.} 
\footnotetext[2]{The initial values are taken at the reference time after junk radiation~\cite{Lovelace2009Reducing}. At an accuracy level of $10^{-4}$, the spin components in $\hat x$ and $\hat y$ directions are zero for all the primary, secondary and remnant black holes listed here.}
\footnotetext[3]{For simplicity, in the main text we use $\chi_f$ to represent $(\vec \chi_f)_z$.}
\end{table}

\section{\label{sec:nonspinning}Benchmark and Nonspinning Binaries}

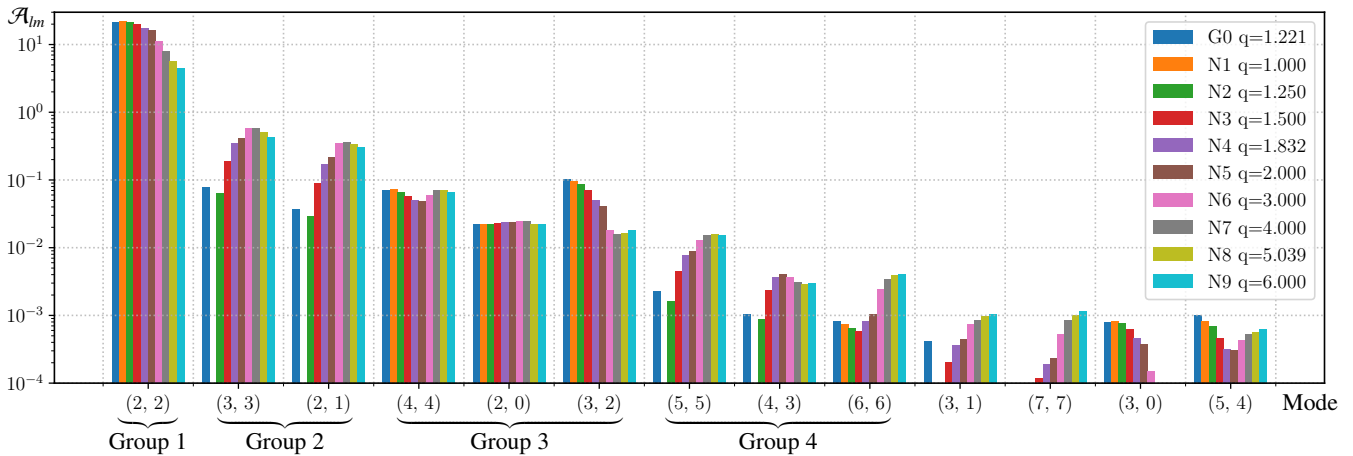
\begin{figure*}[t]
\begin{center}
\begin{overpic}[width=0.995\textwidth]{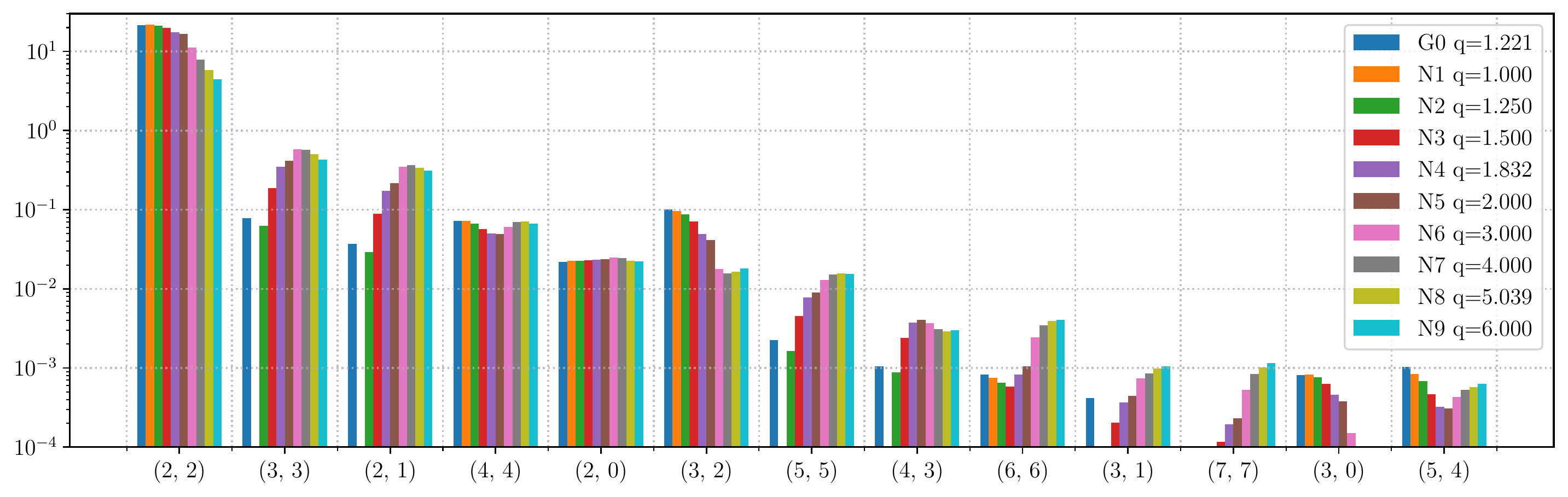}
\put(1,30){{$\mathcal{A}_{lm}$}}
\put(96,1.2){{Mode}}
\put(8.5,-0.5){$\begin{matrix} \underbrace{\hspace{0.8cm}} \\ \textrm{Group 1}\end{matrix}$}
\put(16.6,-0.5){$\begin{matrix} \underbrace{\hspace{1.8cm}} \\ \textrm{Group 2}\end{matrix}$}
\put(30,-0.5){$\begin{matrix} \underbrace{\hspace{3cm}} \\ \textrm{Group 3}\end{matrix}$}
\put(50,-0.5){$\begin{matrix} \underbrace{\hspace{3cm}} \\ \textrm{Group 4}\end{matrix}$}
\end{overpic}
\vspace{0.1cm}
\caption{\label{fig:hpc_Elm_nonspinning} 
The relative importance $\mathcal{A}_{lm}$, defined in Eq.\,\eqref{eq:Elm} as the strain component in spin-weighted spherical mode $(l,m)$ squared and integrated from $t_{\rm peak}$ to $t_{\rm peak}+ 100M$. Groups 1--4 of the $(l,m)$ modes are defined according to their relative importance in the QNM expansion, and are added to the fitting models in order. See details in Sec.~\ref{sec:strategy_lm}.} 
\end{center}
\end{figure*}

In this section, we focus on spin aligned binaries, for which the spin of the remnant black hole is in the same direction as the binary orbital angular momentum. For both the $S$ and $Y$ models of QNM expansion, starting from different times (described by the hyperparameter $t_{\rm offset}$), we compare the optimal distance $\chi^{2}[h,g_{\rm opt}^{(S/Y)}]$ ($\chi^2$ for short in the plot labels) and the corresponding $(M_{f,\rm est},\chi_{f,\rm est})$, when different groups of angular $(l,m)$ modes and overtones are included.
We demonstrate that, as expected, the $S$ model describes the NR waveforms better than the $Y$ model. 
We carry out this study firstly for a ``benchmark binary'' waveform, G0 (with parameters similar to the first GW event GW150914~\cite{LVC2016GW150914,Lovelace2016Modeling,Boyle2019SXS}), and then nine non-spinning binaries with mass ratio $q$ ranging from 1 to 6, as listed in Table~\ref{tab:SXS_N}.

In Sec.~\ref{sec:strategy_lm}, we introduce the our strategy for choosing which angular $(l,m)$ modes to include in the QNM expansion. In Sec.~\ref{sec:results_G0}, we present the fitting results for the benchmark binary waveform G0. In Sec.~\ref{sec:nonsp_qs}, we compare the results for binaries with different mass ratios.

\subsection{\label{sec:strategy_lm}Strategy for choosing angular modes}


Even though the quadrupole $(l,m)=(2,2)$ mode is the dominant component of the inspiral, merger and ringdown waves, gravitational wave detectors at present and in the future are capable of detecting higher multipole modes that are also excited~\cite{Healy2013Template}.  This capability is the foundation for our studies in this paper.  The NR waveforms from the SXS catalog include all $(l,m)$ modes up to $l_{\rm max}=8$~\cite{Boyle2019SXS}. 
However, incorporating too many modes in the QNM expansion will eventually lead to overfitting and numerical noise. We need a strategy to include the appropriate angular modes, which should be based on: (i) the strength of excitation of the modes, and (ii) the accuracy of the NR waveforms, and the sensitivity of our detectors.

Let us now address (i) above, while (ii) will be discussed at the end of Sec.~\ref{sec:results_G0}. To investigate the excitation strength of each $(l,m)$ mode, we can rank them in terms of their \emph{relative importance} for the QNM expansion:
\begin{equation}\label{eq:Elm}
    \mathcal{A}_{lm}=\int_{t_{\rm peak}}^{t_{\rm peak}+100M}dt \left|h_{l m}(t)\right|^{2}.
\end{equation}
For G0 and N1--N9 binaries, we plot their $\mathcal{A}_{lm}$  as vertical bars in Fig.~\ref{fig:hpc_Elm_nonspinning}. According to the order of magnitude of $\mathcal{A}_{lm}$ for all $q$'s, the $(l,m)$ modes can be qualitatively categorized into four groups:
\begin{equation*}
\begin{split}
    &\textrm{Group 1: } (2,2),\\
    &\textrm{Group 2: } (3,3),(2,1),\\
    &\textrm{Group 3: } (4,4),(2,0),(3,2),\\
    &\textrm{Group 4: } (5,5),(4,3),(6,6).
\end{split}
\end{equation*}
In our studies, instead of testing each of the individual modes, we add modes to our QNM expansion by groups, from Group 1 to Group 4 sequentially. The contribution of each group can thus be quantified by comparing the fitting results before and after adding it. 

Before moving on to the fitting, let us comment that the magnitude of the $\mathcal{A}_{lm}$'s can be traced qualitatively to the excitation of the corresponding $(l,m)$ mode during the inspiral stage. The initial magnitudes of $(l,m)$ modes are determined by (a) their post-Newtonian order, i.e., higher $(l,m)$ modes are generally weaker, and (b) the suppression due to symmetry, i.e., the $(l,m)=(3,3),(2,1),(5,5),(4,3)$ modes are prohibited for equal mass ratio binaries (e.g. N1). As $\mathcal{A}_{lm}$ is an integral over time, the ranking of each $\mathcal{A}_{lm}$ is determined jointly by the initial magnitude and the decay rate of that specific mode.

\subsection{\label{sec:results_G0}Fittings results for the benchmark binary G0}

\begin{figure*}
\centering
\includegraphics[width=\textwidth]{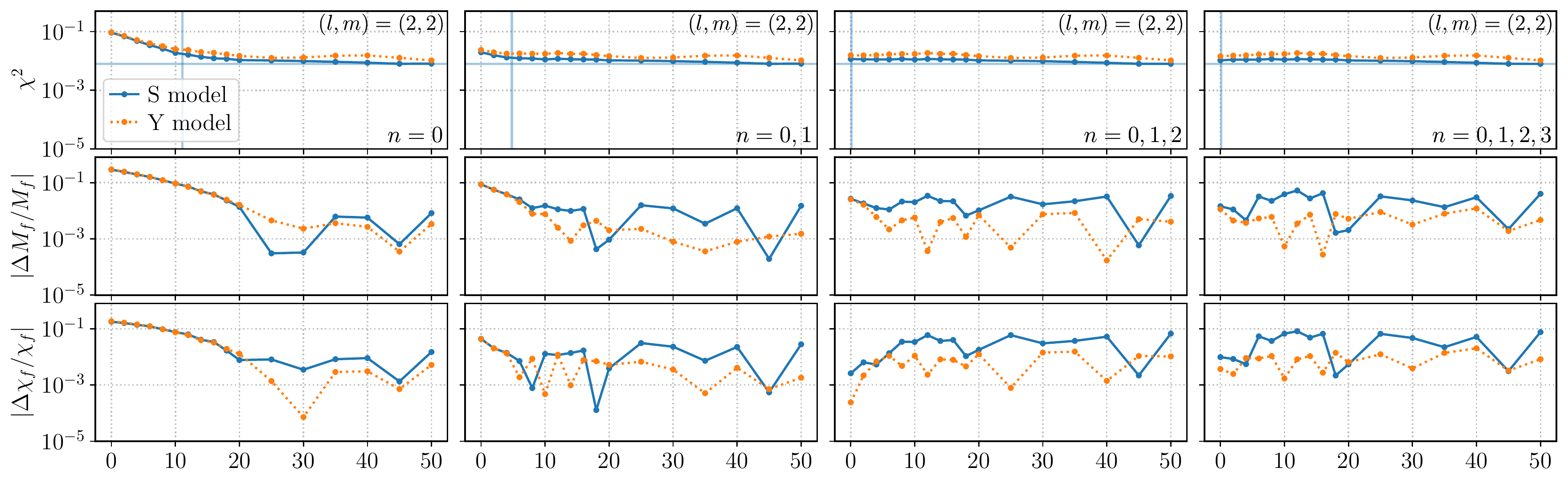}
\includegraphics[width=\textwidth]{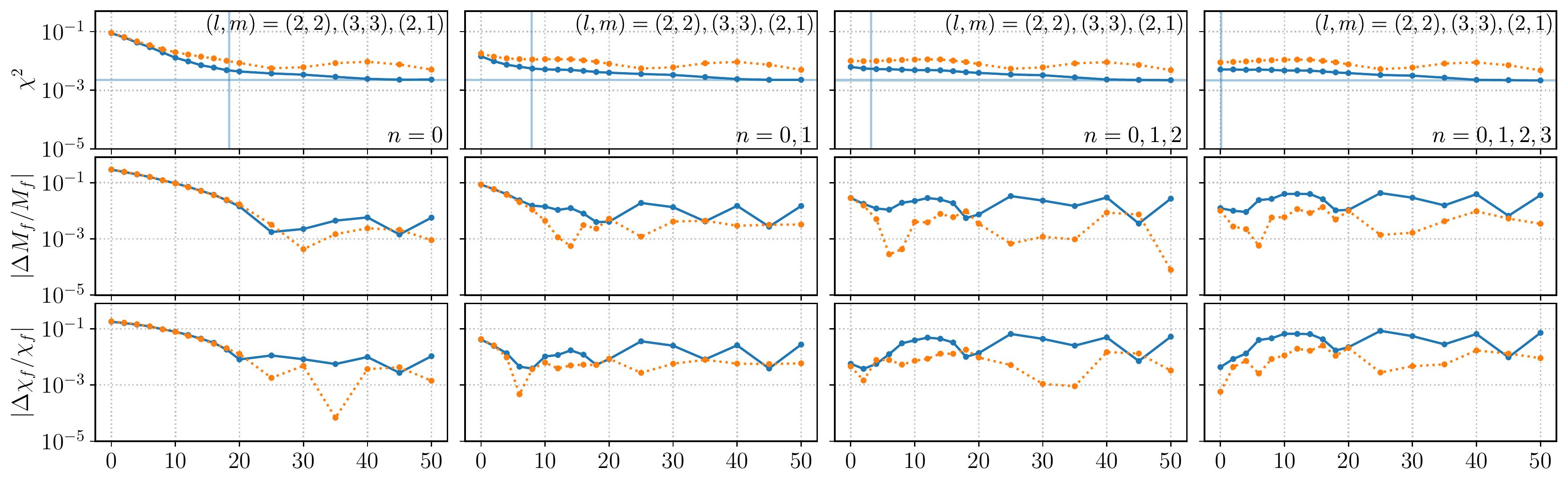}
\includegraphics[width=\textwidth]{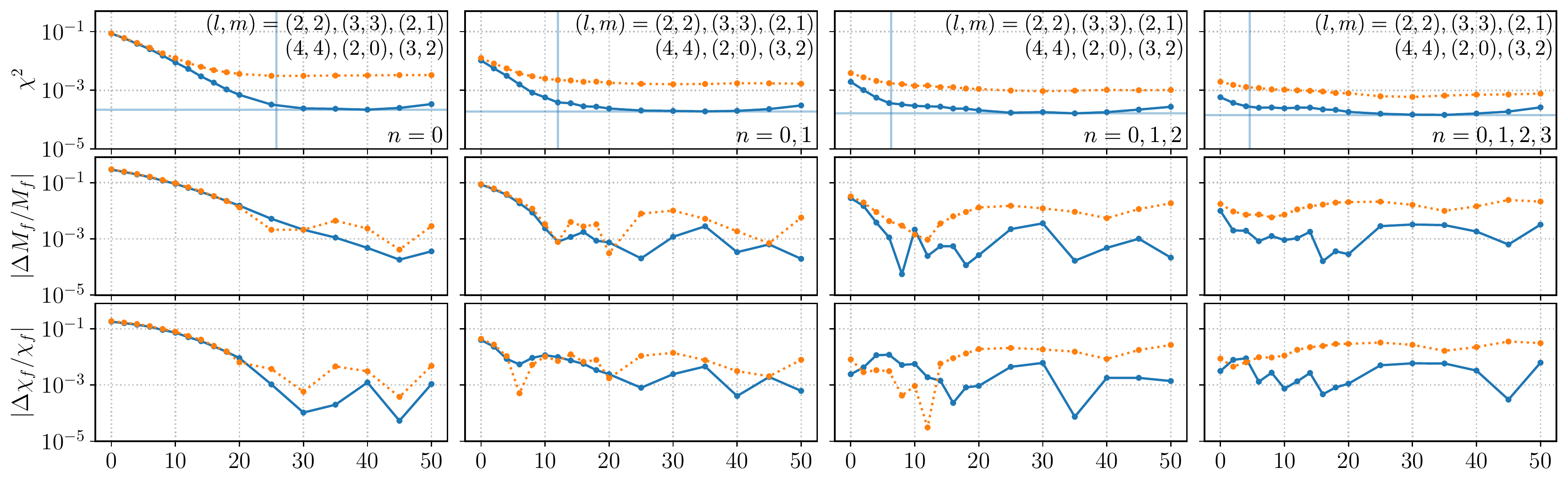}
\includegraphics[width=\textwidth]{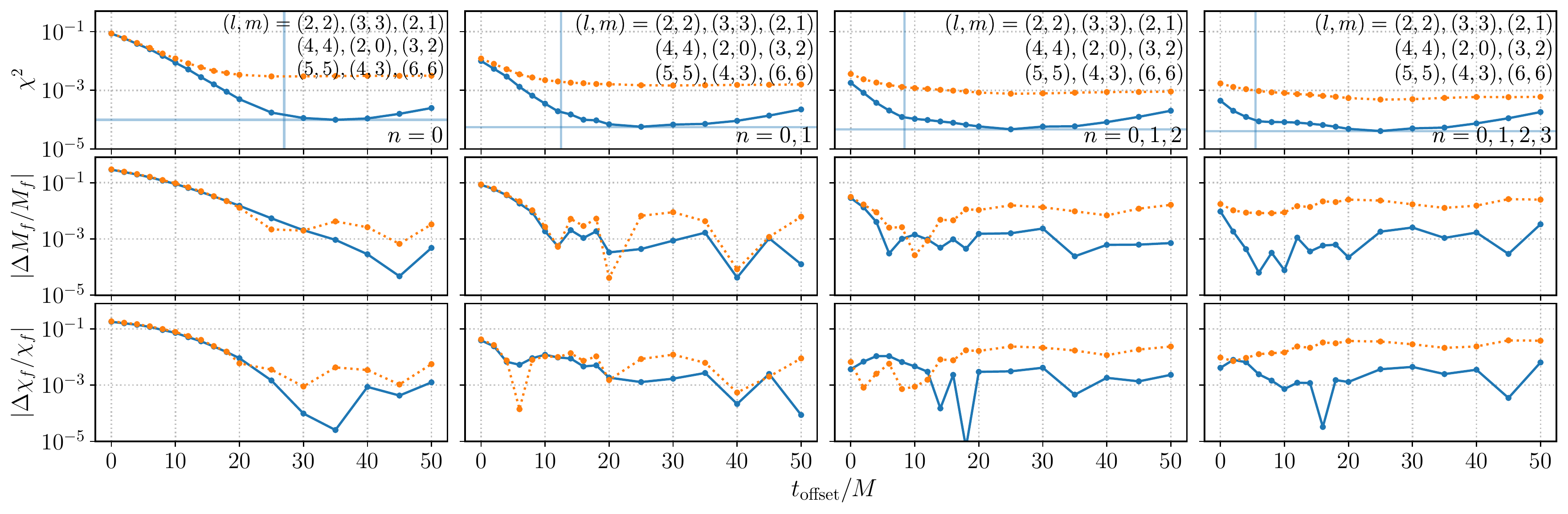}
\vspace{-0.8cm}
\caption{\label{fig:G0_ots}Fitting results for the benchmark binary G0. Each three closely laid blocks show the optimal distance $\chi^2$ and the relative error in estimated $(M_f,\chi_f)$, with respect to $t_{\rm offset}/M$. The solid blue and dotted orange curves correspond to the $S$ and $Y$ models, respectively. Different rows, from top to bottom, correspond to adding $(l,m)$ modes sequentially in groups, as specified in the upper right corner of each $\chi^2$ block. Within each row, different columns, from left to right, correspond to adding overtones for the same set of $(l,m)$ modes, as specified in the lower right corner of each $\chi^2$ block.}
\end{figure*}

\begin{figure*}
\centering
\includegraphics[width=\textwidth]{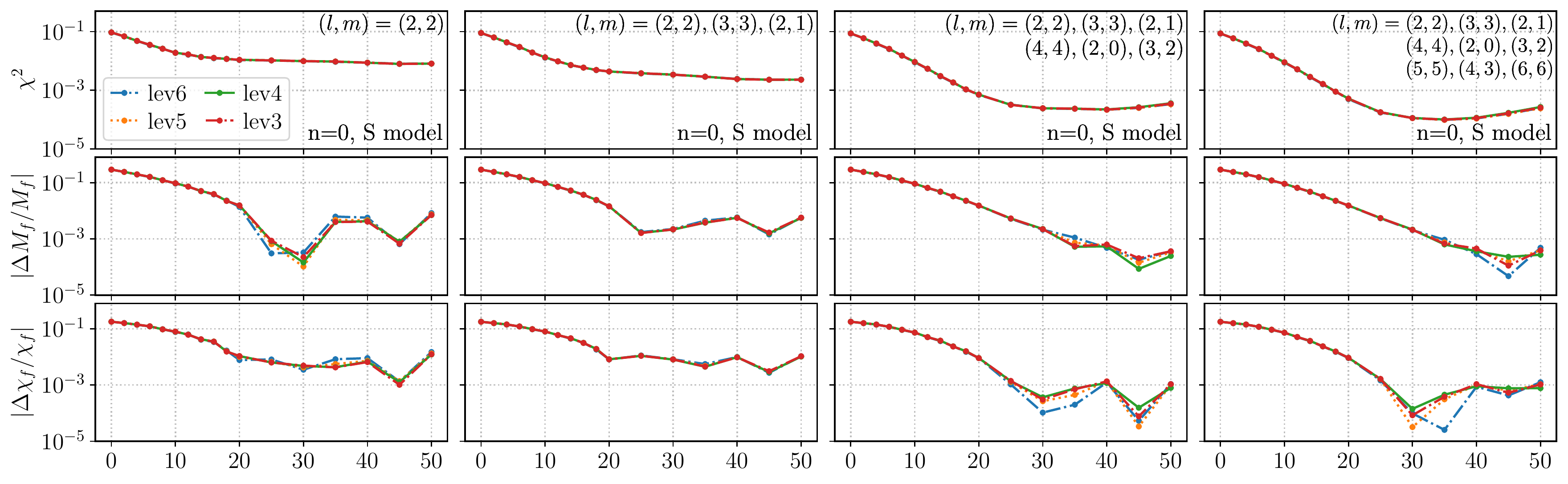}
\includegraphics[width=\textwidth]{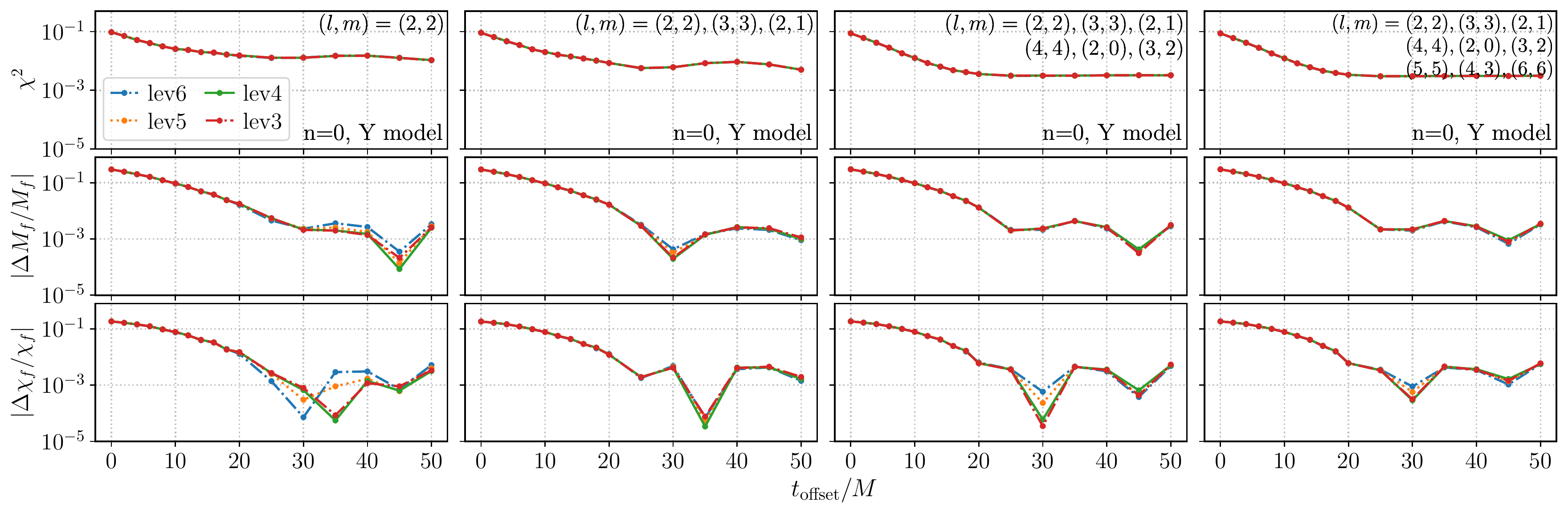}
\caption{\label{fig:G0_ot0_levs} Fitting results for G0 using SXS data with different numerical levels (presented in different colors).
The first and second rows correspond to fitting with the $S$ and $Y$ models, respectively. Within each row, the columns, from left to right, correspond to adding $(l,m)$ modes sequentially in groups. In this comparison, the models only contain the $n=0$ fundamental modes without including overtones.}
\end{figure*}

In this section, we discuss the fitting results for the benchmark binary G0. This binary waveform is used to verify the fitting algorithm described in Sec.~\ref{sec:fitting_method}, as it has the best numerical precision in the SXS catalog to date~\cite{LVC2016GW150914,Lovelace2016Modeling,Boyle2019SXS}. We first investigate the distinguishability of the $S$ and $Y$ models and the contributions of higher-order angular modes and overtones. We then comment on the fitting error by comparing the results obtained using waveforms at different numerical resolution levels. 

To characterize the accuracy of $(M_{f,\rm est},\chi_{f,\rm est})$, we define the relative errors as follows:
\begin{subequations}\label{eq:relative_PE_err}
\begin{align}
  &|\Delta M_f/M_f| \equiv |(M_{f,\rm est}-M_{f,\rm true})/M_{f,\rm true}|,\\
  &|\Delta \chi_f/\chi_f| \equiv |(\chi_{f,\rm est}-\chi_{,\rm true})/\chi_{f,\rm true}|,
\end{align}
\end{subequations}
where the true values $(M_{f,\rm true},\chi_{f,\rm true})$ are taken from the SXS metadata. In Fig.~\ref{fig:G0_ots}, for both the $S$ and $Y$ models, we show results of the optimal distance $\chi^2[h,g^{(S/Y)}_{\rm opt}]$ and the relative estimation errors. The results obtained using the $S$ and $Y$ models are presented by blue solid and orange dotted curves, respectively. All results are shown with respect to $t_{\rm offset}$ defined in Eq.\,\eqref{eq:time_offset}. Each row represents the fitting with the same $(l,m)$ modes but different numbers of overtones. Different rows display the results with different groups of $(l,m)$ modes.

Let us first discuss the influence of varying the hyperparameter $t_0$, or equivalently, $t_{\rm offset}$. As shown in each panel of Fig.~\ref{fig:G0_ots}, with the increasing $t_{\rm offset}$, $\chi^{2}[h,g_{\rm opt}^{(S/Y)}]$ decreases and converges to a stable level after some specific value of $t_{\rm offset}$, which looks like a ``flat tail'' in the plot. We define that specific $t_{\rm offset}$ as \emph{transition time} ($\ttrans$), and the converged value of distance as \emph{minimum distance} ($\chimin$). More discussion about $\ttrans$ and $\chimin$ can be found at the end of Sec.~\ref{sec:nonsp_qs}. Practically, we define $\ttrans$ as the time when 30\% of the maximum slope in logarithmic scale of the fitting distance with respect to $t_{\rm offset}$ is reached. For the $S$ model, in each $\chi^2$ block, $\ttrans$ is marked by a vertical line  and $\chimin$ by a horizontal line, both in translucent blue. The accuracies of $M_{f,\rm est}$ and $\chi_{f,\rm est}$ (shown in blocks below each $\chi^2$ block) oscillate, while in general the levels of accuracy agree with the evolution of $\chi^2$ with respect to $t_{\rm offset}$, i.e., the relative errors decrease significantly before $\ttrans$ and slightly oscillate around a stable level after that. 
Despite the oscillation, after $\ttrans$, the relative errors are generally bounded by some specific small error level (similar to the reaching of $\chimin$ in the $\chi^2$ plot).
As the $\ttrans$ values generally agree between $\chi^2$ and the accuracies of $(M_{f,\rm est},\chi_{f,\rm est})$, we simply refer to the optimal distance when comparing the fitting results in the following discussions.

\begin{table}[b]
\caption{\label{tab:chi2_G0} The values of $\chimin$ when fitting G0 with different $(l,m)$ modes ($n=0$ only).}
\begin{ruledtabular}
\begin{tabular}{cccccccc}
Model & Group 1 & Group 1,2 & Group 1--3 & Group 1--4\\
\hline
$S$ & $7.9\times10^{-3}$ & $2.3\times10^{-3}$ & $2.2\times10^{-4}$ & $9.9\times10^{-5}$ \\
$Y$ & $1.0\times10^{-2}$ & $5.1\times10^{-3}$ & $3.1\times10^{-3}$ & $3.0\times10^{-3}$
\end{tabular}
\end{ruledtabular}
\end{table}

We then discuss the contribution of different $(l,m)$ angular modes. The ranking of each $\mathcal{A}_{lm}$ (Fig.~\ref{fig:hpc_Elm_nonspinning}) indicates its significance in the fitting. The contribution of higher-order angular modes can be observed by comparing among different rows in Fig.~\ref{fig:G0_ots}. Specifically, the four panels in the first column show the results of fitting with different $(l,m)$ mode groups, with the fundamental $n=0$ overtones only. From top to bottom, the $\chimin$ values for the $S$ and $Y$ models are listed in Table~\ref{tab:chi2_G0}. The $\chimin$ values for the $S$ model are consistently smaller than those for the $Y$ model.
In the case of using the $S$ model, adding Group 3, $(l,m)=(4,4),(2,0),(3,2)$, improves the results most significantly. It reduces $\chimin$ from $\sim 10^{-3}$ to $\sim 10^{-4}$ and the relative estimation error from $\sim 10^{-2}$ to $\sim 10^{-3}$. 
On the contrary, no order of magnitude improvement is seen when using the $Y$ model by adding $(l,m)$ groups, as the $Y$ model is not accurate enough such that the errors due to missing higher-order modes are smaller compared to the errors caused by the inaccuracy of the $Y$ model itself.
As shown in Fig.~\ref{fig:hpc_Elm_nonspinning}, for binary G0, $(l,m)=(3,2)$ in Group 3 and $(3,3)$ in Group 2 are the subdominant modes except for the leading mode $(2,2)$ in Group 1; while Group 4 is the least important group among 1--4. 
Accordingly, in Fig.~\ref{fig:G0_ots}, we can see that including Groups 2 and 3 leads to significant improvement while including Group 4 does not. 

On the other hand, the effect of overtones can be observed by comparing different columns within each row. It is shown that adding overtones can bring $\ttrans$ to an earlier time; the more overtones added, the earlier $\ttrans$ becomes. This effect can be explained by referring to Fig.~\ref{fig:freq_branches}: the higher-order overtones have larger decay rates and usually play roles at the time closer to $t_{\rm peak}$. By including overtones, the ringdown model is more accurate at an earlier time after the merger and thus $\chimin$ can be achieved with a smaller $t_{\rm offset}$.

The distinguishability between the $S$ and $Y$ models is influenced jointly by $(l,m)$ modes and overtones.
With the same modes included, the $S$ model can always yield smaller $\chimin$ than the $Y$ model. Moreover, the improvement in $\chimin$ of using $S$ compared to $Y$ is more significant when more $(l,m)$ modes are added, and when more overtones are included. This is illustrated by Eq.\,\eqref{eq:SY_mixing}: there is a certain difference between $S$ and $Y$ bases for each given $lmn$ mode. Specifically, in the $S$ model, different overtones of the $(l,m)$ mode have different angular distributions, while they share the same distribution in the $Y$ model. 
Therefore, adding more $(l,m)$ modes and overtones enlarges the difference between two models. Additionally, by adding higher-order overtones, $\chimin$ generally does not change, because higher-order overtones are damped before $\ttrans$ and have negligible contribution to the converged level of distance. In terms of the relative errors in $(M_{f,\rm est},\chi_{f,\rm est})$, the advantage of using the $S$ model only becomes obvious when adding modes up to Group 3. This will be further discussed in Sec.~\ref{sec:nonsp_qs} with other nonspinning binaries N1--N9. 

Before the extensive comparison of binaries with various mass ratios in the next section, we first comment on possible sources of fitting errors that could potentially impact the conclusion of model distinguishability. In this study, no simulated noise has been added. Thus, apart from the limitation of the model itself, the fitting errors mostly come from numerical noises in the NR waveforms. The SXS waveforms are provided at several numerical resolution levels labeled as Lev1, Lev2, Lev3, etc. For the same binary, a higher level has higher resolution; while between different binaries, level numbers are not necessarily meaningful~\cite{Boyle2019SXS}. The uncertainties in $\chimin$ and $(M_{f,\rm est},\chi_{f,\rm est})$ are supposed to be bounded by the difference between the results obtained from the numerical level used in Table~\ref{tab:SXS_N} and the adjacent lower level~\cite{Boyle2019SXS}. 
For the results in Fig.~\ref{fig:G0_ots}, we used the highest available numerical level of G0, Lev6, as listed in Table~\ref{tab:SXS_N}. We now compare them with the results obtained from lower levels, Lev5--Lev3. If the difference between using the $S$ and $Y$ models at Lev6 is clearly larger than the difference of results obtained among different levels, we are able to state that the two models are distinguishable. 
In Fig.~\ref{fig:G0_ot0_levs}, we plot the results obtained from different levels, with different $(l,m)$ groups ($n=0$ only). It is shown that the $\chi^2$ values obtained at different levels are not distinguishable by eye, definitely smaller than the $S/Y$ difference shown in Fig.~\ref{fig:G0_ots}.
Thus, the conclusion that the $S$ model can be distinguished from the $Y$ model and is a more faithful representation of the NR waveform is not impacted by the numerical errors.
Note that not all SXS waveforms have such a good numerical precision in the ringdown part for the purpose of this study, 
probably due to memory residuals 
in the chosen Bondi-Metzner-Sachs (BMS) frame~\cite{Mitman2021Fixing}. To avoid this impact, we select the SXS waveforms that can produce consistent results from different numerical levels. See Appendix~\ref{app:result_levs} for more details.
When the ringdown waveform has significantly decayed, if the model is sufficiently accurate with Group 3 and Group 4 included, subtle ringdown residuals due to the choice of frame in the numerical data will manifest as slightly increased $\chi^2$ values for $40 \lesssim t_{\rm offset}/M\lesssim 50$ (see Fig.~\ref{fig:G0_ots} and Appendix~\ref{app:result_all}).
Potentially, subtractions of the late ringdown residuals~\cite{Giesler2019Black} could help improve the fitting in the range of $40 \lesssim t_{\rm offset}/M\lesssim 50$. We do not conduct such subtraction because it is not well motivated, and the small impact from the residuals~\cite{Zertuche2021High} does not quantitatively change the resulting transition time and minimum distance in our analysis.
Recent studies show that using the Cauchy-characteristic extraction (CCE) and mapping to the super-rest frame would be a proper approach to obtain the memory-free waveform~\cite{Mitman2021Fixing,Zertuche2021High}.

\subsection{\label{sec:nonsp_qs}Nonspinning binaries with different mass ratios}

\begin{figure}
\centering
\begin{overpic}[width=\columnwidth]{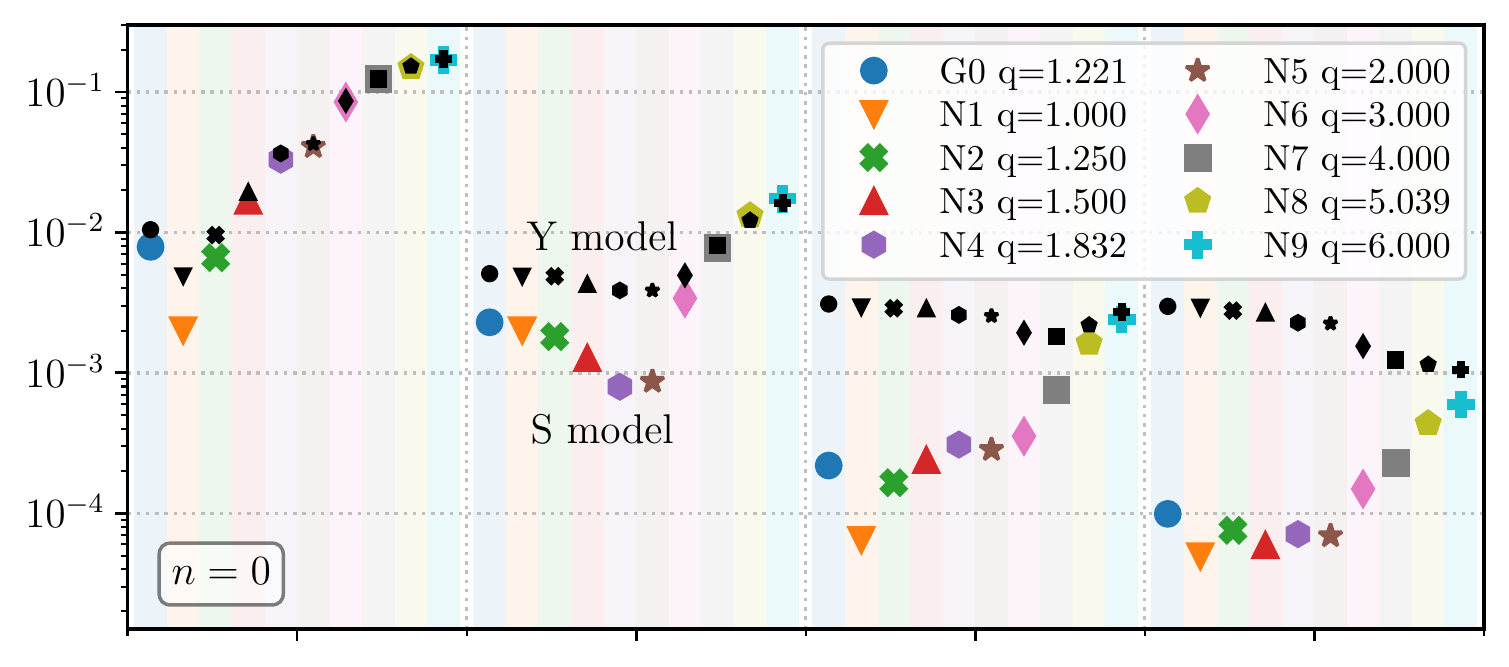}
\put(3,42){{(a)}}
\end{overpic}
\begin{overpic}[width=\columnwidth]{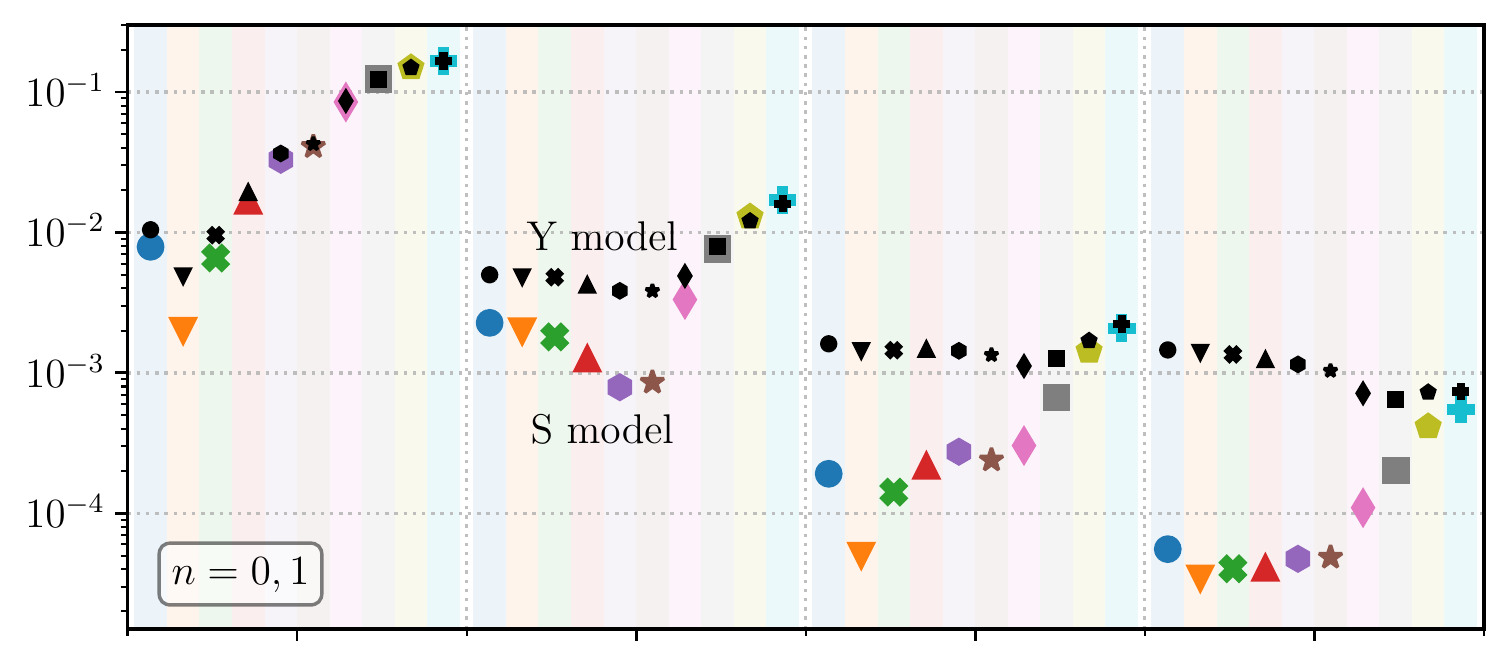}
\put(3,42){{(b)}}
\end{overpic}
\begin{overpic}[width=\columnwidth]{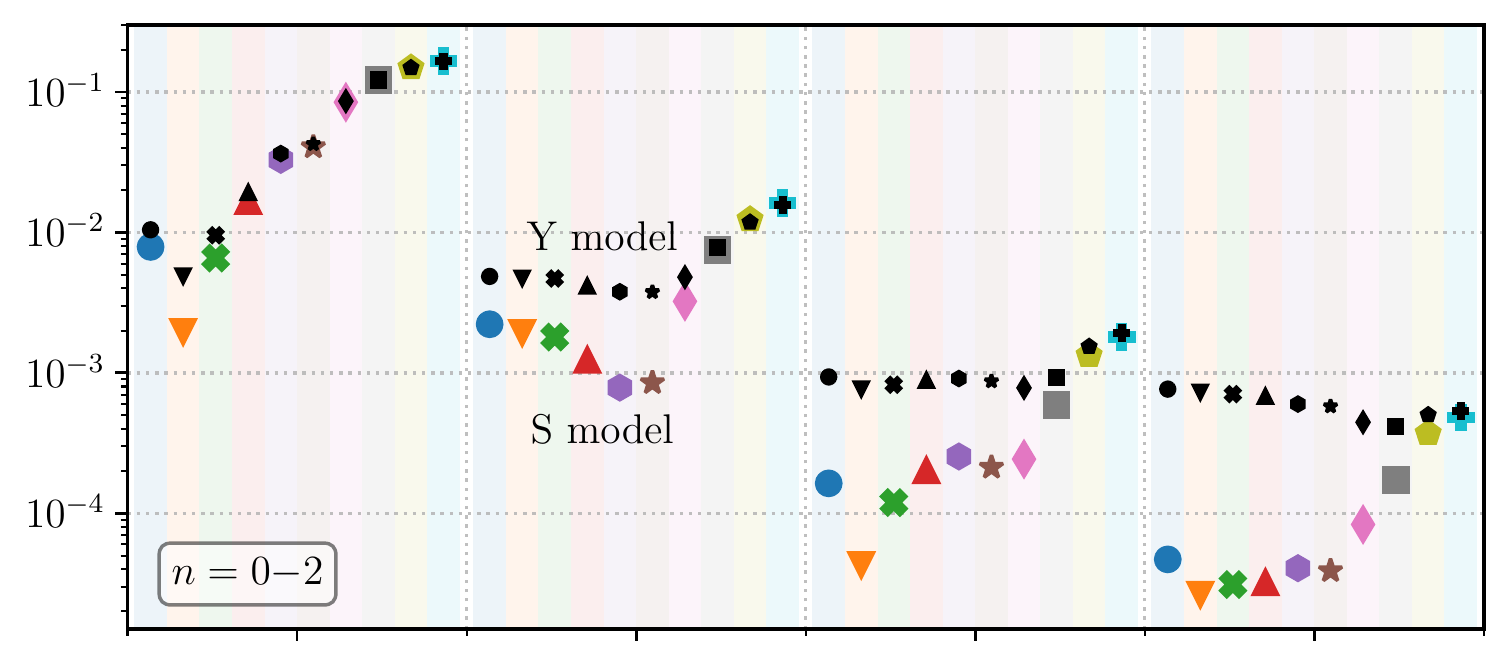}
\put(3,42){{(c)}}
\end{overpic}
\begin{overpic}[width=\columnwidth]{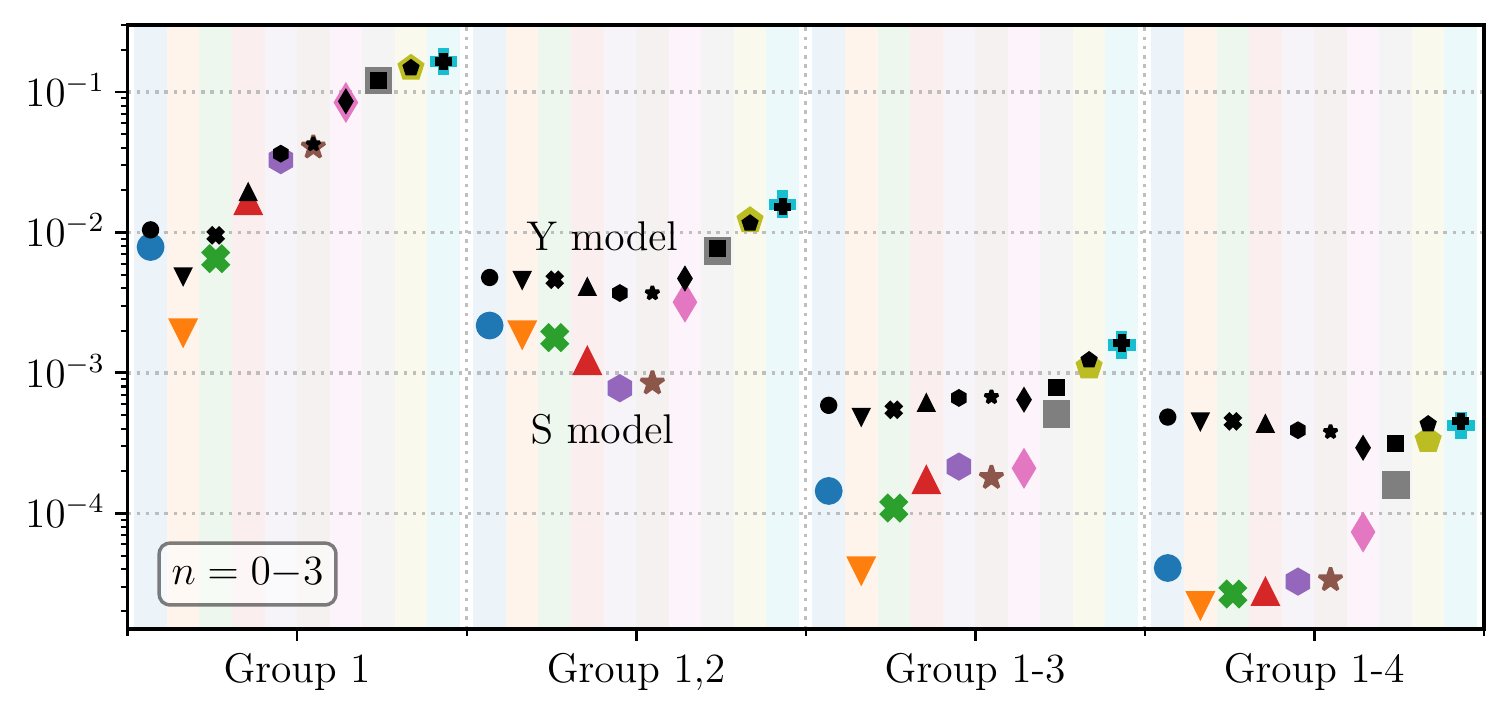}
\put(3,45){{(d)}}
\end{overpic}
\caption{\label{fig:nsp_min_dist} The minimum distance $\chimin$ obtained for binaries G0 and N1--N9 when (a) $n=0$, (b) $n=0,1$, (c) $n=0,1,2$, and (d) $n=0,1,2,3$ are considered. The horizontal axes are arranged by $(l,m)$ groups. Within each group, results for different waveforms are shown by markers in different shapes, as indicated by the legend in (a). For each binary waveform, the big colored and the small black markers indicate $\chimin$ obtained using the $S$ and $Y$ models, respectively. 
}
\end{figure}

\begin{figure}
\centering
\includegraphics[width=\columnwidth]{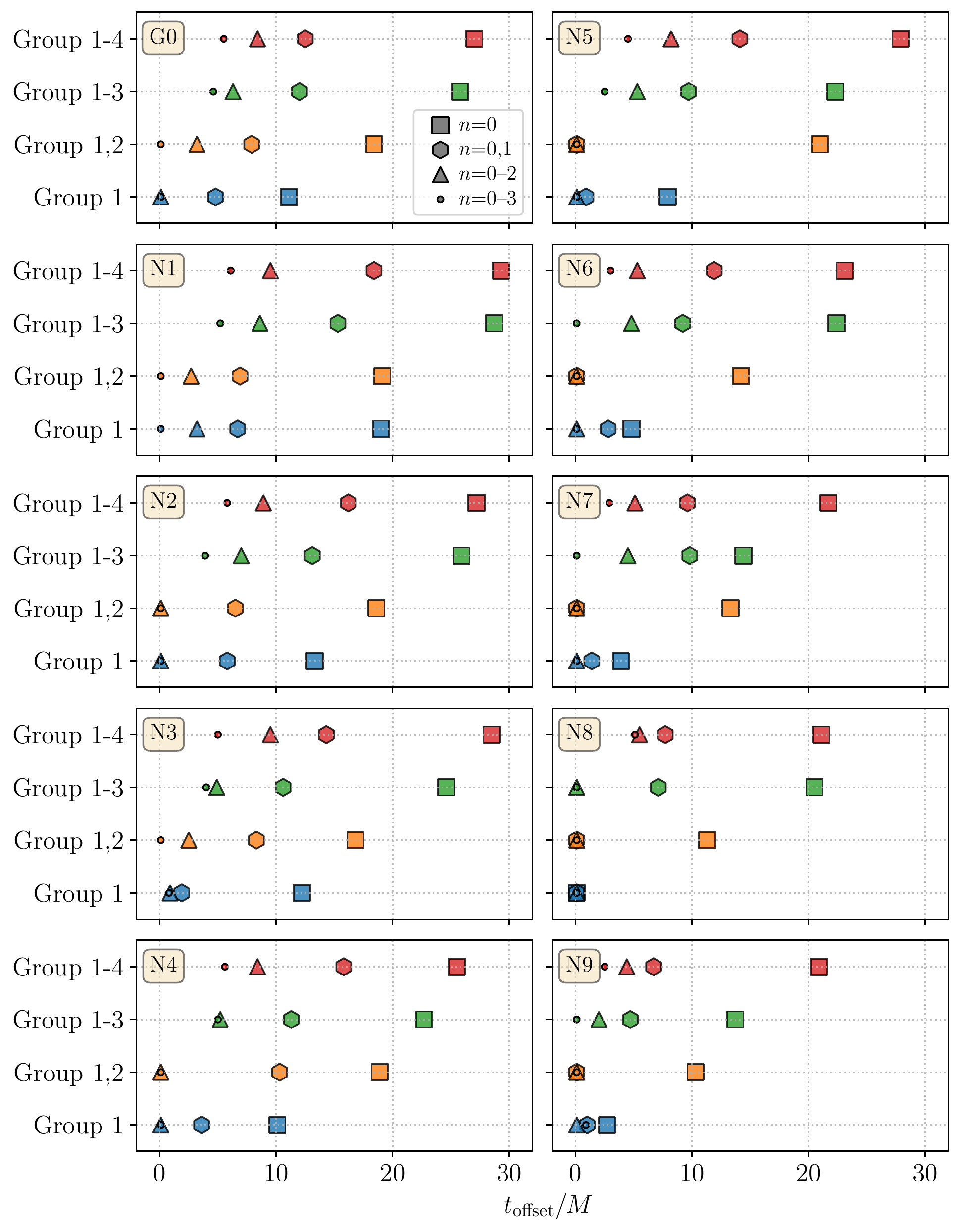}
\caption{\label{fig:S_t_trans} The $S$ model transition time $\ttrans$ for binaries G0 and N1--N9. Each panel corresponds to a specific binary, labeled in the upper left corner. The horizontal axis is $t_{\rm offset}/M$, and the discretized vertical axis specifies the $(l,m)$ groups. Markers in different shapes indicate different numbers of overtones included, shown in the legend in the top left panel.
}
\end{figure}

In Sec.~\ref{sec:results_G0}, we have verified the fitting algorithm with the benchmark binary G0. 
To study the contributions of $(l,m)$ modes and overtones in binaries with various mass ratios~\cite{Forteza2020Spectroscopy}, we apply the same method to a series of nonspinning binary waveforms N1--N9 with mass ratios ranging from $1$ to $6$, as listed in Table~\ref{tab:SXS_N}. In this section, we demonstrate that more modes are needed in order to achieve the same level of $\chimin$ with larger mass ratios, as the higher-order modes are more strongly excited in larger mass-ratio binaries. We summarize the results in two plots grouped by $(l,m)$ groups and numbers of overtones: Fig.~\ref{fig:nsp_min_dist} shows the comparison between the $S$ and $Y$ models in terms of $\chimin$; Fig.~\ref{fig:S_t_trans} displays $\ttrans$ for the $S$ model when different modes and overtones are included. Detailed fitting results in the same format as those of G0 in Fig.~\ref{fig:G0_ots} are presented in Appendix~\ref{app:result_all}. 

In Fig.~\ref{fig:nsp_min_dist}, for both the $S$ and $Y$ models, $\chimin$ decreases when more $(l,m)$ modes are added.
For a binary with larger $q$, more significant improvement is seen when adding more $(l,m)$ modes. In the case of the $S$ model, $\chimin$ for binary N1 ($q=1$) decreases from $2\times 10^{-3}$ (Group 1) to $7\times 10^{-5}$ (Group 1--4), 
with the accuracy level improved by a factor of $\sim 30$; for binary N9 ($q=6$), an improvement by a factor of $\sim 280$ is achieved from Group 1 to Group 1--4.
As indicated in Fig.~\ref{fig:hpc_Elm_nonspinning}, higher angular modes are more strongly excited in binaries with larger mass ratios and thus play a more important role in the QNM expansion. On the other hand, to reach an accuracy of $\chimin <0.01$ in the $S$ model, the fundamental $(2,2)$ mode is enough for $q<1.25$ ; while $(l,m)$ modes up to Group 2 are needed for $1.25<q<4$ binaries and Group 1--3 are needed for $q>5$. 
Comparing panels (a)--(d) in Fig.~\ref{fig:nsp_min_dist}, we see that adding more overtones does not result in any order-of-magnitude improvement to $\chimin$. That is because higher-order overtones decay faster and thus only influence the ringdown waveform at an earlier time while have negligible impact on the converged $\chimin$ after $\ttrans$.

Comparing results of different binaries in Fig.~\ref{fig:nsp_min_dist}, the differences between $S$ and $Y$ are the most and least significant for the $q=1$ and $q=6$ binaries, respectively. Taking the last column of groups (Groups 1--4) in panel (a) for example, for binary N1 ($q=1$), we have $\chimin = 7\times10^{-5}$ and  $3\times10^{-3}$ for the $S$ and $Y$ models, respectively, 
with a factor of $\sim 40$ improvement in accuracy by using $S$ versus $Y$; while for binary N9 ($q=6$), the accuracy is only a factor of $\sim 1.8$ better by using $S$ ($\chimin = 6.0\times10^{-4}$ for $S$ and $1\times10^{-3}$ for $Y$).
This is because the distinction between the $S$ and $Y$ bases depends on the spheroidicities, as shown in Eq.\,\eqref{eq:SY_mixing}, which are proportional to $|\chi_f|$, $M_f$, and $\omega_{lmn}$'s. Meanwhile, when the progenitor black holes are nonspinning, $|\chi_f|$ decreases as $q$ increases, e.g., $\chi_f=0.6864$ for N1 and $\chi_f=0.3725$ for N9.

\begin{figure}
\centering
\includegraphics[width=\columnwidth]{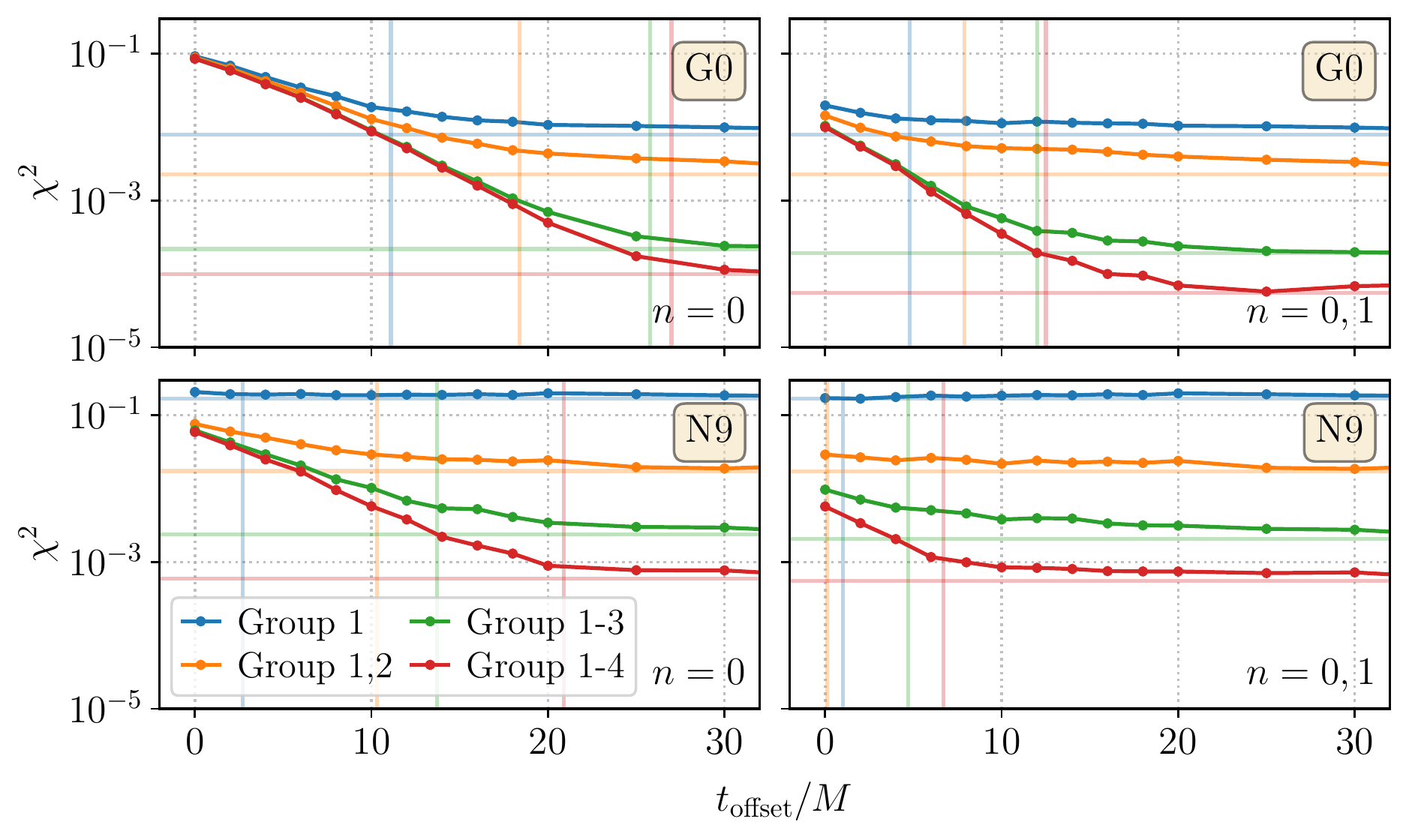}
\caption{\label{fig:example_floor} Example $S$ model distances $\chi^2$ for binaries G0 and N9,  with respect to $t_{\rm offset}$. The left and right columns are for $n=0$ and $n=0,1$, respectively. Curves with different colors stand for different $(l,m)$ Groups, as shown in the legend. The translucent vertical and horizontal lines indicate $\ttrans$ and $\chimin$, respectively.
}
\end{figure}

Even though the $S$ model results in better $\chimin$ than the $Y$ model, it is not always the case for the estimated parameters $M_{f,\rm est}$ and $\chi_{f,\rm est}$. The features of relative errors defined in Eq.\,\eqref{eq:relative_PE_err} are summarized below, with detailed results presented in Appendix~\ref{app:result_all}:

\begin{itemize}
    \item For G0, N1--N4: before adding $(l,m)$ Group 3, $Y$ model behaves better; $S$ turns better after adding Group 3.
    \item For N5: there is no clear $S/Y$ distinction when including only Group 1; $Y$ behaves better after adding Group 2; $S$ behaves better after adding Group 3.
    \item For N6: $Y$ behaves better before adding Group 3; there is no distinction after adding Group 3 but not Group 4; $S$ behaves better after adding Group 4.
    \item For N7: $Y$ behaves better before adding Group 3; there is no distinction after adding Group 3.
    \item For N8--N9: $Y$ behaves better before adding Group 4; there is no distinction after adding Group 4.
\end{itemize}
From the observations above, we notice that when not including enough $(l,m)$ modes, $S$ model is not necessarily better than $Y$ model in estimating parameters of the remnant black hole. It indicates that the more accurate $S$ model with spin-weighted spheroidal harmonics are more impacted by the missing $(l,m)$ modes, while the less accurate $Y$ model with orthogonal spin-weighted spherical harmonics is less impacted.
For binaries with larger $q$, more $(l,m)$ modes have non-negligible contributions to the ringdown waveform, and thus are all required for a precise characterization when using the $S$ model. Once those $(l,m)$ modes are included, the $S$ model is consistently better than the $Y$ model in both $\chimin$ and the accuracy of $(M_{f,\rm est},\chi_{f,\rm est})$. 

Based on the discussion above, we conclude that the spin-weighted spheroidal harmonics ($S$ model) is indeed the better representation of gravitational wave ringdown signals compared to the spin-weighted spherical harmonics ($Y$ model), and that the difference is distinguishable in NR waveforms.

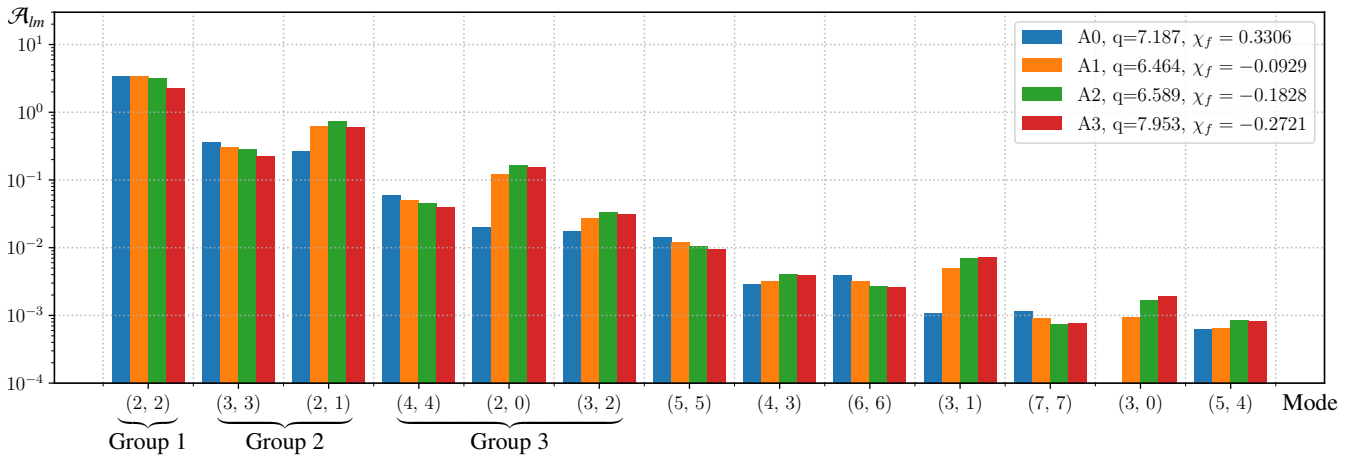
\begin{figure*}
\begin{center}
\begin{overpic}[width=0.995\textwidth]{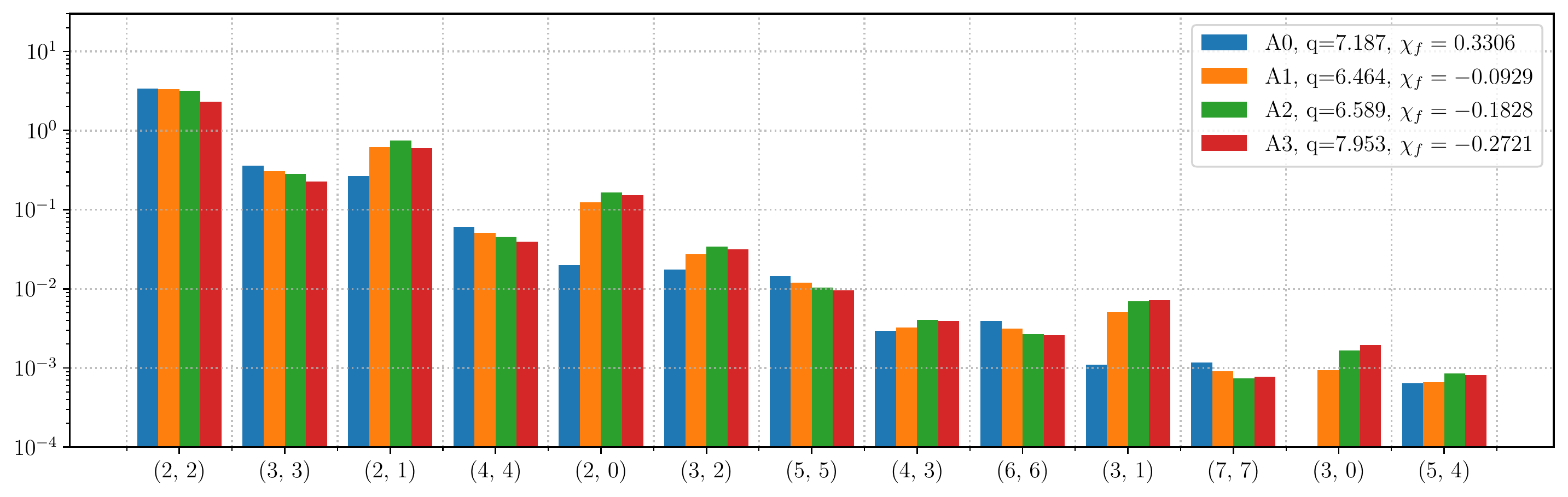}
\put(1,30){{$\mathcal{A}_{lm}$}}
\put(96,1.2){{Mode}}
\put(8.5,-0.5){$\begin{matrix} \underbrace{\hspace{0.8cm}} \\ \textrm{Group 1}\end{matrix}$}
\put(16.6,-0.5){$\begin{matrix} \underbrace{\hspace{1.8cm}} \\ \textrm{Group 2}\end{matrix}$}
\put(30,-0.5){$\begin{matrix} \underbrace{\hspace{3cm}} \\ \textrm{Group 3}\end{matrix}$}
\end{overpic}
\vspace{0.1cm}
\caption{\label{fig:hpc_Elm_aliant} 
The relative importance $\mathcal{A}_{lm}$ of binaries A0--A3. The definitions of Group 1--3 follow those in Fig.~\ref{fig:hpc_Elm_nonspinning}. Considering the computing cost with retrograde modes added, we only include $(l,m)$ modes up to Group 3 in Section~\ref{sec:retrograde}.}
\end{center}
\end{figure*}

Let us comment on the temporal behavior of fitting with different $(l,m)$ modes and overtones. Now we only consider the $S$ model as we have demonstrated that the $S$ model is a better representation. 
In Fig.~\ref{fig:S_t_trans}, we summarize $\ttrans$ for binaries G0 and N1--N9, and further in Fig.~\ref{fig:example_floor}, we show the details of $\chi^2$ with respect to $t_{\rm offset}$ for two example binaries, G0 and N9. Essentially, the transition time $\ttrans$ happens when the missing overtones have mostly decayed in the NR waveform, with $\chi^2$ reaching the minimum distance $\chimin$ determined by the precision that the model can achieve. 
Adding overtones brings forward $\ttrans$, while adding $(l,m)$ modes postpones $\ttrans$ -- when the model becomes more accurate, the achievable $\chimin$ is smaller and thus takes more time to arrive. 
In other words, when the model template includes more $(l,m)$ modes, more overtones are needed accordingly to reach the transition at a similar time. 
Similarly, mass ratios also influence the transition time as a result of different achievable $\chimin$ levels. As shown in Fig.~\ref{fig:S_t_trans}, with the same sets of modes included in the template, $\ttrans$ generally occurs earlier when $q$ is larger, because the achievable $\chimin$ is relatively larger. 


In this section, we have discussed the $S/Y$ model distinguishability and contribution of higher-order $(l,m)$ modes and overtones. In the next section, we will further consider the situations when the progenitor binaries have spins along the $-\hat z$ direction that leads to anti-aligned spins in remnant black holes.

\section{\label{sec:retrograde}Spinning Binaries and Retrograde Excitation}

\begin{table}[b]
\caption{\label{tab:SXS_A}SXS BBH waveforms used in Sec.~\ref{sec:retrograde}.}
\begin{ruledtabular}
\begin{tabular}{cccccccc}
Label\footnotemark[1] & SXS ID/Lev &$q_{\rm ref}$ &$(\vec{\chi}_{\rm ref,1})_z$ &  $(\vec{\chi}_{\rm ref,2})_z$ & $\chi_{\rm eff}$ & $(\vec \chi_f)_z$\\
\hline
A0 & 0188/Lev3 & 7.187 & 0.0000 & 0.0000 & 0.0000 & 0.3306\\
A1 & 1424/Lev3 & 6.464 & $-$0.6566 & $-$0.7991 & $-$0.6757 & $-$0.0929\\
A2 & 1435/Lev3 & 6.589 & $-$0.7893 & 0.0673 & $-$0.6764 & $-$0.1828\\
A3 & 1422/Lev3 & 7.953 & $-$0.8001 & $-$0.4588 & $-$0.7620 & $-$0.2721
\end{tabular}
\end{ruledtabular}
\footnotetext[1]{Omit the same notes as in Table~\ref{tab:SXS_N}.} 
\end{table}

For nonspinning binaries, only the orbital angular momentum contributes to the remnant spin. While in general cases, the spin angular momentum of each individual progenitor black hole also leaves imprints in the ringdown waveform. Specifically, when the spins of the progenitor black holes are anti-aligned with the orbital angular momentum~\cite{Campanelli2006Spinning,Berti2008Cosmological,Rodriguez2016Illuminating} and are large enough, retrograde modes could be excited in the remnant black hole. Retrograde excitations have been studied in the case of extreme mass ratio inspirals~\cite{Taracchini2014Small,Hughes2019Learning,Apte2019Exciting1,Lim2019Exciting2} using black hole perturbation theory~\cite{Teukolsky1973Perturbations}. Features of the ringdown waveforms have also been numerically studied in superkick BBH systems with equal mass~\cite{Ma2021Universal}. While not many studies have been done in the intermediate mass ratio inspirals~\cite{Mandel2008Rates,Huerta2011Thesis}. 


To study the retrograde excitations, we apply the fitting method described above to three binaries, A1--A3, with $\chi_f<0$ ($|\chi_f|$ increases from A1 to A3), as listed in Table~\ref{tab:SXS_A}. A nonspinning binary A0 with $\chi_f>0$ is included for comparison purposes. The orbital angular momentum and the spin angular momentum of the primary black hole, when in opposite directions, will cancel out to some extent in the merger stage~\cite{Campanelli2006Spinning,Berti2008Cosmological}. Becasue of that, retrograde QNMs, excited when the anti-aligned spin dominates, have smaller frequencies compared with the corresponding prograde modes. 
During the merger, the inspiral polarization pattern transitions smoothly to the dominating prograde or retrograde ringdown modes for remnant spin $\chi_f>0$ and $\chi_f<0$, respectively. This will be discussed in more detail in Sec.~\ref{subsec:pro_retro}. 

In this section, we describe the fitting strategy, again, based on the relative importance of different modes in these waveforms (A0--A3) and implement the fitting with both prograde and retrograde modes included, or with prograde modes only. We then analyze the results and discuss the features of QNM frequencies and polarization patterns in the spin anti-aligned case.

The relative importance $\mathcal{A}_{lm}$ (defined in Eq.\,\eqref{eq:Elm}) of binaries A0--A3 are plotted in Fig.~\ref{fig:hpc_Elm_aliant}. 
We follow the convention in Fig.~\ref{fig:aliant_coord} to extend the parameter space of the orbital frame spin $\chi_f$ to include negative values. 
In Sec.~\ref{sec:nonspinning}, we have demonstrated that the $S$ model is more accurate compared to the $Y$ model. In this section, we focus on the fitting using the $S$ model. 
We follow the grouping and ranking of $(l,m)$ modes discussed in Sec.~\ref{sec:strategy_lm}. Given that the fitting including retrograde modes is more computationally expensive (the number of modes doubled) and that the main purpose here is to study the retrograde modes, we directly compare the results between including only the prograde modes up to Group 3, i.e., $(l,m)=(2,2),(3,3),(2,1),(4,4),(2,0),(3,2)$, and the results including the corresponding retrograde modes $(l,m)=(2,-2),(3,-3),(2,-1),(4,-4),(3,-2)$ as well.
To confirm the contribution of overtones, we implement two sets of fittings with $n=0$ and $n=0,1$ for each of the two scenarios above. Also, since $\chimin$ will be stabilized at some given level after $\ttrans$, we carry out the fitting up to $t_{\rm offset} = 35M$ in this section.

\begin{figure}
\includegraphics[width=0.98\columnwidth]{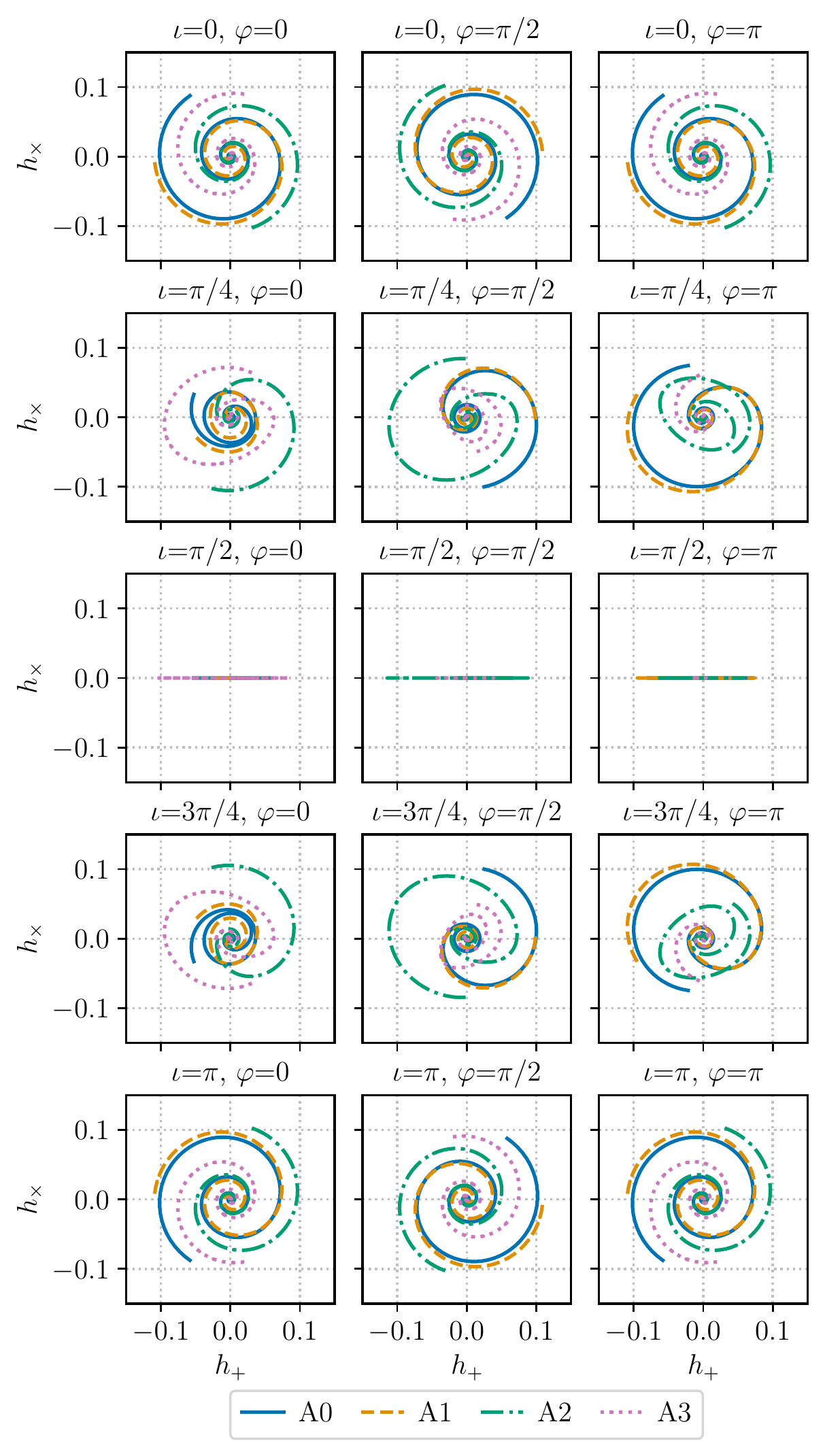}
\caption{Polarization contents ($h_+(t), h_\times(t)$) of the ringdown waveform of binaries A0--A3, with $t=0$ starting at their own $t_{\rm peak}$. The plotting convention follows that of Fig.~\ref{fig:Bpm_afpm}.
}
\label{fig:polar_aliant}
\end{figure}

The polarization patterns for A0--A3 have similar features at different emission directions, as shown in Fig.~\ref{fig:polar_aliant} -- they all look counterclockwise when viewed from the north side ($\iota<\pi/2$) and clockwise when viewed from the south side ($\iota>\pi/2$). Combined with Fig.~\ref{fig:Bpm_afpm}, we can see that the dominant excitations are either characterized by $B_{m>0}^{(S\pm)}$ for spin aligned binaries, or by $B_{m<0}^{(S\pm)}$ for spin anti-aligned binaries. Thus the dominant QMNs should be prograde modes for A0 but retrograde modes for A1--A3.

Another feature we expect to see when comparing the fitting results with and without the retrograde modes is associated with $|\chi_f|$. 
As shown in Fig.~\ref{fig:freq_branches}, the QNM frequencies of the retrograde modes correspond to the dotted curves, extending towards lower frequencies when $|\chi_f|$ increases from the points with $\chi_f=0$, while the curves of prograde modes extend towards higher frequencies when $|\chi_f|$ increases. 
For larger $|\chi_f|$, the frequencies of a prograde mode and the corresponding retrograde mode becomes more separated in the spectrum. Thus the difference in fittings with and without the retrograde modes becomes more distinct.

\begin{figure*}
\begin{overpic}[width=0.98\textwidth]
{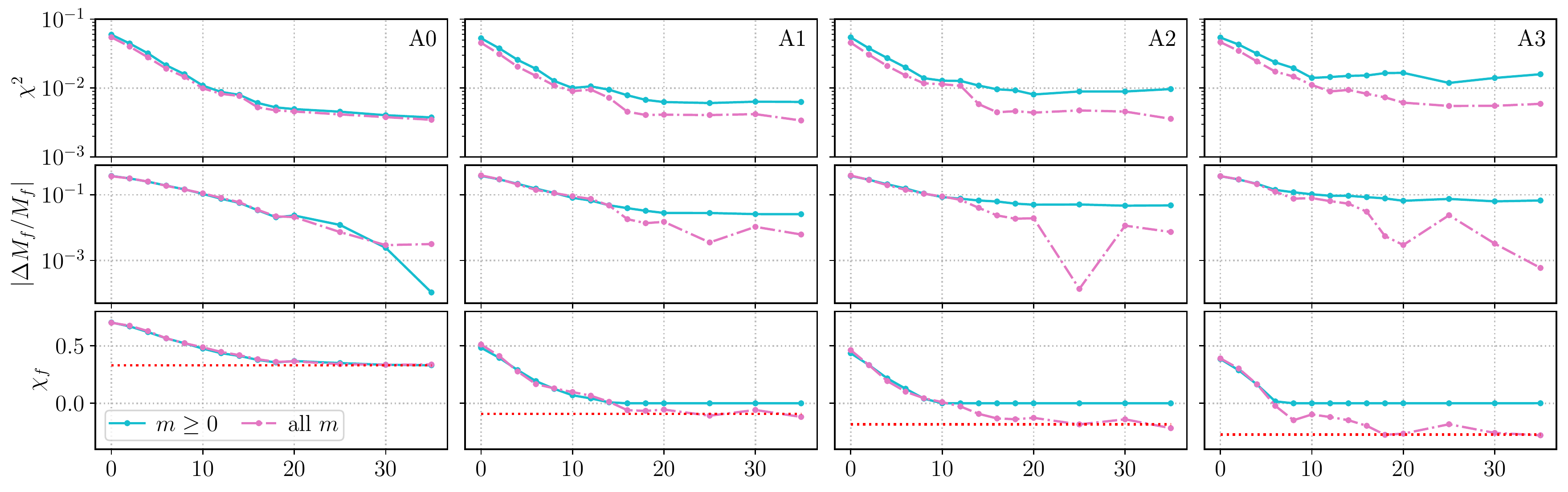}
\put(-1,29.5){{(a1)}}
\put(7,22){{$n=0$}}
\put(30.5,22){{$n=0$}}
\put(54,22){{$n=0$}}
\put(77.5,22){{$n=0$}}
\end{overpic}
\begin{overpic}[width=0.99\textwidth]{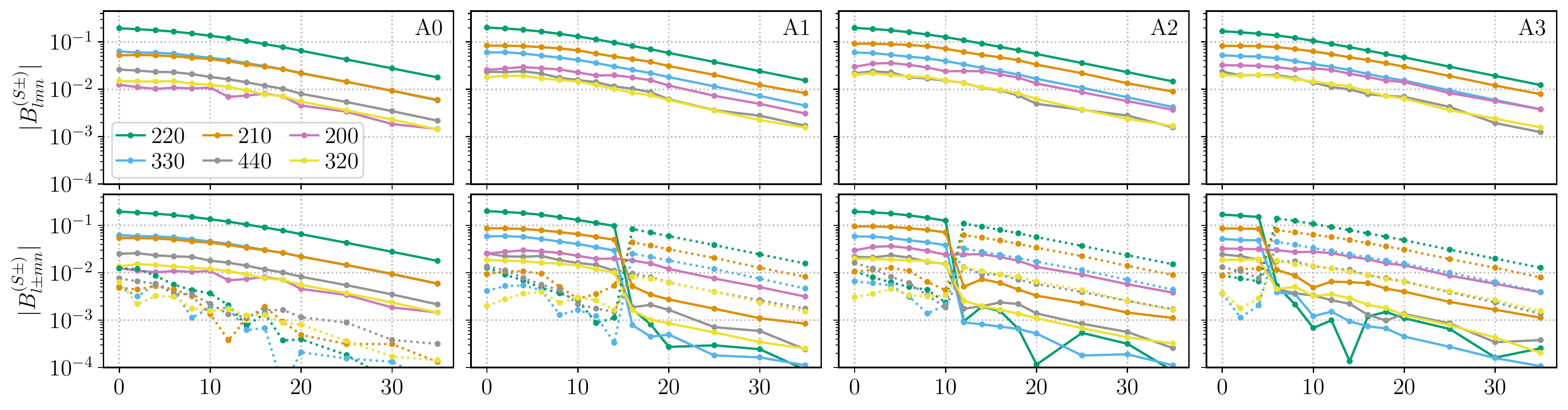}
\put(-1,24.5){{(a2)}}
\end{overpic}
\begin{overpic}[width=0.98\textwidth]
{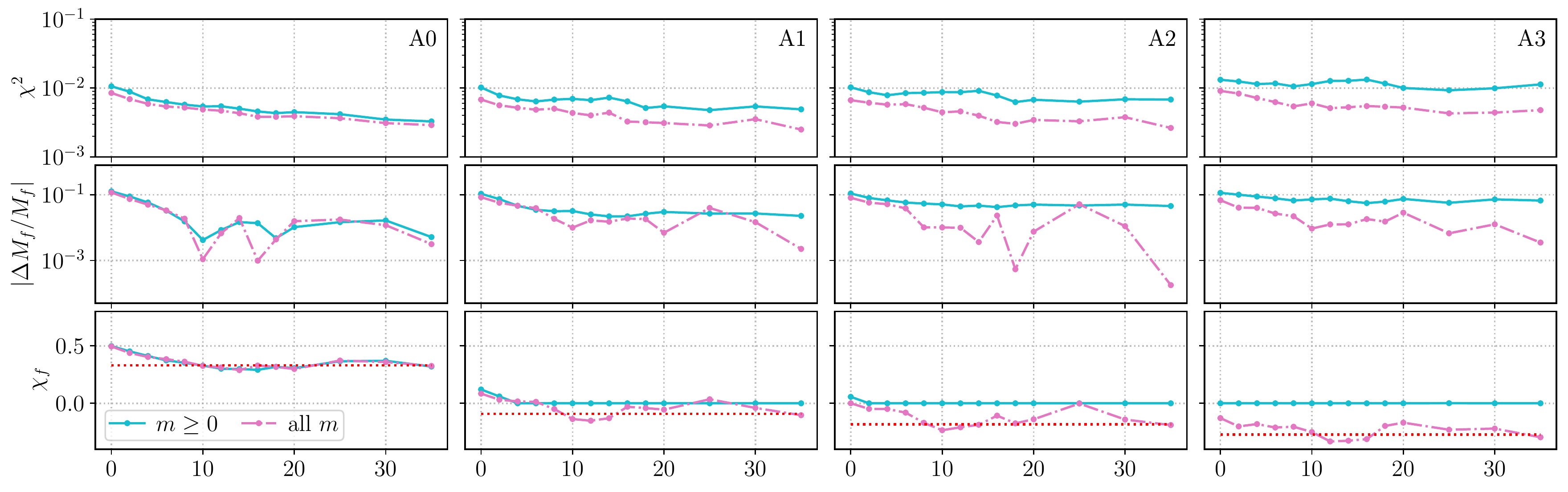}
\put(-1,29.5){{(b1)}}
\put(7,22){{$n=0,1$}}
\put(30.5,22){{$n=0,1$}}
\put(54,22){{$n=0,1$}}
\put(77.5,22){{$n=0,1$}}
\end{overpic}
\begin{overpic}[width=0.99\textwidth]{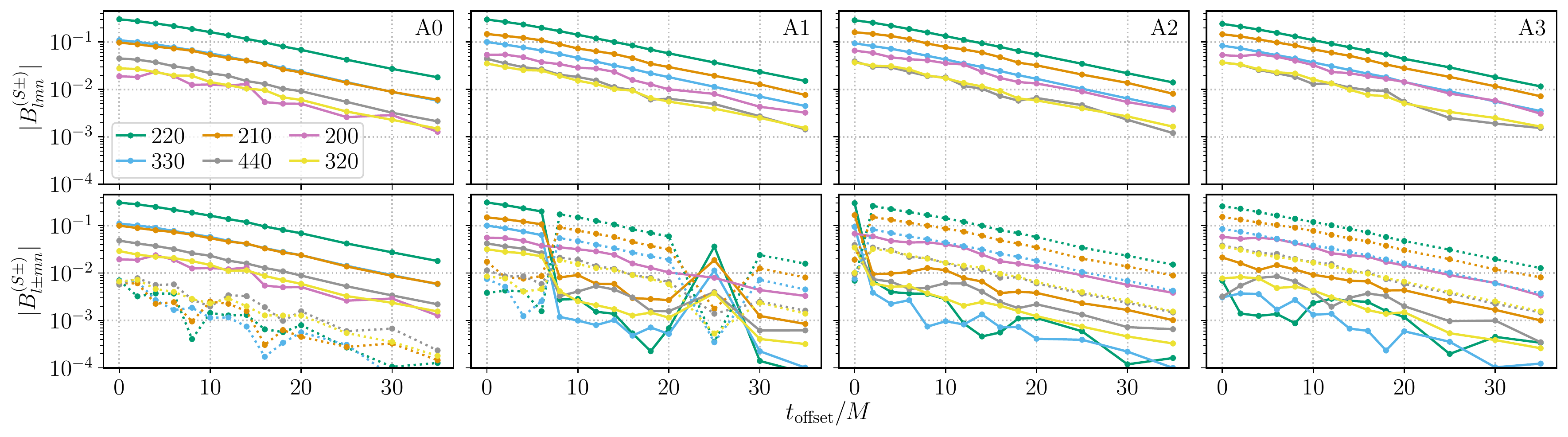}
\put(-1,26){{(b2)}}
\end{overpic}
\caption{
Fitting results for binaries A0--A3 using the $S$ model with (a) only the fundamental modes $n=0$ and (b) overtones $n=0,1$. For each binary (in each column), (a1) and (b1) show the optimal distance $\chi^2$, the relative error in $M_{f,\rm est}$, and $\chi_{f,\rm est}$; (a2) and (b2) show the magnitudes of the optimal coefficients. 
In (a1) and (b1), the solid cyan and dot-dashed magenta curves correspond to the results without and with retrograde modes included, respectively. The dotted red line in each $\chi_f$ block indicates the $\chi_{f, \rm true}$.
In (a2) and (b2), the upper and lower blocks show optimal coefficients when fitting without and with retrograde modes included, respectively. The solid and dotted curves correspond to prograde and retrograde modes, respectively.
Note that $B_{lmn}^{\,(S+)}$ and $B_{lmn}^{\,(S-)}$ with the same $lmn$ indices are roughly conjugate to each other (in the spin aligned or spin anti-aligned cases). For brevity, we only plot their absolute values.
}
\label{fig:aliant_result}
\end{figure*}

\begin{figure*}
\begin{overpic}[width=0.98\textwidth]
{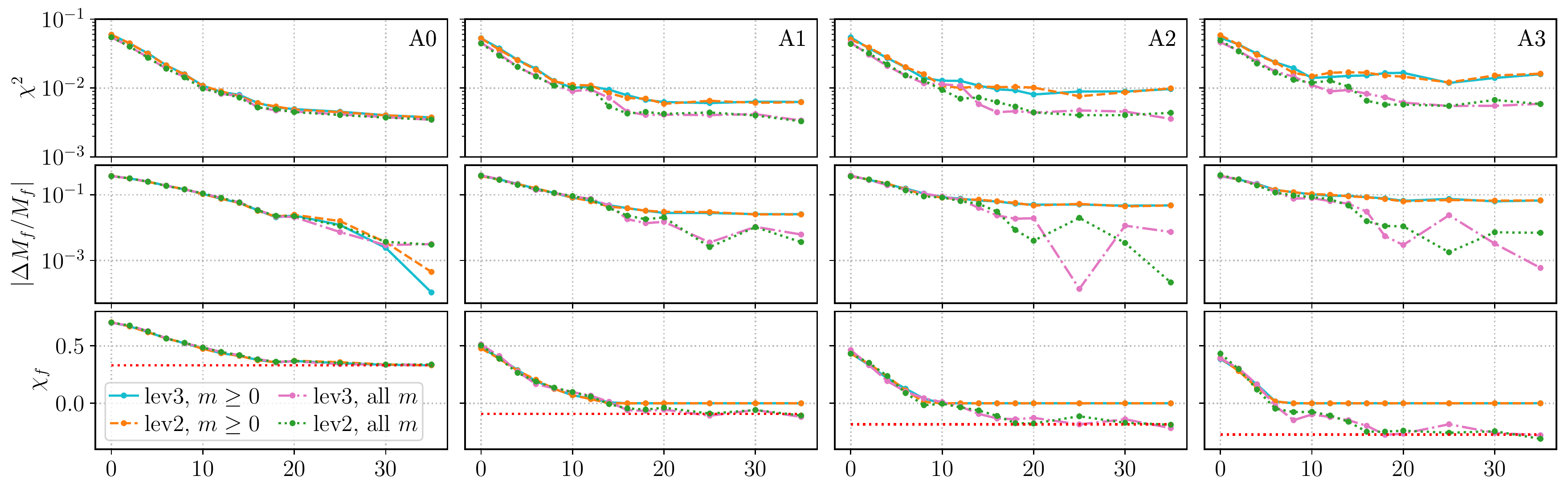}
\put(7,22){{$n=0$}}
\put(30.5,22){{$n=0$}}
\put(54,22){{$n=0$}}
\put(77.5,22){{$n=0$}}
\put(12,24){\includegraphics[width=0.14\textwidth]{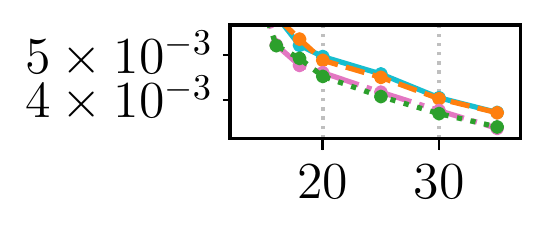}}
\end{overpic}
\caption{Fitting results for the binary waveforms A0--A3 using SXS data with different numerical levels. The $S$ model used for fitting only contains the fundamental $n=0$ modes. Plot settings are the same as Fig.~\ref{fig:G0_ot0_levs}. The inset in the $\chi^2$ block of the first column shows the results in a zoomed-in range.}
\label{fig:aliant_ot0_levs}
\end{figure*}

The fitting results are shown in Fig.~\ref{fig:aliant_result}, where (a) and (b) correspond to the cases with overtones $n=0$ and $n=0,1$, respectively. 
When showing $\chi_{f, \rm est}$, we no longer use the relative error defined in Eq.\,\eqref{eq:relative_PE_err}, as it is not a good measure when the true value is small and comparable to the level of oscillation, e.g., A1 has $\chi_{f, \rm true}=-0.0929$, while $\chi_{f, \rm est}$ oscillates up to $\sim 0.1$. Instead, we directly compare $\chi_{f, \rm est}$ to $\chi_{f, \rm true}$ (the dotted red lines in Fig.~\ref{fig:aliant_result}).

\begin{table}[b]
\caption{\label{tab:chi2_A0123} The values of $\chimin$ when fitting with prograde modes only (P), versus fitting with both prograde and retrograde modes (PR) (all modes fundamental $n=0$). The last row lists the ratio between the $\chimin$ values obtained with and without including retrograde modes.}
\begin{ruledtabular}
\begin{tabular}{cccccccc}
Model & A0 & A1 & A2 & A3\\
\hline
P & $0.0037$ & $0.0063$ & $0.0090$ & $0.0130$ \\
PR & $0.0035$ & $0.0041$ & $0.0046$ & $0.0055$ \\
Ratio (PR/P) & $0.93$ & $0.65$ & $0.51$ & $0.42$
\end{tabular}
\end{ruledtabular}
\end{table}

In Fig.~\ref{fig:aliant_result} both (a) and (b), the difference between excluding and including retrograde modes becomes more significant from left (A0) to right (A3).
Taking the $n=0$ case for example, we list the $\chimin$ values in Table~\ref{tab:chi2_A0123}.  
For the nonspinning binary A0, the ratio between the $\chimin$ values obtained with models including and excluding retrograde modes is close to unity, i.e., the difference is small. For binaries A1--A3, including retrograde modes largely improves the fitting accuracy and the improvement becomes more significant when $|\chi_f|$ is larger.
Again, we estimate the fitting error by comparing the results obtained from different numerical resolution levels. As shown in Fig.~\ref{fig:aliant_ot0_levs}, for all binaries A0--A3, the difference of results obtained between Lev3 and Lev2 are much smaller compared to the difference between using models including and excluding retrograde modes. Thus the results are not impacted by the numerical errors in the waveforms.

In terms of $M_{f,\rm est}$ and $\chi_{f,\rm est}$, there is tiny difference in the fitting results when adding the retrograde modes for A0. While for A1--A3, including retrograde modes improves a lot the accuracy of $M_{f,\rm est}$ and $\chi_{f,\rm est}$. Specifically, when $\chi_{f, \rm true} <0 $ while retrograde modes are not included, the algorithm tends to find a lower spin value but does not go below $\chi_f=0$, even though we allow negative $\chi_f$ values in the fitting in both cases. This is because the prograde modes for $\chi_f<0$ cannot satisfy the polarization patterns of the NR waveforms shown in Fig.~\ref{fig:polar_aliant}. 

The magnitudes of the optimal coefficients $B_{lmn}^{(S\pm)}$ and $B_{l \pm mn}^{(S\pm)}$ are also plotted in Fig.~\ref{fig:aliant_result}, characterizing the contribution of each mode. 
In (a2) and (b2), it is shown that adding retrograde modes barely influences the prograde mode coefficients (solid curves) at all discrete $t_{\rm offset}$ values for A0 (first column), and the prograde mode coefficients are consistently larger than the coefficients of the corresponding retrograde modes (dotted curves) by order(s) of magnitude. 
While for A1--A3, the retrograde modes appear to be dominant (with larger coefficients) when they are included in the fitting. In the case of A3 with $n = 0,1$, as show in the last column of (b2), the retrograde mode coefficients are dominant at all $t_{\rm offset}$ values. 
In other cases, the retrograde mode coefficients are larger at some $t_{\rm offset}$ values but switch to be smaller at other $t_{\rm offset}$'s -- and such switch always happens when $\chi_{f, \rm est}$ flips its sign. 
That is understandable from the data waveform polarization patterns and the fact that the prograde modes dominate for $\chi_f>0$ while retrograde modes dominate for $\chi_f<0$, but both cases have similar polarization patterns and frequencies and thus generate similar waveforms. 
At earlier times, the incorrect results with dominant prograde modes are caused by the lack of overtones that have non-negligible contributions before $\ttrans$, and thus the model is not accurate enough to represent the waveform, and more specifically, the correct frequencies. The switch happens at an earlier time for binaries with a larger mass ratio (A3). For A1, there is another switch at $t_{\rm offset}=25M$ as shown in the second column of (b2). This could be caused by numerical errors or the random jumps in the $(M_f,\chi_f)$ parameter space -- when $|\chi_{f, \rm true}|$ has such a small value of $0.0929$, the estimates can easily oscillate around zero.

The results and analysis in this section are all consistent with our expectations discussed in Sec.~\ref{subsec:pro_retro}.
An interesting follow-up study is to map the progenitor black hole properties to the ringdown QNM excitations. We leave it to future work.  


\section{\label{sec:conclusion}Conclusions}

In this work, using results from numerical simulations, we verified predictions of black-hole perturbation theory for gravitational waves emitted by remnant black holes of binary mergers.
In particular, we simultaneously fitted the temporal and spatial dependences of the NR ringdown waveforms to models of QNM expansion. 
Comparing between the spin-weighted spheroidal harmonics ($S$ model) and the spin-weighted spherical harmonics ($Y$ model), we have demonstrated that the $S$ model, as predicted by the Teukolsky equation, is the more faithful representation of the ringdown waveform. The combination of temporal and spatial behaviors allowed a more comprehensive study of the linearized Einstein's equations in the background of a Kerr black hole and complemented existing black hole spectroscopy studies. 

With spatial dependence included, we reinforced conclusions in previous studies of black-hole spectroscopy about higher-order $(l,m)$ modes and overtones. When the progenitor binary has asymmetric masses, higher-order angular modes are required to accurately represent the ringdown waveform. For nonspinning binaries, the necessity of taking into account the higher-order modes depends on the mass ratio and the resulting remnant spin magnitude. The $(l,m)=(2,2)$ fundamental mode is enough to achieve an accuracy above $99\%$ (i.e. $\chimin <0.01$) when modeling the ringdown with the $S$ model for $q<1.25$, while for $1.25<q<4$ and $q>5$, about three and six modes are needed, respectively, to achieve that level of accuracy (with changes subject to the grouping strategy, see Sec.~\ref{sec:strategy_lm}). On the other hand, adding overtones improves the accuracy of the model at an earlier time of the ringdown. For binaries with higher mass ratios, when more $(l,m)$ modes are included, more overtones are needed accordingly to accurately represent the early stage ringdown waveform. The fact that overtones can improve the QNM expansion when both temporal and angular patterns are matched to numerical waveforms, and the fact that the $S$ model works better than the $Y$ model, provide stronger evidence that overtones are truly excited, and that such improvement is not due to overfitting. 

During the transition from inspiral to ringdown, the magnitudes, spatial dependence, and polarization patterns of the gravitational waves during the inspiral stage are transferred to the ringdown stage. Our study confirmed this transfer. The magnitudes of the initial excitation of the $(l,m)$ ringdown QNMs are determined by the leading post-Newtonian order  of the mode during inspiral, with possible suppression due to symmetry. The polarization content of the mode (i.e., left- versus right-hand) is determined by the direction of the orbital angular momentum of the binary. 
The remnant spin direction and the excitation of prograde or retrograde mode are determined by the binary dynamics. 
When the remnant spin is aligned (anti-aligned) with the orbital angular momentum, the prograde (retrograde) modes are dominant. Including the retrograde modes is necessary to build an accurate model of the ringdown waveform in the spin anti-aligned case.
The more general cases with the remnant spin misaligned with the orbital angular momentum will be left to future work.


Under the sensitivity of the currently working detectors, higher-order angular modes, overtones, or retrograde modes generally do not play an important role in the detection or parameter estimation for most of the events. 
However, events with high signal-to-noise ratios (especially the high signal-to-noise ringdown) are expected to be observed regularly with the next generation detectors~\cite{Divyajyoti2021Detectability}. Features discussed in this work will be important for future studies of the source properties.

Finally, this work also provided a theoretical and analytical foundation for developing strategies for testing the temporal-spatial emission patterns of the ringdown. Even though, practically, each binary is only observed from one particular wave-emission direction, angular emission patter can be reconstructed by collecting multiple events.


\section*{Acknowledgements}
We thank Sizheng Ma for helpful comments on our selection of waveform from the SXS database. We also thank Sizheng Ma and Keefe Mitman for discussions about the gravitational wave memory effect and BMS frame. 
We thank Leo Stein, Maximiliano Isi, Katerina Chatziioannou and Geraint Pratten for discussions about ringdown emission. We also thank Arnab Dhani and Bangalore Sathyaprakash for discussions about mirror modes. 
X.L., and Y.C.'s research is funded by the US National Science Foundation (Grants PHY--2011968, PHY--2011961 and PHY--1836809), the Simons Foundation (Award Number 568762), and the Brinson Foundation. L.S. and E.P. acknowledge the support of the Australian Research Council Centre of Excellence for Gravitational Wave Discovery (OzGrav), Project No. CE170100004. L.S., R.K.L., and E.P. acknowledge the support of the United States National Science Foundation and the LIGO Laboratory. LIGO was constructed by the California Institute of Technology and Massachusetts Institute of Technology with funding from the United States National Science Foundation, and operates under cooperative agreement PHY--1764464. Advanced LIGO was built under Grant No. PHY--0823459. This manuscript carries LIGO Document number DCC--P2100103. We thank Matthew Giesler for reviewing the manuscript during LIGO PNP.

\onecolumngrid

\appendix

\section{\label{app:notation}Notation and terminology}

In Table~\ref{tab:notation}, we summarize the notation and terminology specifically defined and used in this paper. 

\begin{table*}[tbh!]
\caption{\label{tab:notation} Notations and terminology used in this paper.}
\begin{ruledtabular}
\begin{tabular}{ll}
Notation & Definition and description\\
\hline
$S$ & The QNM decomposition in spin-weighted spheroidal harmonics, defined in Eq.~\eqref{eq:S_model}, often labeled in superscripts. \\
$Y$ & The QNM decomposition in spin-weighted spherical harmonics, defined in Eq.~\eqref{eq:Y_model}, often labeled in superscripts. \\
$t_{\rm offset}$ & The offset of starting time from the peak of $\sum_{lm}|h_{lm}(t)|^2$ ($t_{\rm peak}$), defined in Eq.~\eqref{eq:time_offset}, the control hyperparameter in fitting.\\ 
$\chi^{2}[h,g_{\rm opt}^{(S/Y)}]$ & The optimal distance over the searched $(M_f,\chi_f)$ parameter space, defined in Eq.~\eqref{eq:chi2_opt}, shortened as $\chi^2$ in the figures. \\
$\ttrans$ & Transition time, the value of $t_{\rm offset}$ after which the optimal distance converges to a stable level, defined in Sec.~\,\ref{sec:results_G0}. \\
$\chimin$ & Minimum distance, the converged value of the optimal distance after the transition time, defined in Sec.~\,\ref{sec:results_G0}. \\
$|\Delta M_f/M_f|$, $|\Delta\chi_f/\chi_f|$ & Relative errors, the quantities used to characterize the accuracy of the estimated parameters, defined in Eqs.~\eqref{eq:relative_PE_err}.
\end{tabular}
\end{ruledtabular}
\end{table*}

\section{\label{app:QNM_convention} QNM expansion conventions}

Solutions to the Teukolsky equation can be written in the form of:
\begin{equation}
    \psi(t,r,\tilde \iota,\tilde \varphi) = e^{-i\omega_{lmn} t} R_{lmn}(r) S_{lmn}(\tilde \iota) e^{im\tilde \varphi}.
\end{equation}
Here $(l,m)$ are the angular quantum numbers, $n$ is the overtone number, and $R_{lmn}(r)$ is the radial function. 
To avoid confusion, we continue using the notations defined in the main text, i.e., the coordinates in the final spin frame $(\tilde \iota,\tilde \varphi)$ here and the remnant black hole parameters $(a_f,M_f)$, although the solutions to the Teukolsky equation 
apply to all Kerr black holes.

As it turns out, for each overtone number $n$, there is a family of modes with $\mathrm{Re}(\omega_{lmn})>0$ and another family with $\mathrm{Re}(\omega_{lmn})<0$.  They correspond to modes with polarization patterns that either rotate counterclockwise or clockwise, when the wave comes toward the observer directly face-on. We denote 
$\mathrm{Re}(\omega_{lmn}^R)>0$ as right-handed (R) and $\mathrm{Re}(\omega_{lmn}^L)<0$ as left-handed (L). At $r\rightarrow +\infty$, for both families, we can write:
\begin{align}
    \psi^R_{lmn}(t,r\rightarrow +\infty,\tilde \iota,\tilde \varphi) \sim  e^{-i\omega^R_{lmn} (t-r^*)} r^{-1} S^R_{lmn}(\tilde \iota) e^{im\tilde \varphi},\\
    \psi^L_{lmn}(t,r\rightarrow +\infty,\tilde \iota,\tilde \varphi)
    \sim e^{-i\omega_{lmn}^L (t-r^*)} r^{-1} S^L_{lmn}(\tilde \iota) e^{im\tilde \varphi},
    \label{eqLmode}
\end{align}
where $r^*$ is the tortoise coordinate~\cite{Cook2014Gravitational}.

Note that $\omega_{lmn}^R$ corresponds to the usual tabulated values of QNM frequencies, so we write 
\begin{equation}
    \omega^R_{lmn} = \omega_{lmn}. 
\end{equation}
Here $m>0$ are prograde, and $m<0$ are retrograde modes. 
We also have
\begin{equation}
    S^R_{lmn}(\tilde \iota) e^{im\tilde \varphi}  ={}_{-2}S_{lmn}(\chi_f M_f \omega_{lmn},\tilde \iota,\tilde \varphi).
\end{equation}
The frequencies and the angular mode functions of the $L$ and $R$ modes are related, and we wish to make this relation explicit. We notice that if $\psi(t,r,\tilde \iota,\tilde \varphi)$ is an outgoing solution to the Teukolsky equation, then $\psi^*(t,r,\pi-\tilde \iota,\tilde \varphi)$ is also an outgoing solution, with $*$ denoting the complex conjugate. In this way, $\psi_{lmn}^{R*}(t,r,\pi-\tilde \iota,\tilde \varphi)$ is also a QNM, with 
\begin{align}
\psi_{lmn}^{R*}(t,r,\pi-\tilde \iota,\tilde \varphi) 
= e^{i\omega^{*}_{lmn} (t-r^*)}
r^{-1} {}_{-2}S_{lmn}^*(\chi_f M_f \omega_{lmn},\pi-\tilde \iota,\tilde \varphi) .
\end{align}
We can further write~\cite{Cook2014Gravitational}
\begin{eqnarray}
{}_{-2}S_{lmn}^*(\chi_f M_f \omega_{lmn},\pi-\tilde \iota,\tilde \varphi) &=& {}_{-2}S_{lmn}^*(\chi_f M_f \omega_{lmn},\pi-\tilde \iota) e^{-im\tilde \varphi} \\
&=& {}_{-2}S_{lmn}(\chi_f M_f \omega_{lmn}^*,\pi-\tilde \iota) e^{-im\tilde \varphi} \\
&=& (-1)^l {}_{-2}S_{l-mn}(-\chi_f M_f \omega_{lmn}^*,\iota) e^{-im\tilde \varphi},
\end{eqnarray}
leading to 
\begin{align}
\psi_{lmn}^{R*}(t,r,\pi-\tilde \iota,\tilde \varphi) 
= (-1)^l e^{i\omega^{*}_{lmn}(t-r^*)}
r^{-1} {}_{-2}S_{l-mn}(-\chi_f M_f \omega^*_{lmn},\tilde \iota,\tilde \varphi).
\label{psi_R*}
\end{align}
By comparing Eqs.~(\ref{psi_R*}) and (\ref{eqLmode}) and replacing $m$ with $-m$, we can see that
\begin{equation}
    \omega^L_{lmn} =-\omega_{l-mn}^*\,, \quad 
\psi^L_{lmn}(t,r,\tilde \iota,\tilde \varphi) = (-1)^l \psi_{l-mn}^{R*}(t,r,\pi-\tilde \iota,\tilde \varphi).
\end{equation}

Using the symmetry above, we can expand the QNM in two different ways. To avoid confusion, we intend to write out all arguments explicitly. The first way is to group in terms of $(l,m)$ spin-weighted spherical harmonics (see Table~\ref{tab:QNM_convention} for corresponding notations [a][b][c][d]):
\begin{align}
h(t,r,\tilde \iota,\tilde \varphi) &\sim \sum_{l=2}^{l_{\rm max}}\sum_{m=-l}^{l}\sum_{n=0}^{n_{\rm max}}
\left[A^R_{lmn} \psi^R_{lmn}(t,r,\tilde \iota,\tilde \varphi)+A^L_{lmn} \psi^L_{lmn}(t,r,\tilde \iota,\tilde \varphi) \right] \\
&=\frac{M_f}{r} \sum_{l=2}^{l_{\rm max}}\sum_{m=-l}^{l}\sum_{n=0}^{n_{\rm max}} \Big[A^R_{lmn} e^{-i\omega_{lmn}^R (t-r^*)} S^R_{lmn}(\tilde \iota)e^{im \tilde \varphi}+A^L_{lmn} (-1)^l \psi_{l-mn}^{R*}(t,r,\pi-\tilde \iota,\tilde \varphi)\Big] \\ 
&=\frac{M_f}{r} \sum_{l=2}^{l_{\rm max}}\sum_{m=-l}^{l}\sum_{n=0}^{n_{\rm max}} \Big[\underbrace{A^R_{lmn} e^{-i\omega_{lmn} (t-r^*)} {}_{-2} S_{lmn}(\chi_f M_f \omega_{lmn},\tilde \iota,\tilde \varphi)}_{[a]\text{ for }m>0;\ [c]\text{ for }m<0;} +
\underbrace{A^L_{lmn} e^{i\omega_{l-mn}^* (t-r^*)} {}_{-2}  S_{lmn}(-\chi_f M_f \omega_{l-mn}^*,\tilde \iota,\tilde \varphi)}_{[d]\text{ for }m>0;\ [b]\text{ for }m<0.}\Big] .\label{eq:sum_way_1}
\end{align}
In this way, $A_{lmn}^R$ and $A_{lmn}^L$ correspond to the excited modes with different absolute oscillation frequencies and decay rates; there is one prograde and one retrograde mode in each $lmn$ group. The two terms have different polarization patterns (R and L), but the same emission direction: for $m>0$, both terms emit toward the north, while for $m<0$, both terms emit toward the south.  

Alternatively, switching $m$ and $-m$ for the $A^L_{lmn}$ term in Eq.~\eqref{eq:sum_way_1}, we can regroup the summation as follows: 
\begin{align}
h(t,r,\tilde \iota,\tilde \varphi) &= \frac{M_f}{r} \sum_{l=2}^{l_{\rm max}}\sum_{m=-l}^{l}\sum_{n=0}^{n_{\rm max}} \left[A^R_{lmn} e^{-i\omega_{lmn} (t-r^*)}  {}_{-2} S_{lmn}(\chi_f M_f \omega_{lmn},\tilde \iota,\tilde \varphi)+
A^L_{l-mn} e^{i\omega_{lmn}^* (t-r^*)} {}_{-2} S_{l-mn}(-\chi_f M_f \omega_{lmn}^*,\tilde \iota,\tilde \varphi)\right] \\
&= \frac{M_f}{r} \sum_{l=2}^{l_{\rm max}}\sum_{m=-l}^{l}\sum_{n=0}^{n_{\rm max}} \Big[A^R_{lmn}  e^{-i\omega_{lmn} (t-r^*)} {}_{-2} S_{lmn}(\chi_f M_f \omega_{lmn},\tilde \iota,\tilde \varphi)+
A^L_{l-mn} e^{i\omega_{lmn}^* (t-r^*)}  (-1)^l {}_{-2}S_{lmn}^*(\chi_f M_f \omega_{lmn},\pi-\tilde \iota,\tilde \varphi) \Big]\\
&= \frac{M_f}{r} \sum_{l=2}^{l_{\rm max}}\sum_{m=-l}^{l}\sum_{n=0}^{n_{\rm max}} \Big[\underbrace{A^{(+)}_{lmn} e^{-i\omega_{lmn} (t-r^*)} {}_{-2} S_{lmn}(\chi_f M_f \omega_{lmn},\tilde \iota,\tilde \varphi)}_{[a]\text{ for }m>0;\ [c]\text{ for }m<0;}+
\underbrace{A^{(-)}_{lmn} e^{i\omega_{lmn}^* (t-r^*)}  {}_{-2}S_{lmn}^*(\chi_f M_f \omega_{lmn},\pi-\tilde \iota,\tilde \varphi)}_{[b]\text{ for }m>0;\ [d]\text{ for }m<0.} \Big].\label{eq:sum_way_2}
\end{align}
Here we have defined
\begin{equation}
A^{(+)}_{lmn} = A^{R}_{lmn} \,,\quad 
A^{(-)}_{lmn} =  (-1)^l A^{L}_{l-mn}\,.
\end{equation}
In this way of grouping, $A^{(+)}_{lmn}$ and $A^{(-)}_{lmn}$ correspond to the excited modes with different angular emission patterns (in terms of both polarization and direction), but the same absolute oscillation frequency and decay rates. Prograde and retrograde modes are not mixed into the same $lmn$ group. Eq.~\eqref{eq:sum_way_2} has the same form as we defined in the main text Eq.~\eqref{eq:S_model}. We believe this is more convenient, since modes in the same group tends to be either both excited or both not excited. For example, the entire retrograde groups can be ignored in many situations, as is done in e.g., Ref.~\cite{LVC2020Tests}.

\section{\label{app:single_direction}Limitation of single-direction fittings}

Here we briefly comment on the limitation of single-direction fittings, which is one of the motivations of implementing a temporal-spatial fitting strategy in this paper. In brief, the $S/Y$ models cannot be distinguished using the signal of a single event without prior information about QNM excitations. 

The excitation amplitudes $\{B^{(S\pm/Y\pm)}_{lmn}\}$ are governed by progenitor binary dynamics, and the emission strength varies with the spherical coordinate $\vec{\Omega}$, especially the inclination angle $\iota$~\cite{Divyajyoti2021Detectability}. The waveform observed from a single direction cannot reveal angular distribution in the source frame, and thus we can only fit the waveform with a single-direction model:
\equ{\label{eq:singledirection}
    h^{\rm SD} (t)=\frac{M_f}{r}
    \sum_{l=2}^{l_{\rm max}}\sum_{m=-l}^{m=l} \sum_{n=0}^{n_{\rm max}}\left[ C^{(+)}_{lmn} e^{-i\omega_{lmn}t} + C^{(-)}_{lmn} e^{i\omega^*_{lmn}t}\right],
}
with $\{C^{(\pm)}_{lmn}\}$ being the relative amplitudes of different frequency components. While we do find that single-direction fitting is capable of finding the dominant frequency components, the relative amplitudes $\{C^{(\pm)}_{lmn}\}$ do not help in the $S/Y$ model selection unless the binary inclination is known from the inspiral stage. Taking the spin aligned case for an example, the $\{C^{(\pm)}_{lmn}\}$ are related to  $\{B^{(S\pm/Y\pm)}_{lmn}\}$ via:
\begin{subequations}
\be
C^{(+)}_{lmn}=B^{(S+)}_{lmn}{}_{-2}S_{lmn}(\gamma_{lmn},\tilde \iota,\tilde \varphi),\ \  B^{(Y+)}_{lmn}{}_{-2}Y_{lm}(\iota,\varphi),
\ee
\be
C^{(-)}_{lmn}=B^{(S-)}_{lmn}{}_{-2}S^*_{lmn}(\gamma_{lmn},\pi-\tilde \iota,\tilde \varphi),\ \  B^{(Y-)}_{lmn}{}_{-2}Y^*_{lm}(\pi-\tilde \iota,\tilde \varphi),
\ee
\end{subequations}
for the $S/Y$ models, respectively. However, in practice, the inclination angle $\iota$ is usually not well constrained, even from the full inspiral-merger-ringdown waveform fitting. Thus, the $S$ and $Y$ models are not distinguishable from a single event. 

In addition, due to the parameter degeneracy in single-direction fitting, it is likely that different $(M_f,\chi_f)$ values with different dominate QNMs could result in the same set of dominant frequencies. Thus, for non-face-on emission or more complicated binaries with precession and/or misaligned spin~\cite{Healy2013Template,Finch2021modelling}, the lack of spatial information might lead to incorrect estimation of the source parameters. This degeneracy could in principle be broken if we know the relative amplitude of each QNM \emph{a priori}, and we can, in turn, use the relative amplitudes of different frequency components to estimate the inclination angle. 


\section{\label{app:regression_algorithm}Numerical implementation for optimizing $\{B_{lmn}^{(\pm)}\}$}

In the fitting, the NR ringdown waveform is treated as data, $h(\Vec{\Omega},t)$. The template waveform $g^{(S/Y)}(\Vec{\Omega},t)$ is built from Eq.\,\eqref{eq:S_model} and Eq.\,\eqref{eq:Y_model} for the $S$ and $Y$ models, respectively, with excitation coefficients $\{B^{(S\pm/Y\pm)}_{lmn}\}$ to be determined. In numerical realization, we use discretized representation to express temporal-spatial functions and their inner products. We define a $N_t \times N_{\Omega}$ matrix $\mathbb{M}_h$ to represent the temporal-spatial function $h(\Vec{\Omega},t)$, where $N_t$ is the number of discretized time steps and $N_{\Omega}$ is the number of spatial points that include reasonably sampled $\iota$ and $\varphi$ values:
\begin{equation}\label{eq:h_matrix0}
\mathbb{M}_h = \left(\begin{array}{cccc}
\bar h(\Vec{\Omega}_1,t_1) & \bar h(\Vec{\Omega}_1,t_2) & ... & \bar h(\Vec{\Omega}_1,t_{N_t})\\ 
\bar h(\Vec{\Omega}_2,t_1) & \bar h(\Vec{\Omega}_2,t_2) & ... & \bar h(\Vec{\Omega}_2,t_{N_t})\\ 
... & ... & ... & ... \\
\bar h(\Vec{\Omega}_{N_\Omega},t_1) & \bar h(\Vec{\Omega}_{N_\Omega},t_2) & ... & \bar h(\Vec{\Omega}_{N_\Omega},t_{N_t})
\end{array}\right).
\end{equation}
In the matrix \eqref{eq:h_matrix0}, $\bar h(\Vec{\Omega}_i,t_j)\equiv\sqrt{\sin{\iota_i}}\,h(\Vec{\Omega}_i,t_j)$ is the strain value at a specific spatial point $\Vec{\Omega}_i=(\iota_i,\varphi_i)$ and a specific time step $t_j$, weighted by $\sqrt{\sin{\iota_i}}$, the square root of the Jacobian for a unit sphere. This weight factor is introduced because we are going to represent the temporal-spatial inner product (Eq.\,\eqref{eq:2D_innerproduct}) using the vector product in Eq.\,\eqref{eq:2D_vectorproduct}, and want to have the Jacobian $\sin{\iota_i}$ shared equally by the data and template waveforms. We further express the matrix in a $1\times N_t N_{\Omega}$ row vector:
\begin{equation}\label{eq:h_vector0}
 \vec{\mathbb{V}}_h = 
\Big( \bar h(\Vec{\Omega}_1,t_1), \ ...\,,\ \bar h(\Vec{\Omega}_1,t_{N_t}), \ \bar h(\Vec{\Omega}_2,t_1), \ ...\,, \ \bar h(\Vec{\Omega}_{N_\Omega-1},t_{N_t}), \ \bar h(\Vec{\Omega}_{N_\Omega},t_1),\  ...\,,\  \bar h(\Vec{\Omega}_{N_\Omega},t_{N_t})
\Big).
\end{equation}
For the template waveform $g(\vec\Omega,t)$, using the similar matrix and vector representations as \eqref{eq:h_matrix0} and \eqref{eq:h_vector0}, the temporal-spatial inner product in Eq.\,\eqref{eq:2D_innerproduct} can be represented by the vector product as follows (c.f. Eq.\eqref{eq:gh_gg_inner}):
\begin{equation}\label{eq:2D_vectorproduct}
    \left\langle g \mid h\right\rangle = \vec{\mathbb{V}}_h \vec{\mathbb{V}}_g^\dagger.
\end{equation}

With a given set of parameters $(M_f,\chi_f)$, each $lmn$ mode contributes to two $1\times N_t N_{\Omega}(M_f,\chi_f)$ vectors $\vec{\mathbb{V}}_{lmn}^{(S\pm/Y\pm)}(M_f,\chi_f)$, for $S/Y$ model respectively. Combined with coefficients $B^{(S\pm/Y\pm)}_{lmn}$, we can construct the discretized template waveform $g^{(S/Y)}(\vec\Omega,t;M_f,\chi_f)$ in the \eqref{eq:h_vector0} vector form:
\begin{equation}\label{eq:template_vector}
   \vec{\mathbb{V}}_g^{(S/Y)}(M_f,\chi_f) = \sum_{lmn} \Big( B^{(S+/Y+)}_{lmn} \,\vec{\mathbb{V}}_{lmn}^{(S+/Y+)}(M_f,\chi_f) + \ \ B^{(S-/Y-)}_{lmn} \,\vec{\mathbb{V}}_{lmn}^{(S-/Y-)}(M_f,\chi_f)\Big).
\end{equation}
Considering the $\pm$ components, there are $2N_{lmn}$ modes in total. We form a large matrix for these $2N_{lmn}$ modes:
\begin{equation}\label{eq:template_matrix}
\mathcal{M}_g^{(S/Y)}(M_f,\chi_f) = \left(\begin{array}{cccc}
\vspace{0.1cm}
\vec{\mathbb{V}}_{(lmn)_1}^{(S+/Y+)}(M_f,\chi_f)\\
\vspace{0.1cm}
...\\
\vspace{0.1cm}
\vec{\mathbb{V}}_{(lmn)_{N_{lmn}}}^{(S+/Y+)}(M_f,\chi_f)\\
\vspace{0.1cm}
\vec{\mathbb{V}}_{(lmn)_1}^{(S-/Y-)}(M_f,\chi_f)\\
\vspace{0.1cm}
...\\
\vspace{0.1cm}
\vec{\mathbb{V}}_{(lmn)_{N_{lmn}}}^{(S-/Y-)}(M_f,\chi_f)
\end{array}\right),
\end{equation}
and assemble the matrix with the coefficient vector,
\begin{equation}
\vec{B}^{\,(S/Y)} = \left(B_{(lmn)_1}^{(S+/Y+)},\ ...\,,\ B_{(lmn)_{N_{lmn}}}^{(S+/Y+)},\ B_{(lmn)_1}^{(S-/Y-)},\ ...\,,\ B_{(lmn)_{N_{lmn}}}^{(S-/Y-)}\right),
\end{equation}
such that the discretized template waveform \eqref{eq:template_vector} can be expressed as: 
\begin{equation}
\vec{\mathbb{V}}_g^{(S/Y)}(M_f,\chi_f) = \vec{B}^{\,(S/Y)} \mathcal{M}_g^{(S/Y)}(M_f,\chi_f).
\end{equation}

The fitting procedure is as follows: For each set of remnant parameters $(M_f,\chi_f)$, we obtain the QNM frequencies $\{\omega_{lmn}\}$ using the \texttt{qnm} package~\cite{Stein:2019mop} and compose the template waveform \eqref{eq:template_matrix} for $S$ and $Y$ models, separately. We then apply multi-variable linear regression~\cite{freedman2009statistical} to determine the least-squares excitation coefficients $B^{(S\pm/Y\pm)}_{lmn}$ (c.f. Eq.~\eqref{eq:coef_ls}): 
\begin{equation}\label{eq:vls}
\vec{B}_{\operatorname{l-s}}^{\,(S/Y)} (M_f,\chi_f) =
\big[\vec{\mathbb{V}}_{h} \mathcal{M}_{g}^{(S/Y)\dagger}(M_f,\chi_f)\big] \big[\mathcal{M}_{g}^{(S/Y)}(M_f,\chi_f) \mathcal{M}_{g}^{(S/Y)\dagger}(M_f,\chi_f)\big]^{-1},
\end{equation}
such that the distance between the data waveform $\vec{\mathbb{V}}_h$ and the template waveform $\vec{\mathbb{V}}_{g,\,\operatorname{l-s}}^{(S/Y)}(M_f,\chi_f)=\vec{B}_{\operatorname{l-s}}^{\,(S/Y)}(M_f,\chi_f)\mathcal{M}_g^{(S/Y)}(M_f,\chi_f)$ is minimized (c.f. Eq.~\eqref{eq:chi2_ls}):
\begin{equation}\label{eq:chi_opt}
\chi^2[h, \ g_{\operatorname{l-s}}^{(S/Y)}(M_f,\chi_f)]= 
\frac{\big[\vec{\mathbb{V}}_{h}-\vec{\mathbb{V}}_{g,\,\operatorname{l-s}}^{(S/Y)}(M_f,\chi_f)\big] \big[\vec{\mathbb{V}}_{h}-\vec{\mathbb{V}}_{g,\,\operatorname{l-s}}^{(S/Y)}(M_f,\chi_f)\big]^\dagger}{\vec{\mathbb{V}}_h \vec{\mathbb{V}}_h^\dagger}.
\end{equation}
Note that Eqs.\,\eqref{eq:template_vector}--\eqref{eq:chi_opt} are defined under values of $M_f$ and $\chi_f$. Then we carry out a search in the 2D parameter space of $(M_f,\chi_f)$ to find the minimum point of $\chi^2[h,g_{\operatorname{l-s}}^{(S/Y)}(M_f,\chi_f)]$ and define it as the optimal distance:
\begin{equation}\label{eq:chi2_opt}
\chi^{2}[h, g_{\mathrm{opt}}^{(S / Y)}] \equiv \min_{(M_f,\chi_f)} \chi^2[h,g_{\operatorname{l-s}}^{(S/Y)}(M_f,\chi_f)].
\end{equation}
The mass and spin that yield this optimal distance are denoted by $(M_{f,\rm est},\chi_{f,\rm est})$ (for $S$ and $Y$ models separately). The corresponding excitation coefficients \eqref{eq:vls} computed with $(M_{f,\rm est},\chi_{f,\rm est})$ are the optimal coefficients and are labeled as $\vec{B}_{\rm opt}^{\,(S/Y)}$. Comparing results between the $S$ and $Y$ models, the one that yields a smaller optimal distance demonstrates a better fit.

\section{\label{app:result_all}Full fitting results for N1--N9}

\begin{figure*}
\centering
\includegraphics[width=\textwidth]{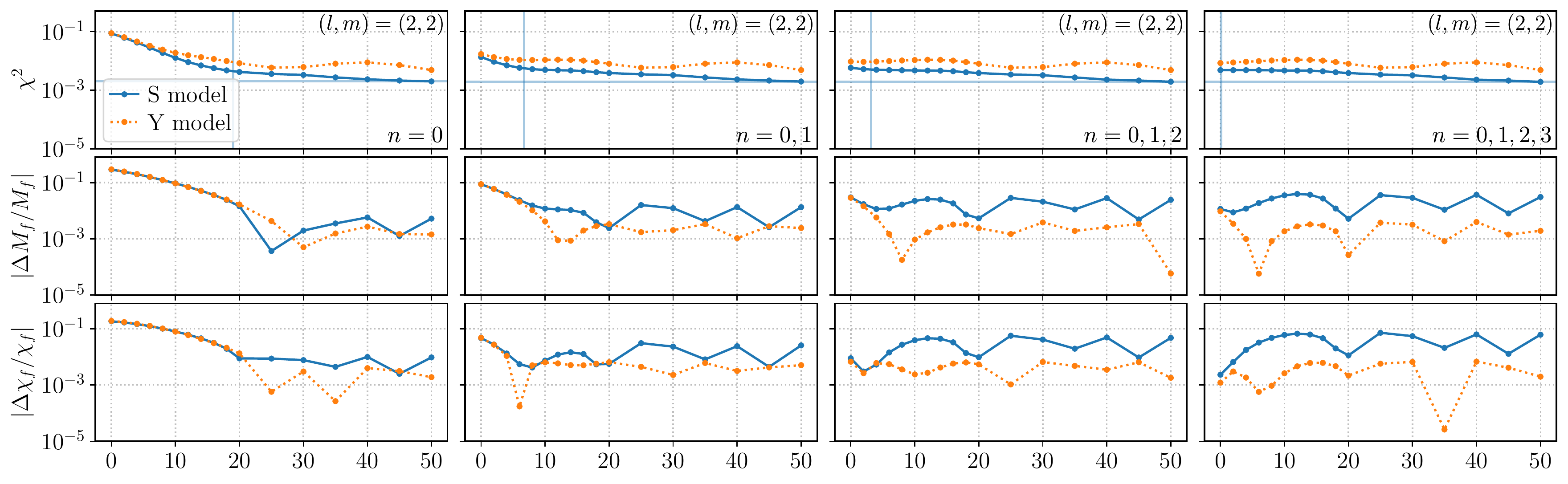}
\includegraphics[width=\textwidth]{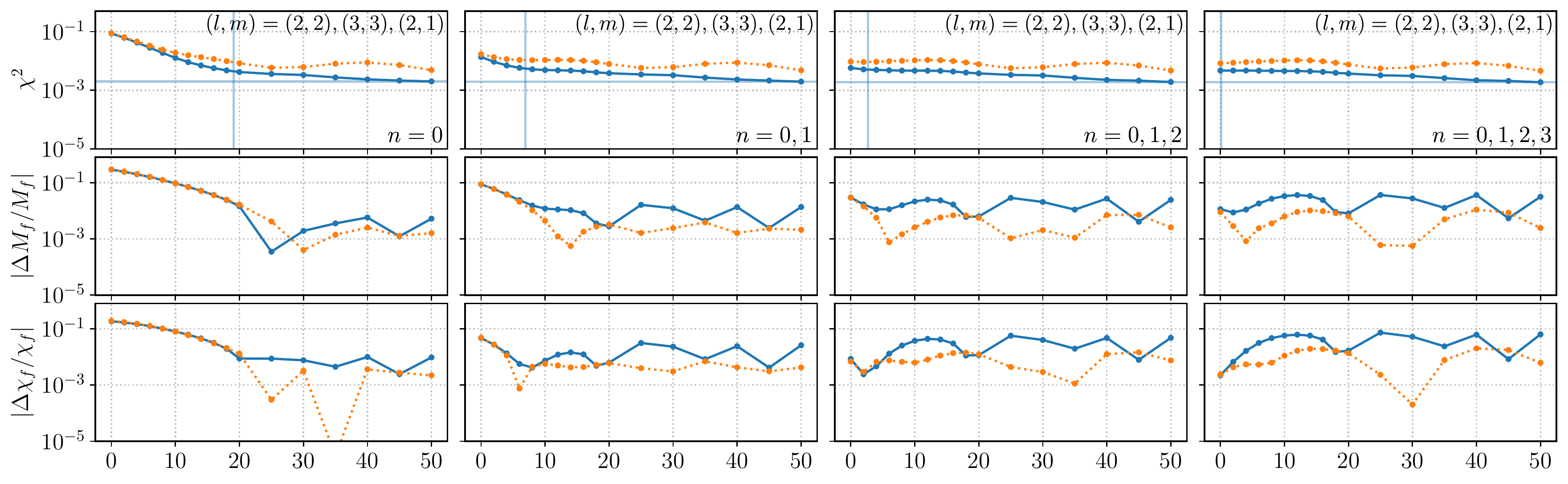}
\includegraphics[width=\textwidth]{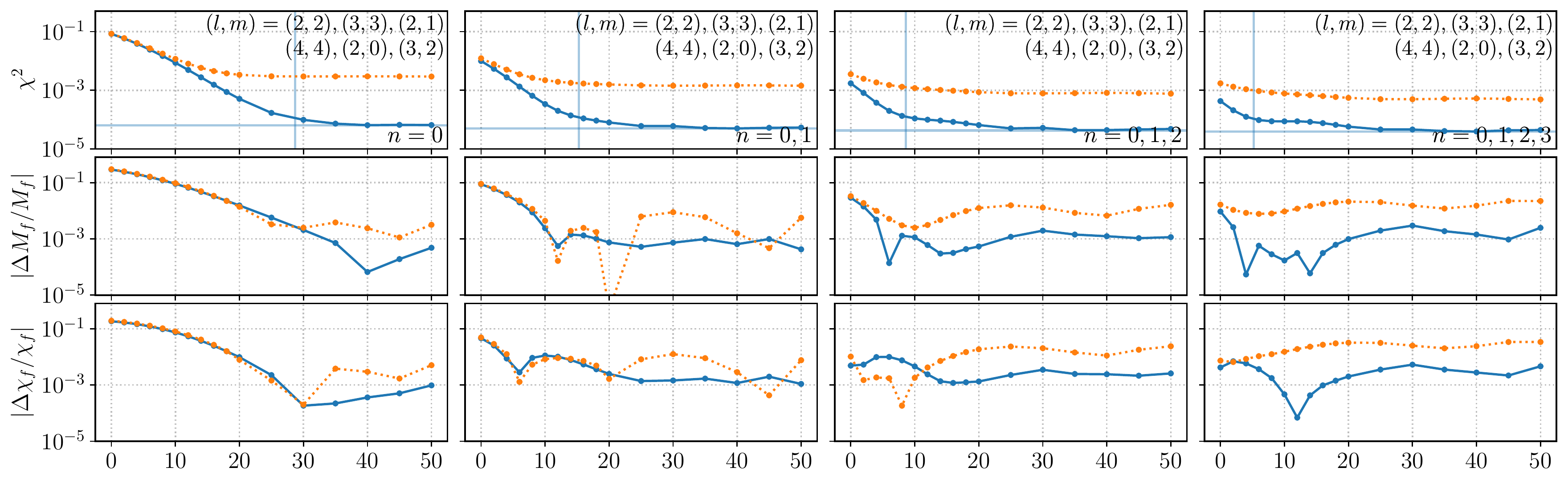}
\includegraphics[width=\textwidth]{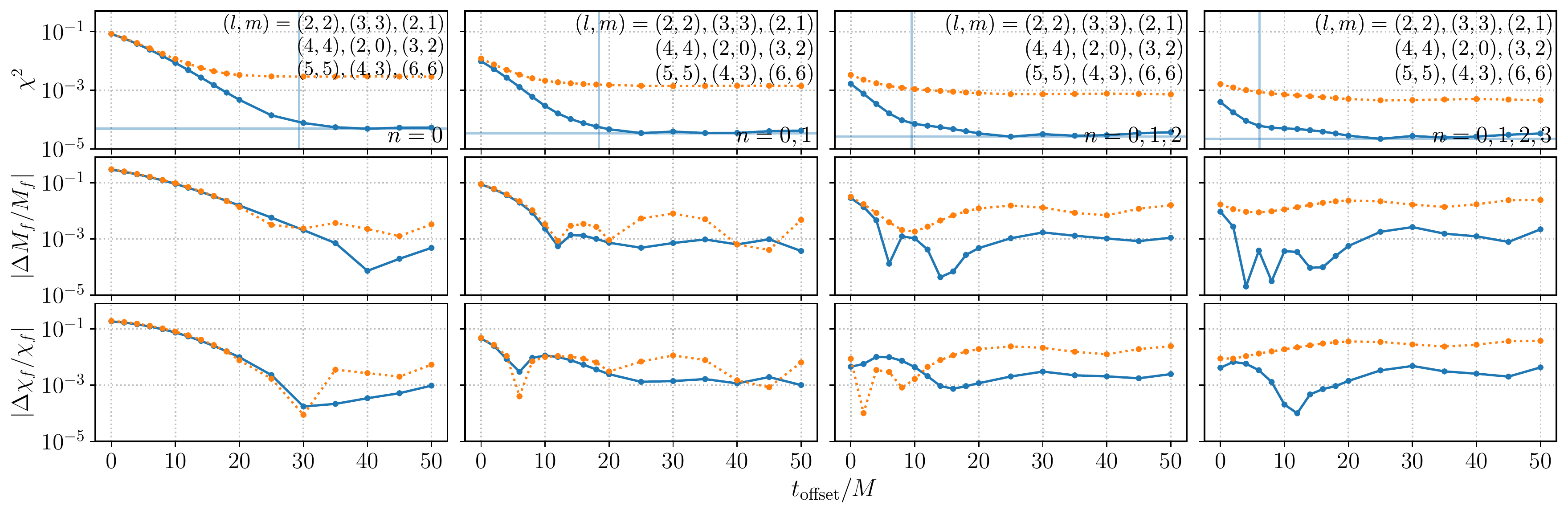}
\caption{\label{fig:N1_ots}Fitting results for binary waveform N1. Plot settings are the same as Fig.~\ref{fig:G0_ots}.}
\end{figure*}

\begin{figure*}
\centering
\includegraphics[width=\textwidth]{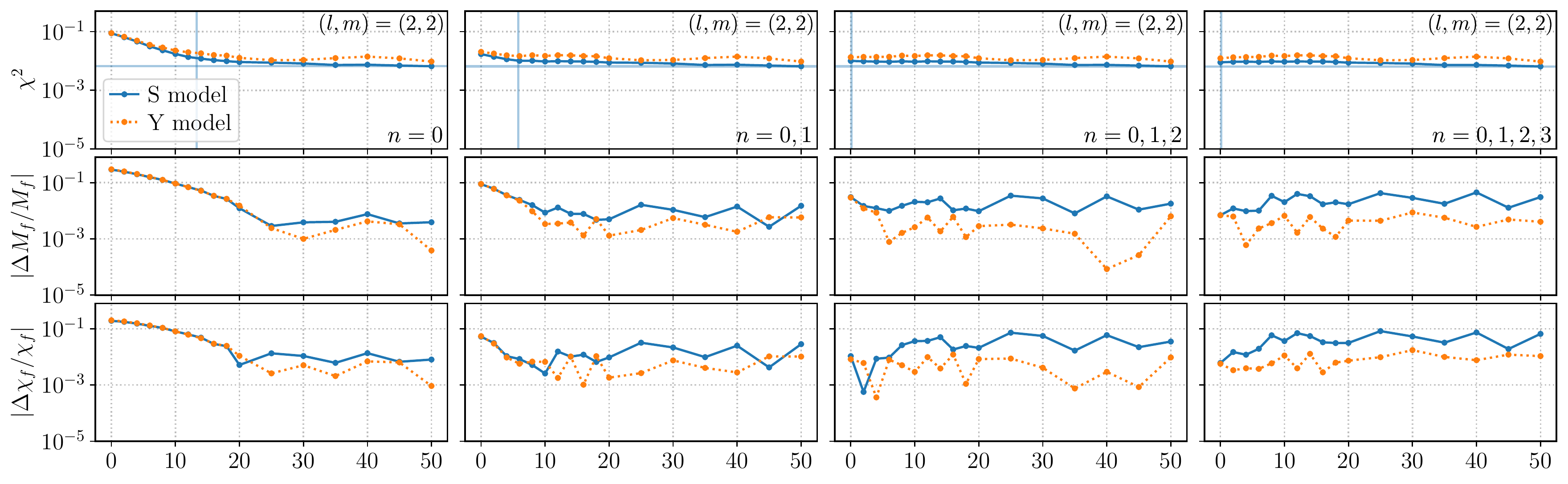}
\includegraphics[width=\textwidth]{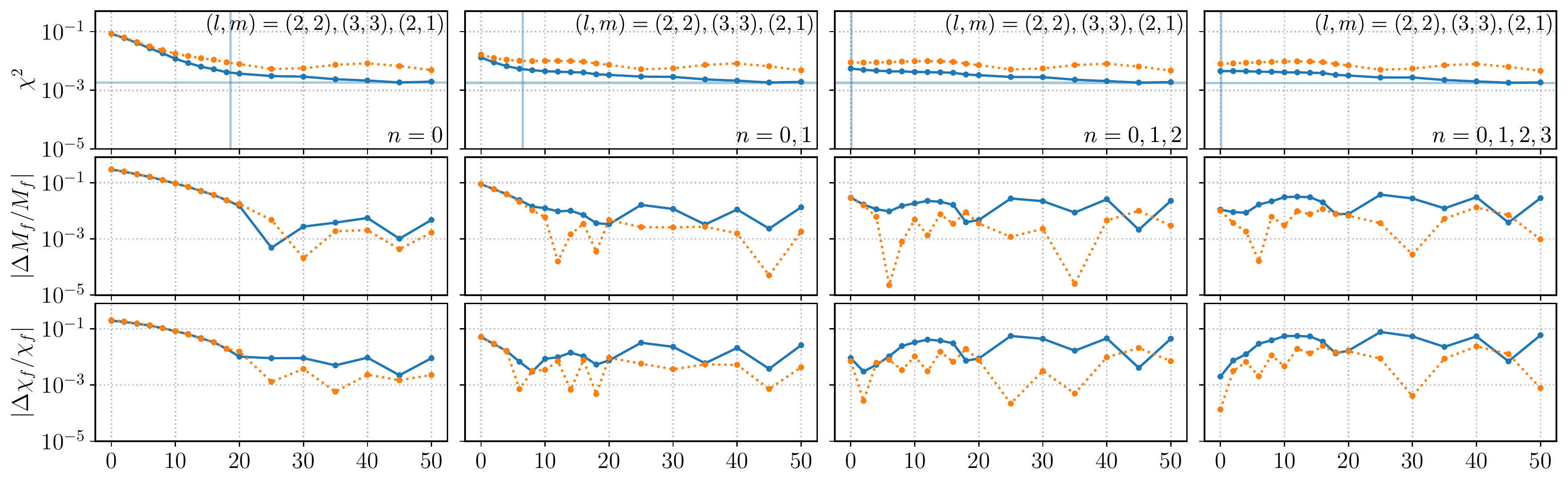}
\includegraphics[width=\textwidth]{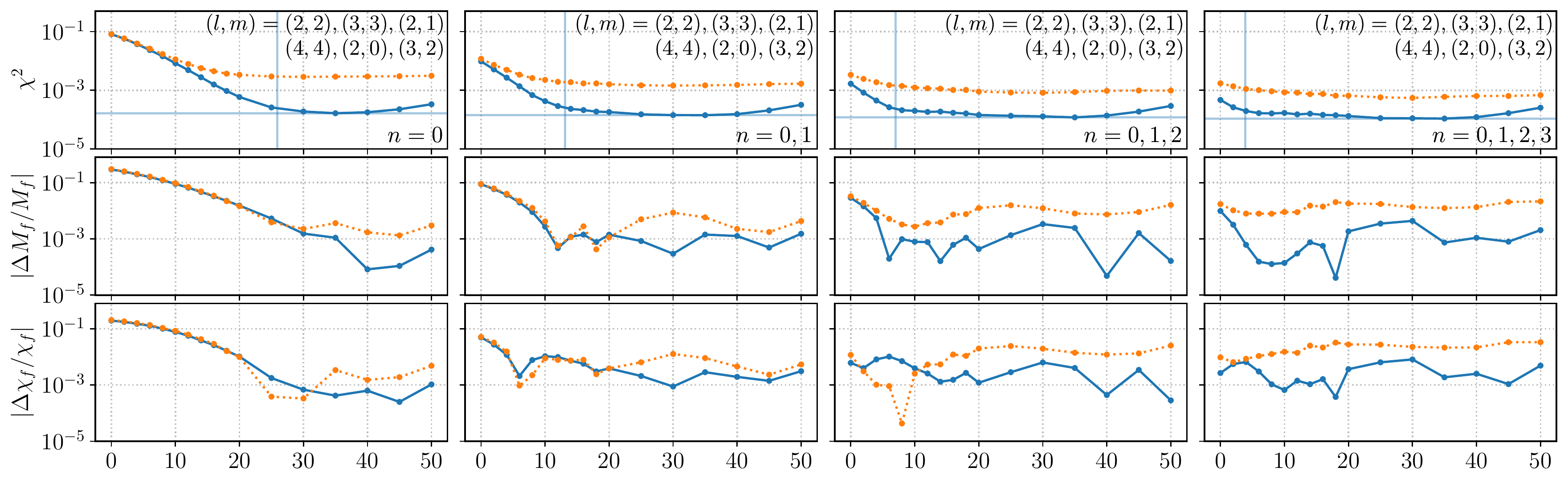}
\includegraphics[width=\textwidth]{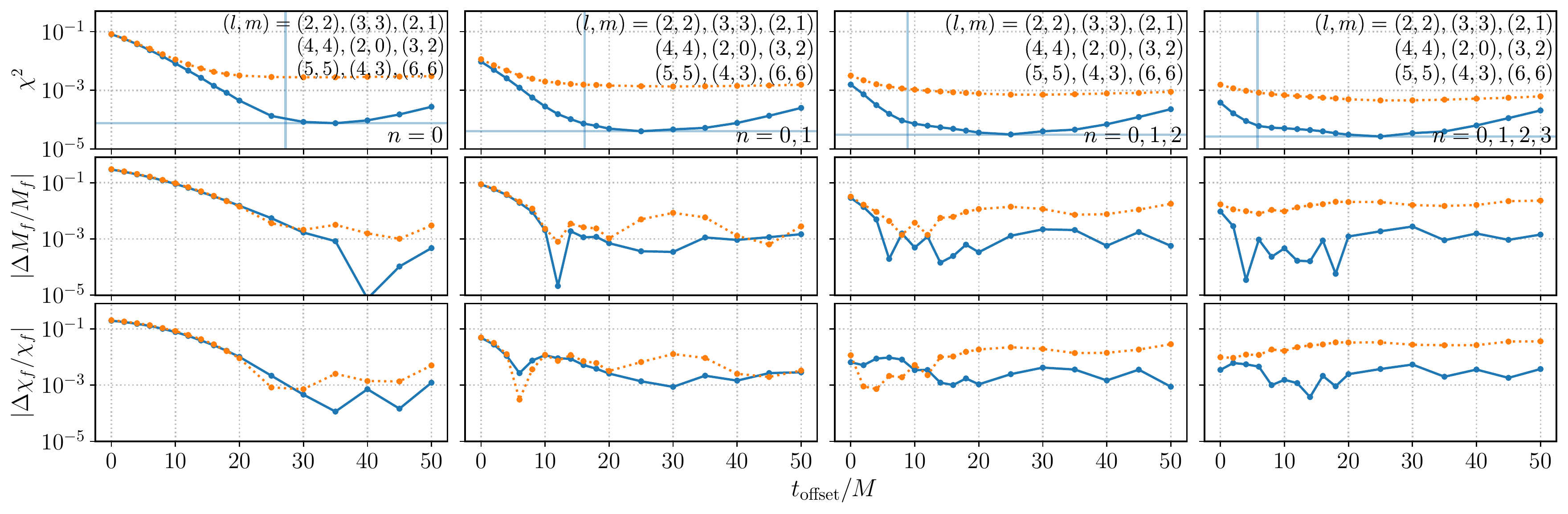}
\caption{\label{fig:N2_ots}Fitting results for binary waveform N2. Plot settings are the same as Fig.~\ref{fig:G0_ots}.}
\end{figure*}

\begin{figure*}
\centering
\includegraphics[width=\textwidth]{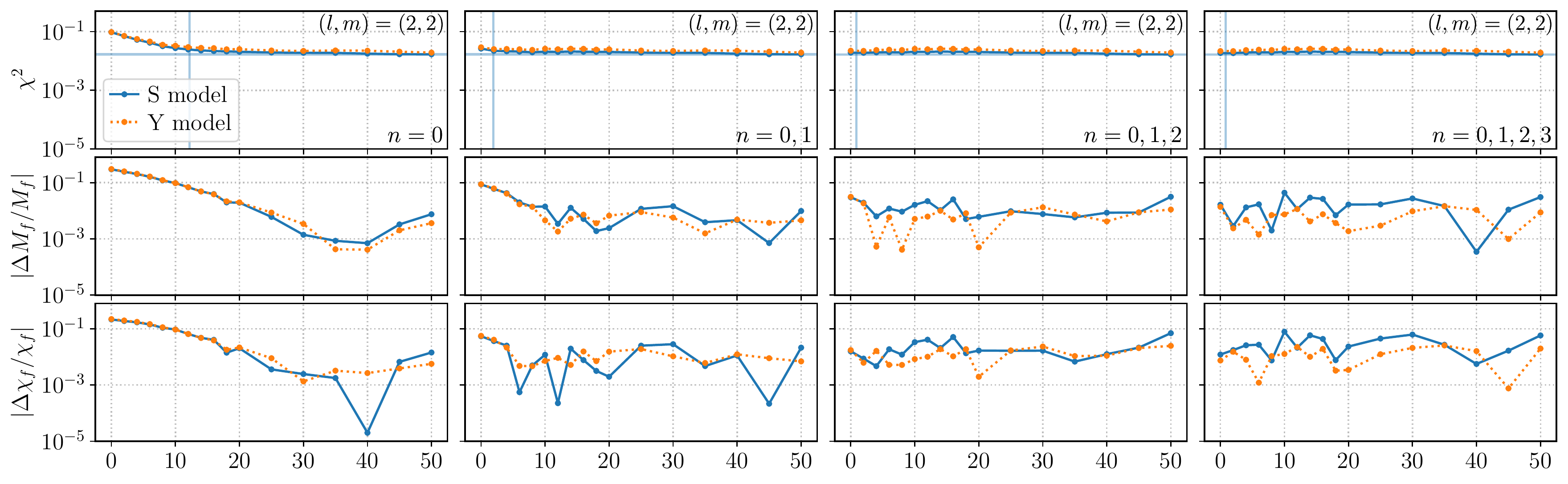}
\includegraphics[width=\textwidth]{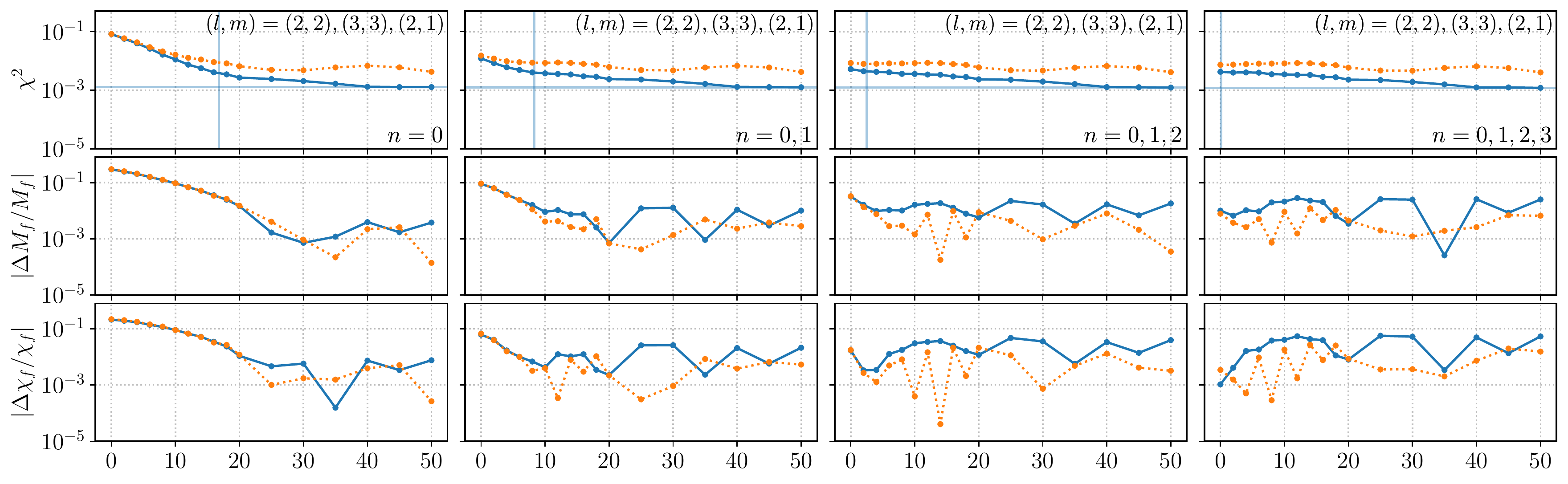}
\includegraphics[width=\textwidth]{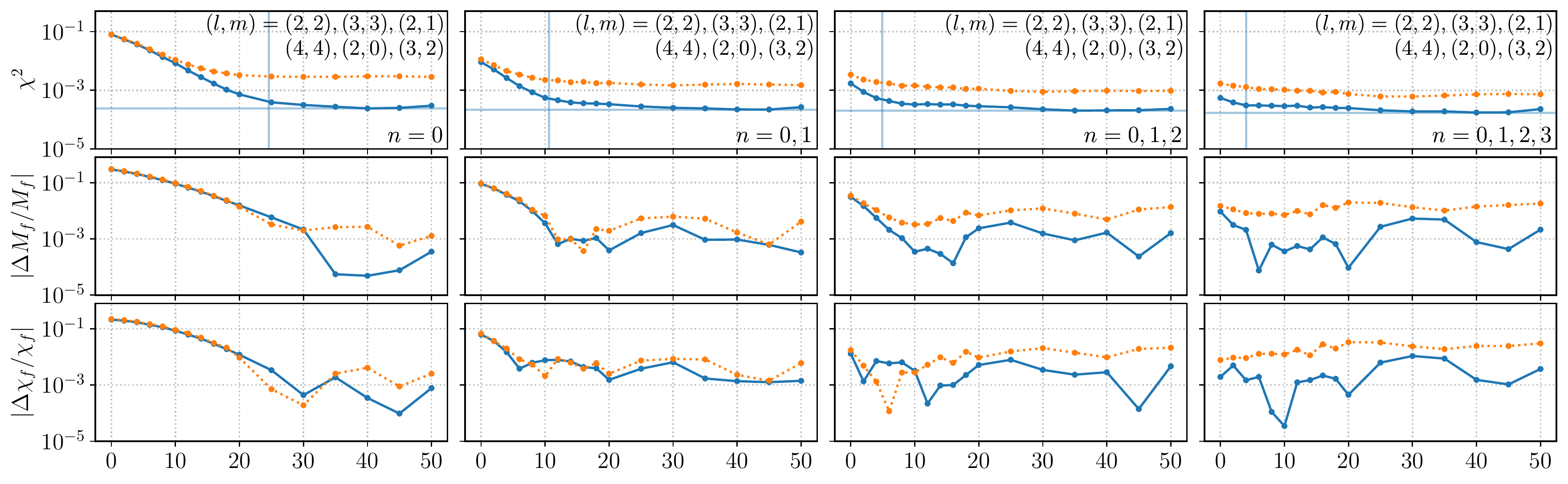}
\includegraphics[width=\textwidth]{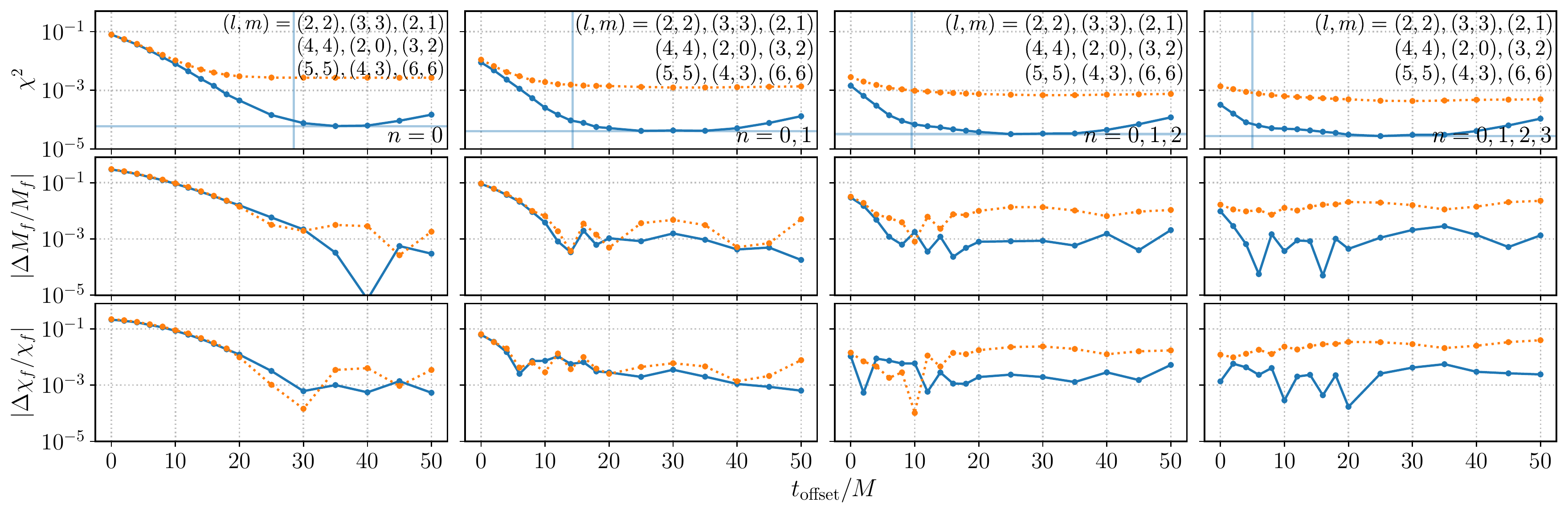}
\caption{\label{fig:N3_ots}Fitting results for binary waveform N3. Plot settings are the same as Fig.~\ref{fig:G0_ots}.}
\end{figure*}

\begin{figure*}
\centering
\includegraphics[width=\textwidth]{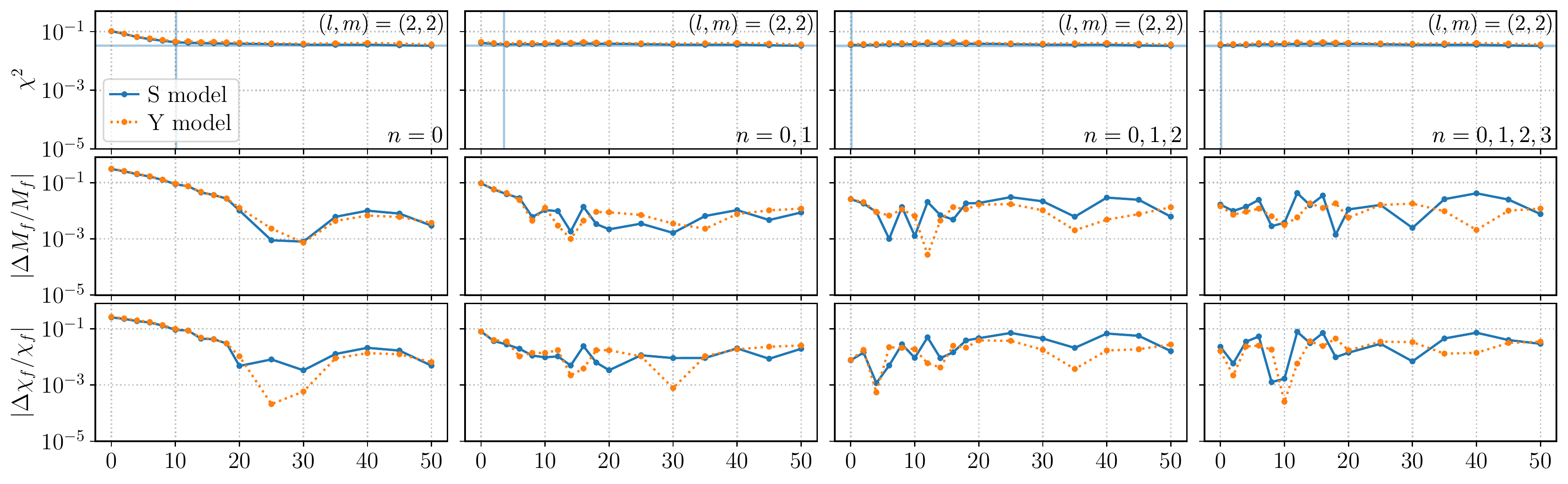}
\includegraphics[width=\textwidth]{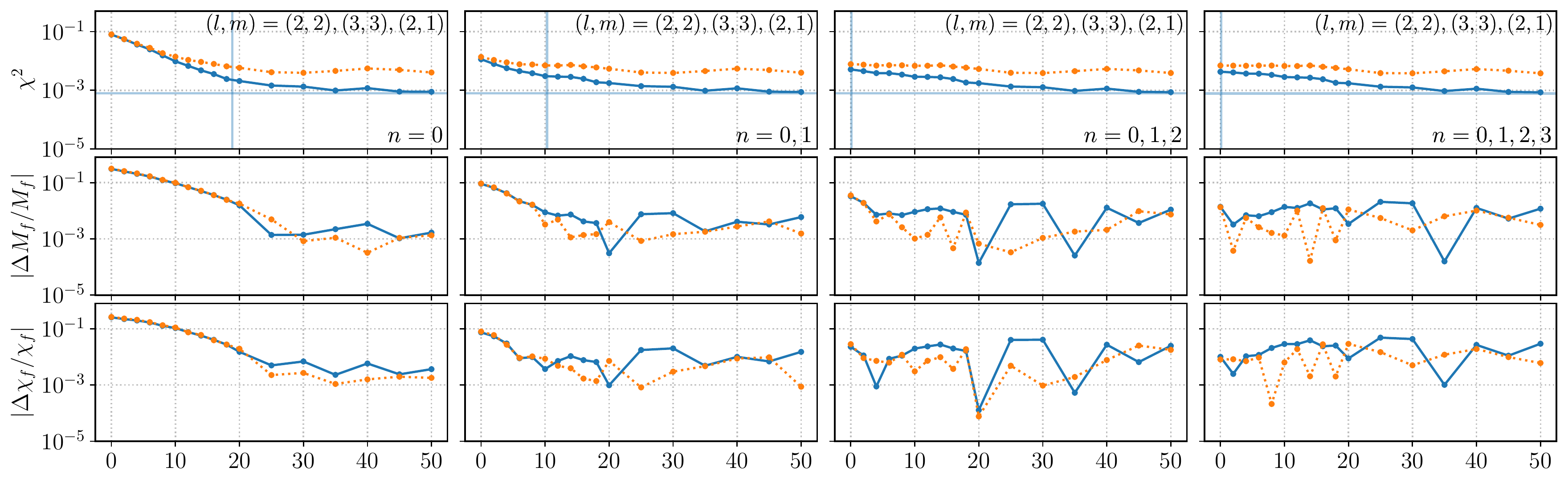}
\includegraphics[width=\textwidth]{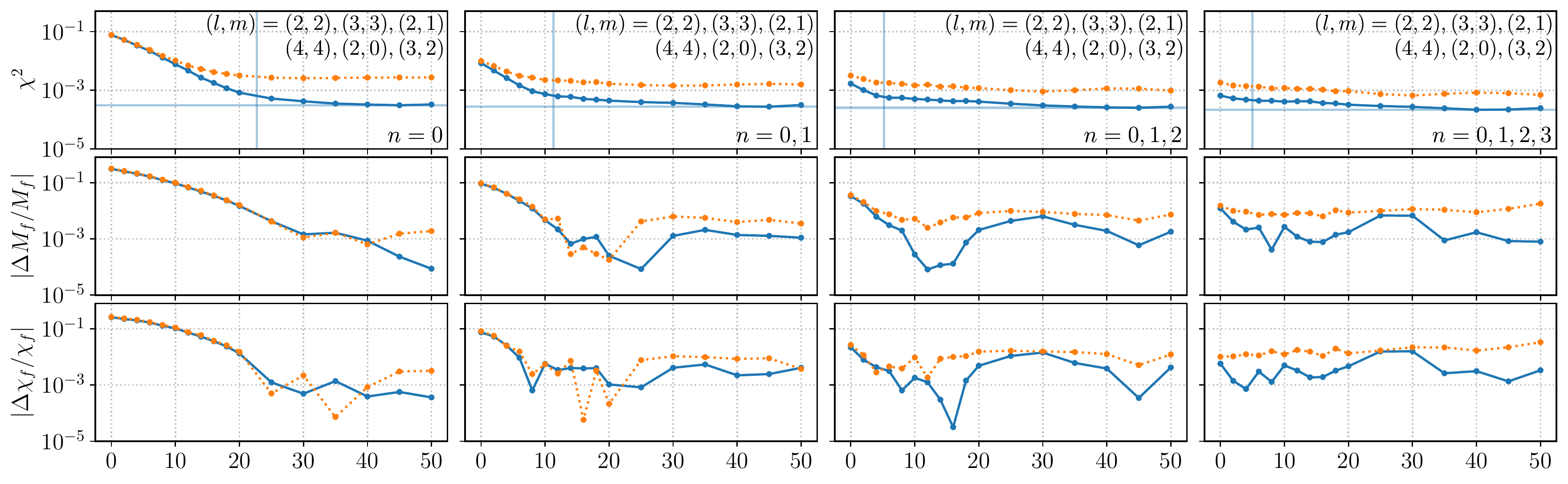}
\includegraphics[width=\textwidth]{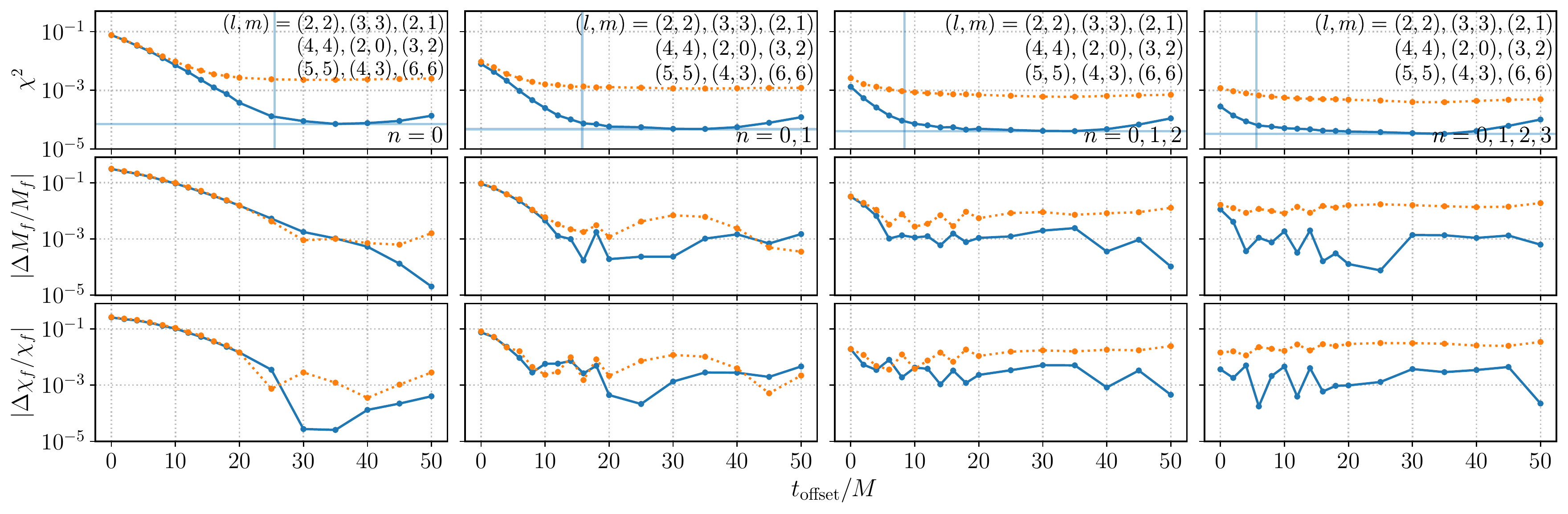}
\caption{\label{fig:N4_ots}Fitting results for binary waveform N4. Plot settings are the same as Fig.~\ref{fig:G0_ots}.}
\end{figure*}

\begin{figure*}
\centering
\includegraphics[width=\textwidth]{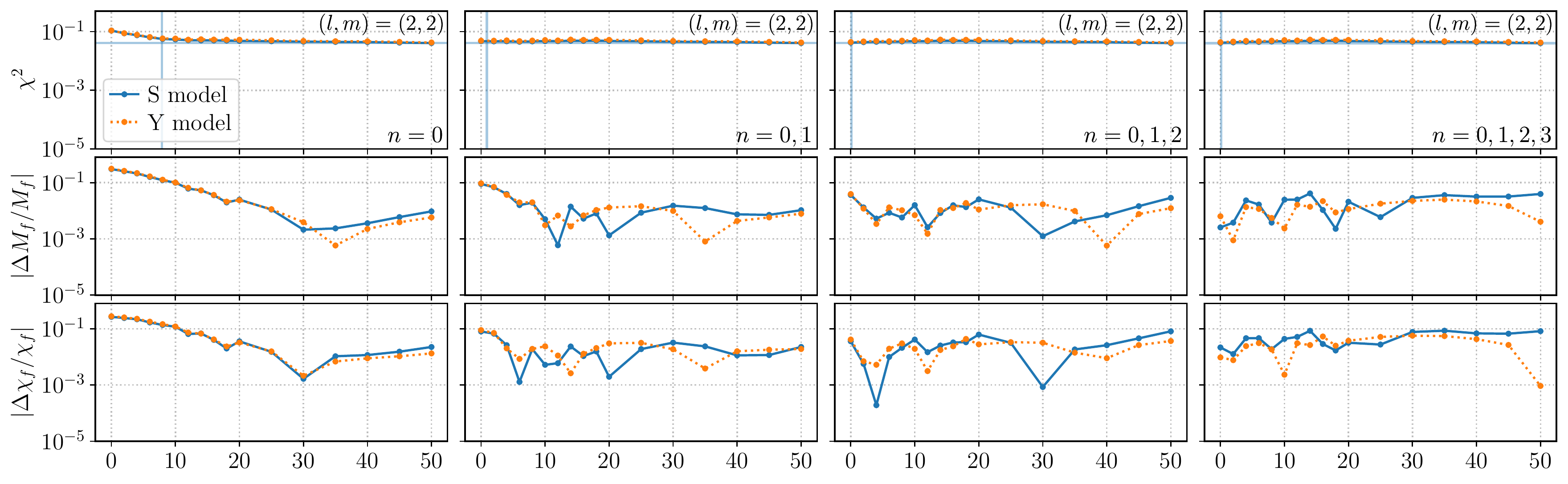}
\includegraphics[width=\textwidth]{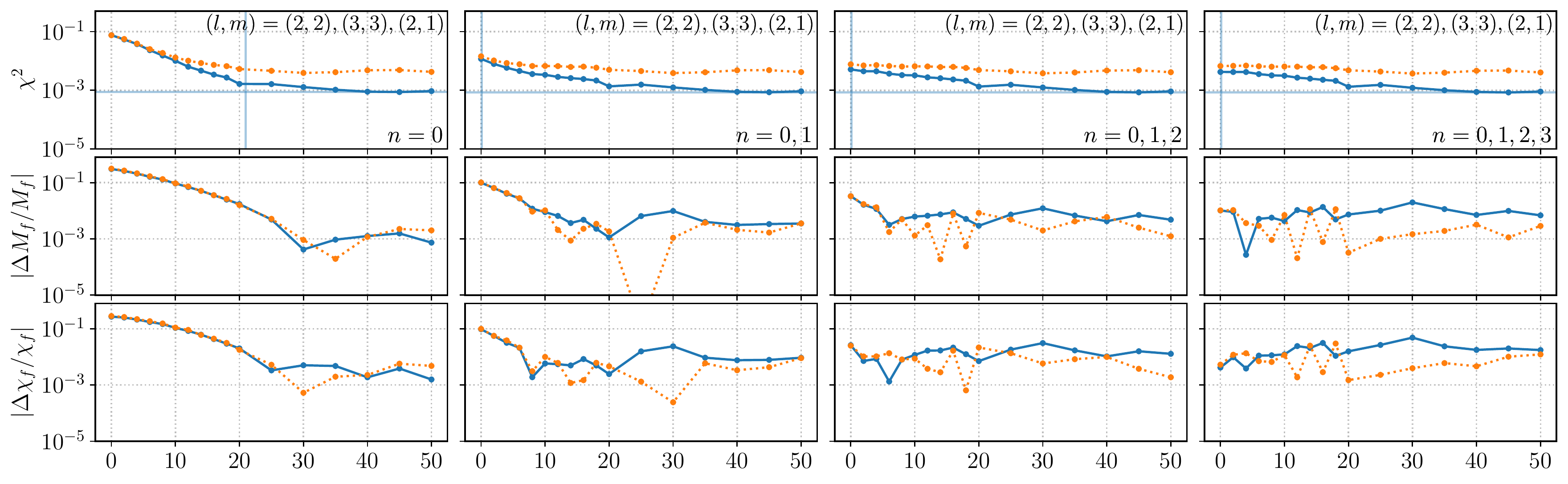}
\includegraphics[width=\textwidth]{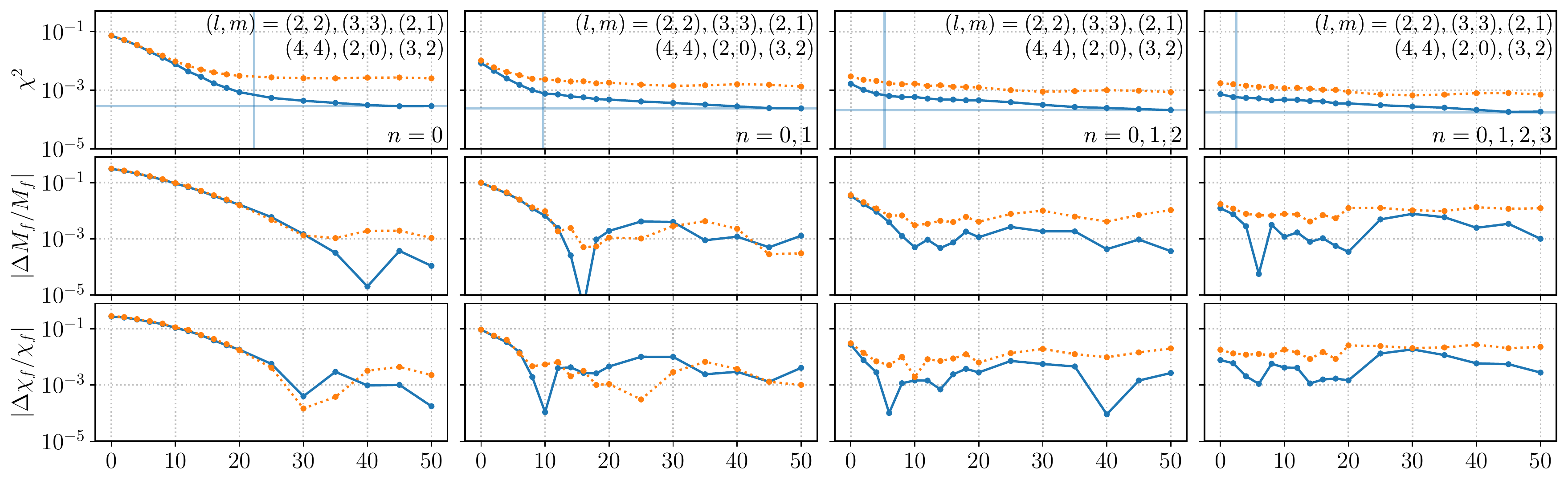}
\includegraphics[width=\textwidth]{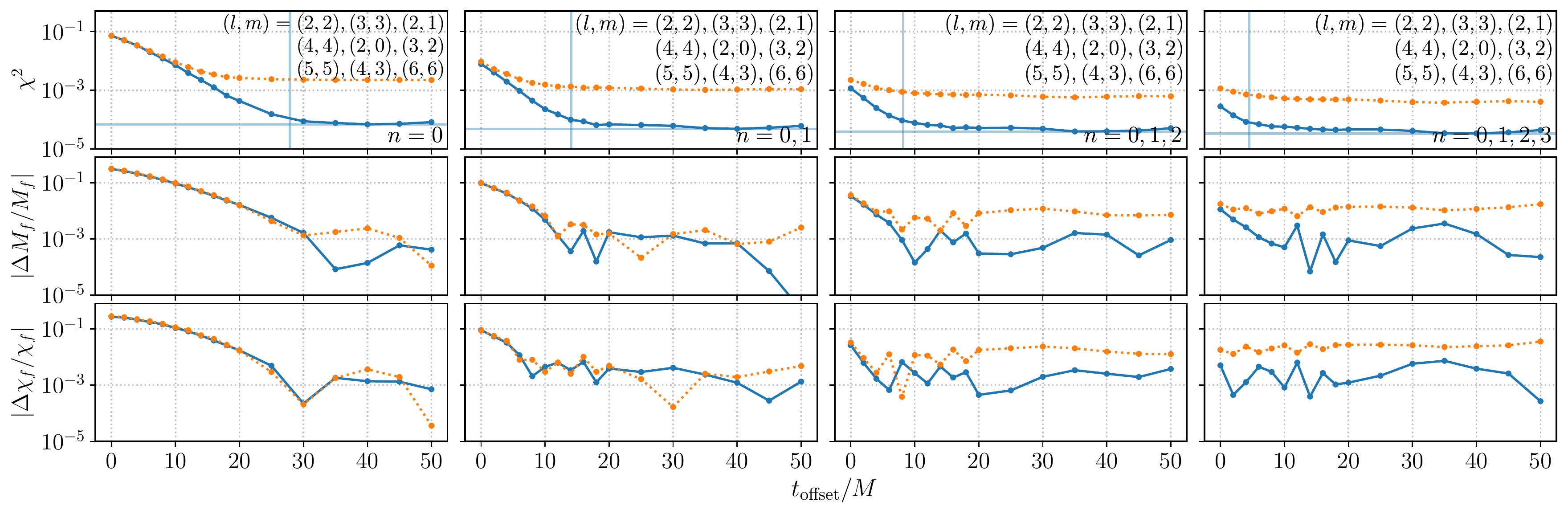}
\caption{\label{fig:N5_ots}Fitting results for binary waveform N5. Plot settings are the same as Fig.~\ref{fig:G0_ots}.}
\end{figure*}

\begin{figure*}
\centering
\includegraphics[width=\textwidth]{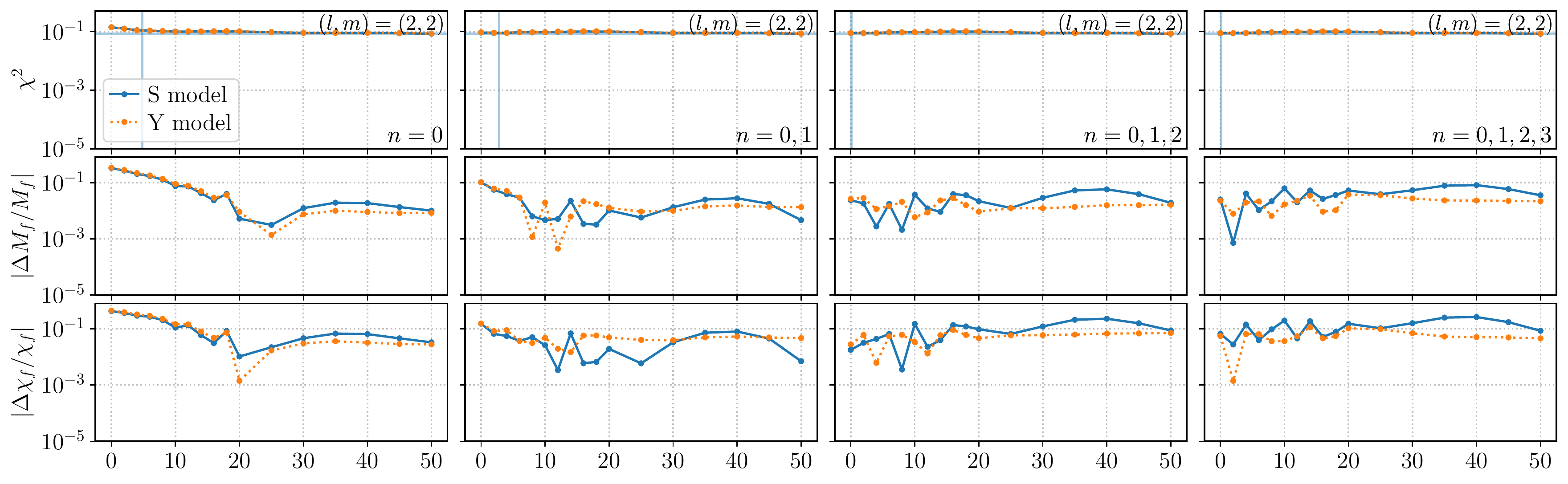}
\includegraphics[width=\textwidth]{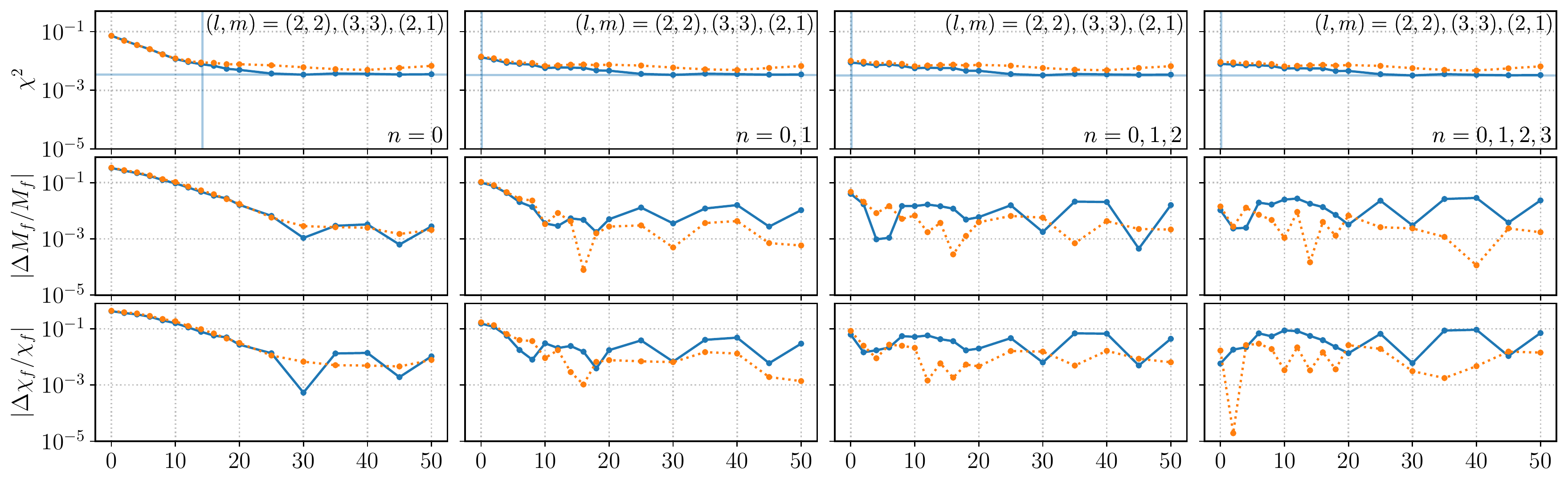}
\includegraphics[width=\textwidth]{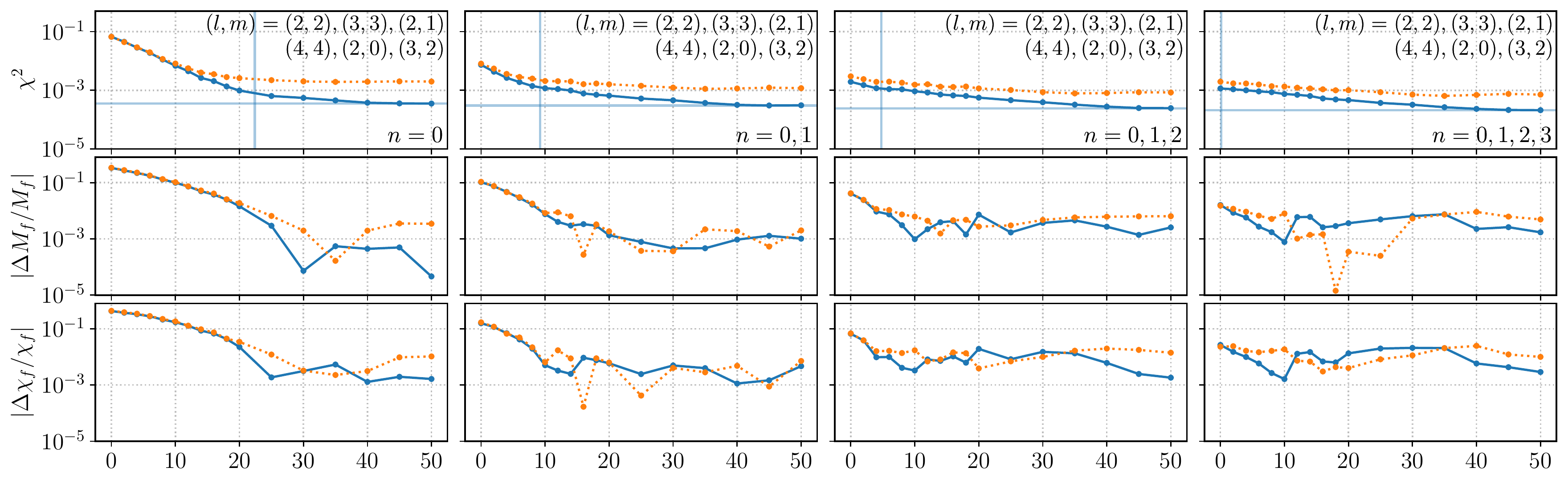}
\includegraphics[width=\textwidth]{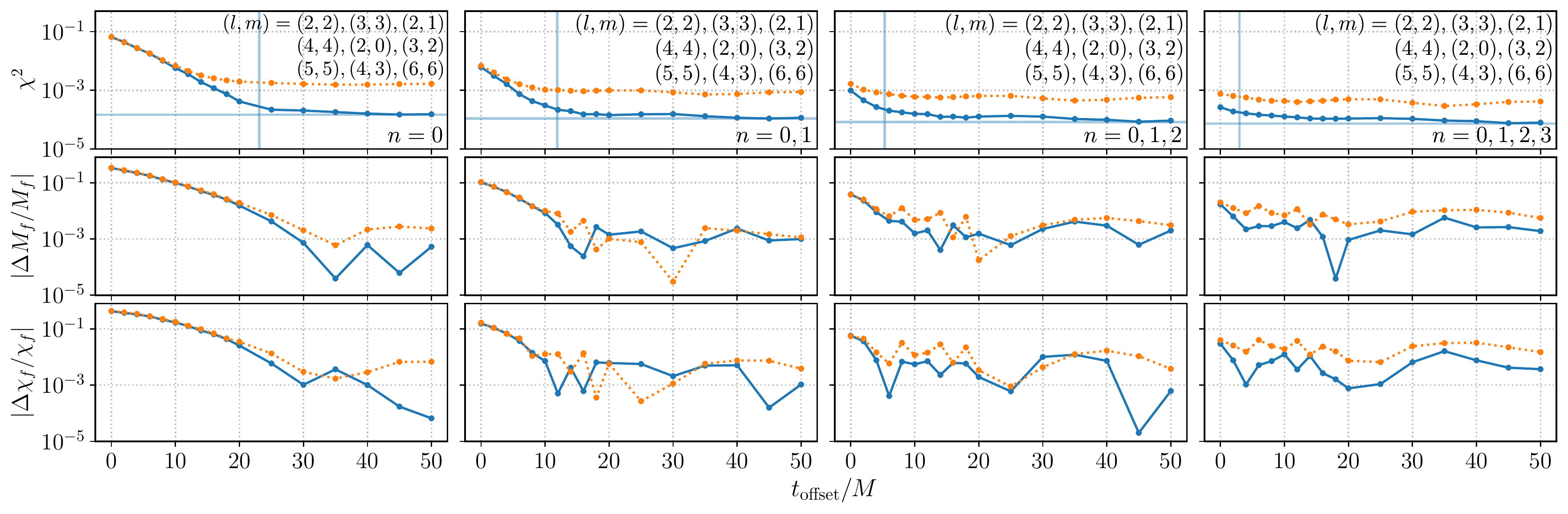}
\caption{\label{fig:N6_ots}Fitting results for binary waveform N6. Plot settings are the same as Fig.~\ref{fig:G0_ots}.}
\end{figure*}

\begin{figure*}
\centering
\includegraphics[width=\textwidth]{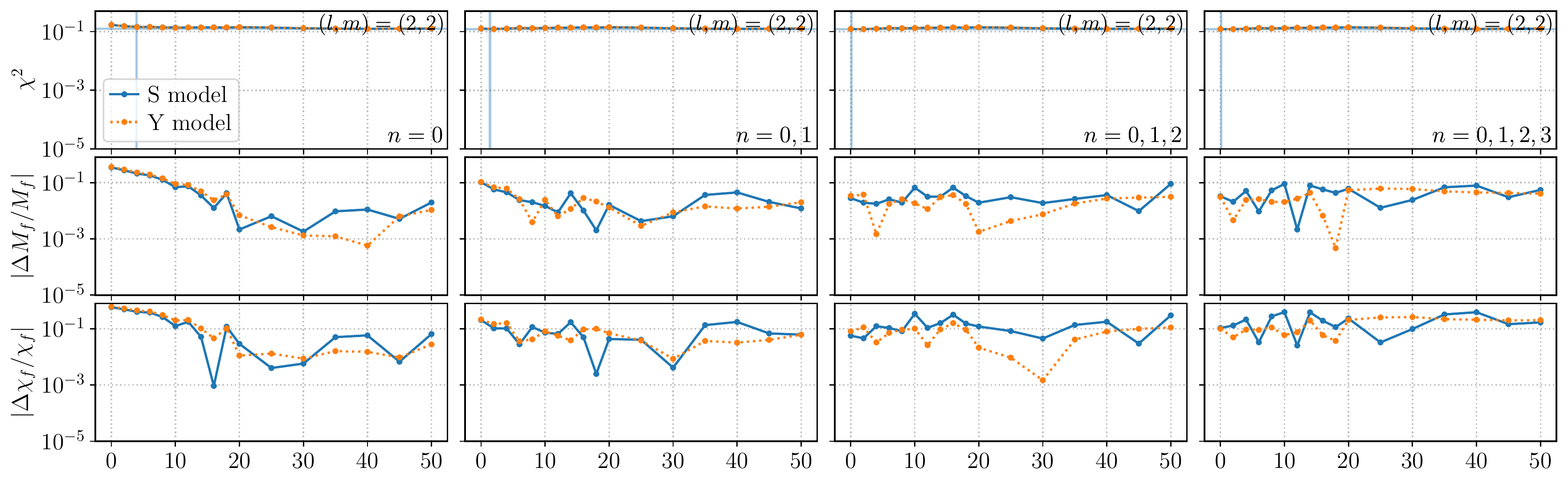}
\includegraphics[width=\textwidth]{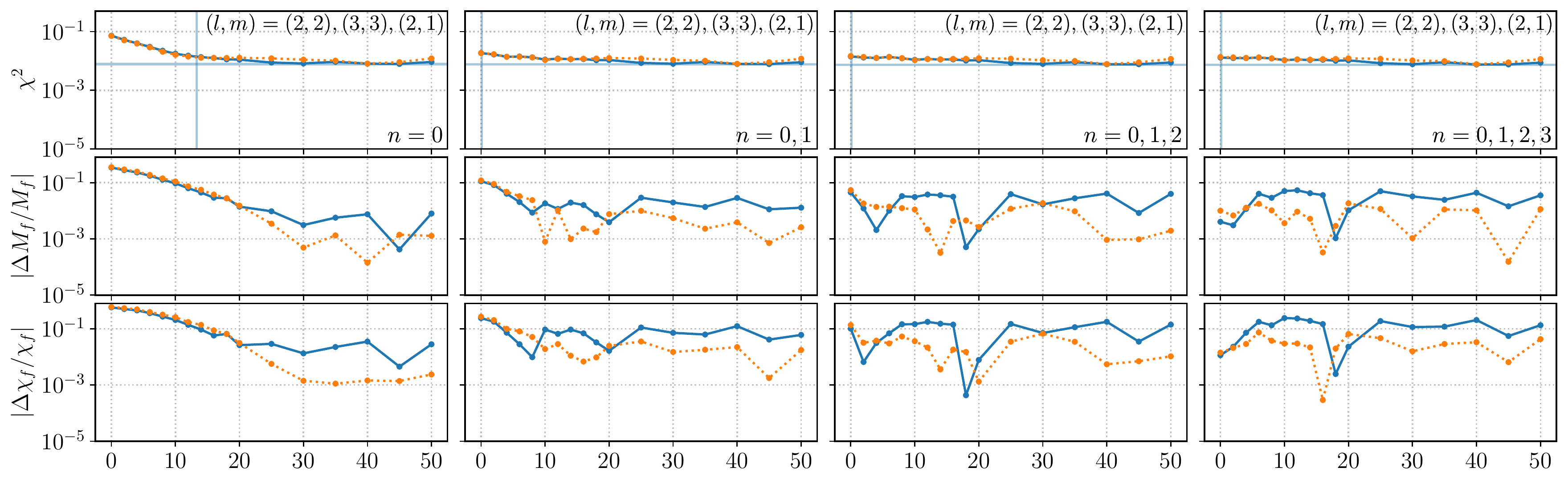}
\includegraphics[width=\textwidth]{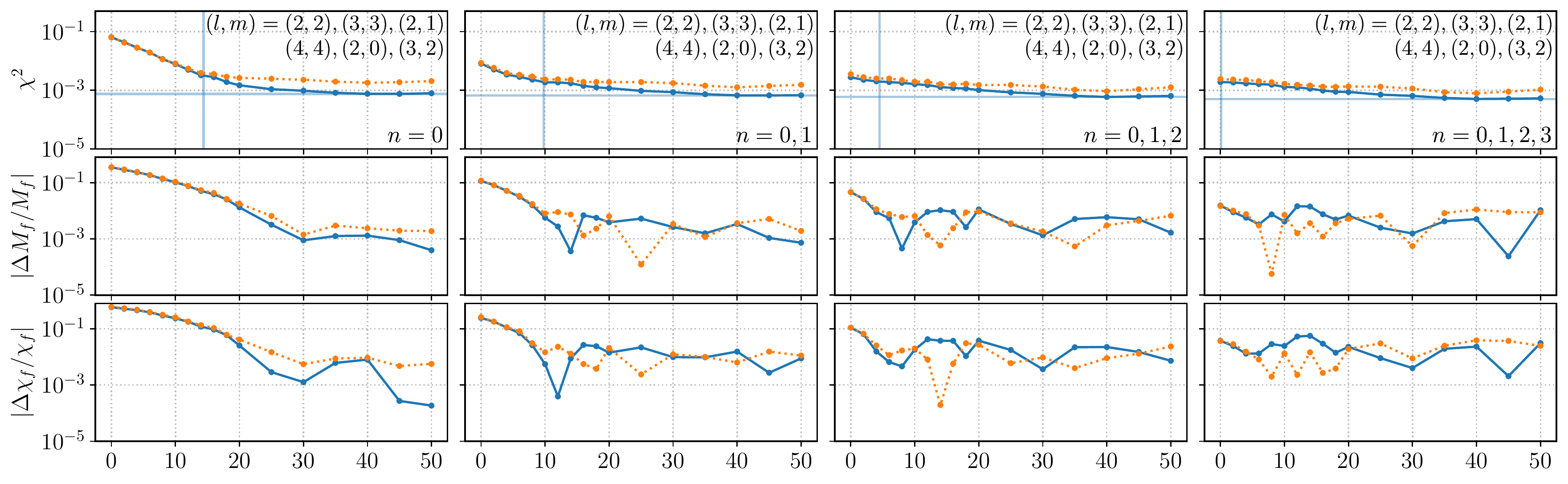}
\includegraphics[width=\textwidth]{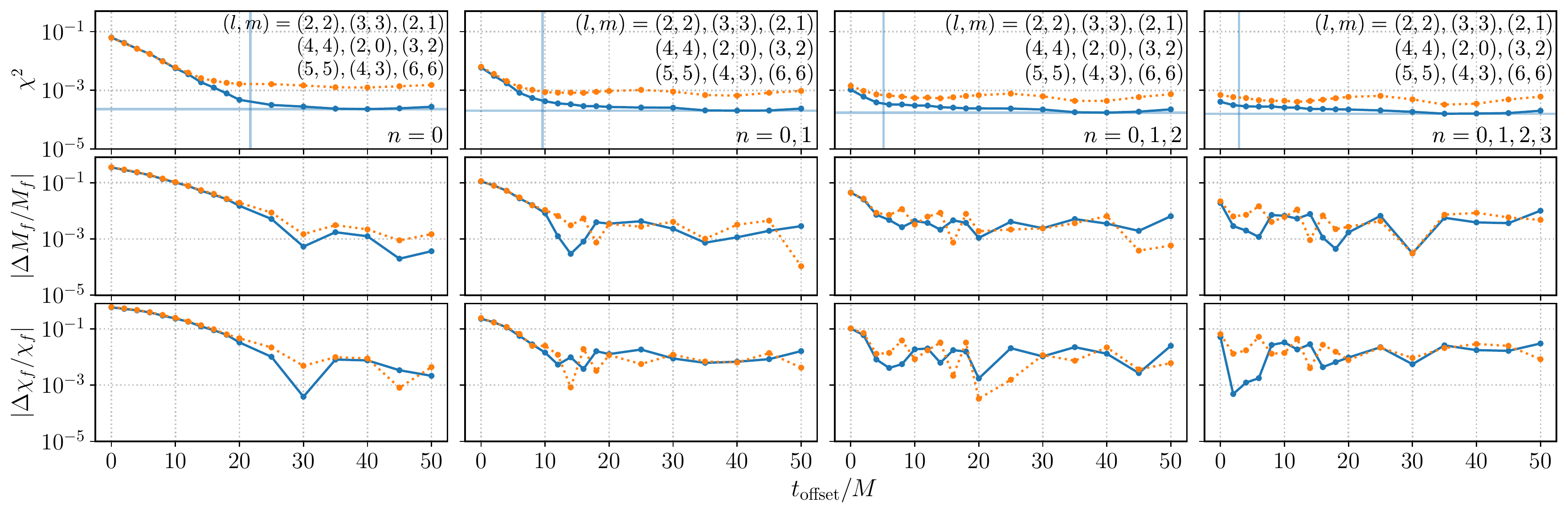}
\caption{\label{fig:N7_ots}Fitting results for binary waveform N7. Plot settings are the same as Fig.~\ref{fig:G0_ots}.}
\end{figure*}

\begin{figure*}
\centering
\includegraphics[width=\textwidth]{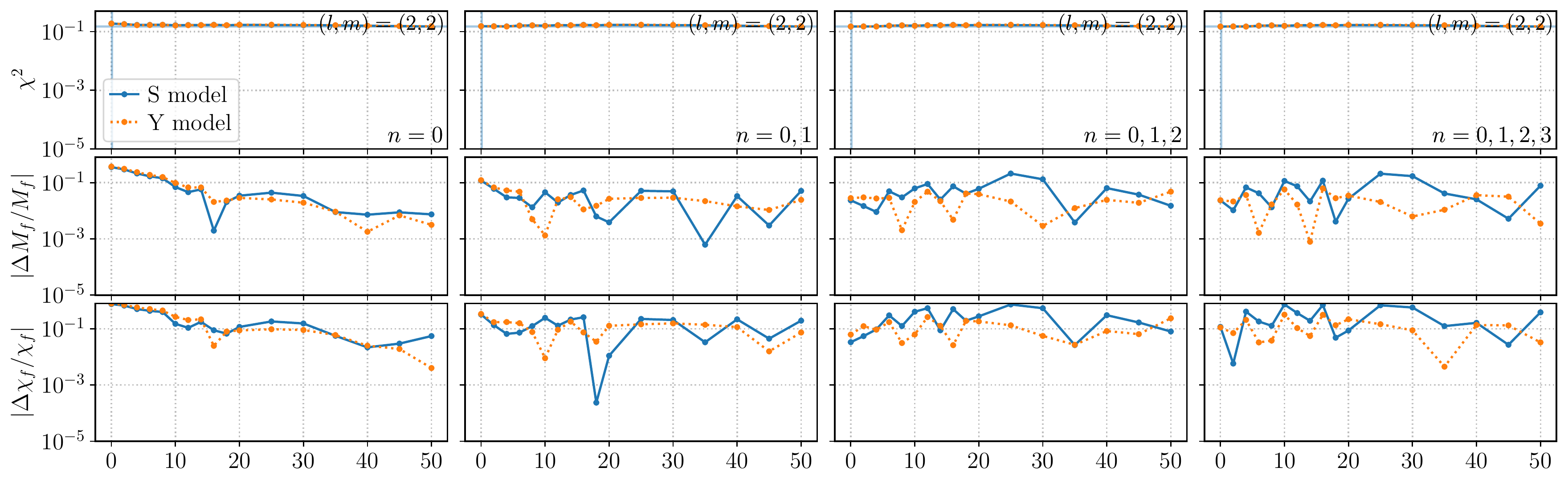}
\includegraphics[width=\textwidth]{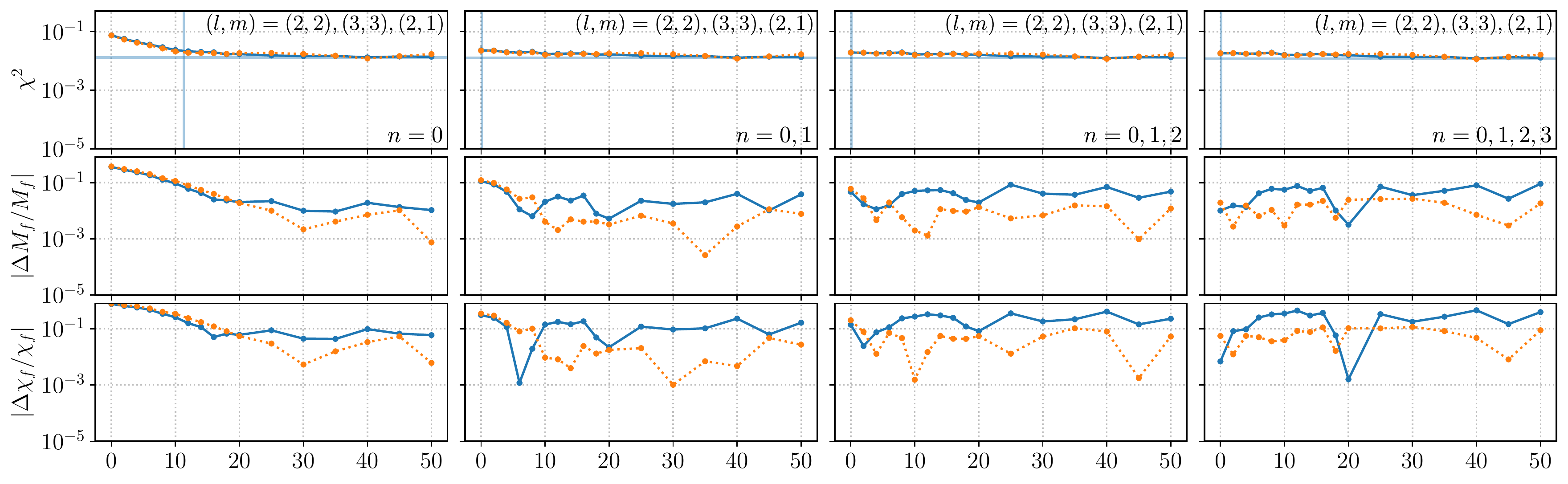}
\includegraphics[width=\textwidth]{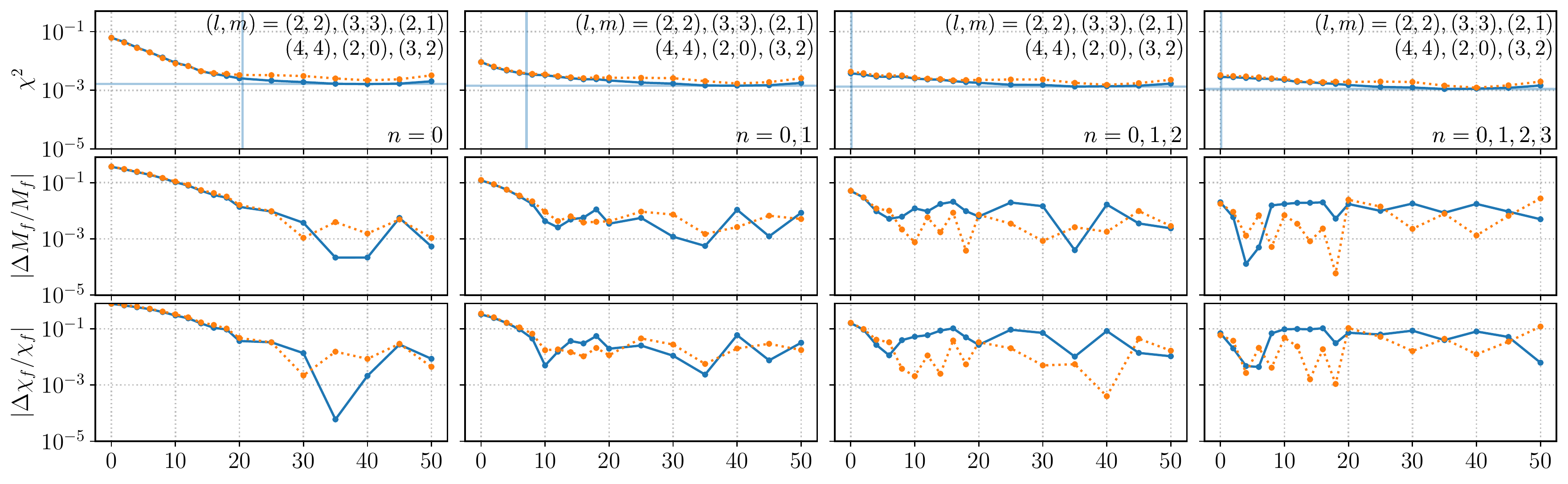}
\includegraphics[width=\textwidth]{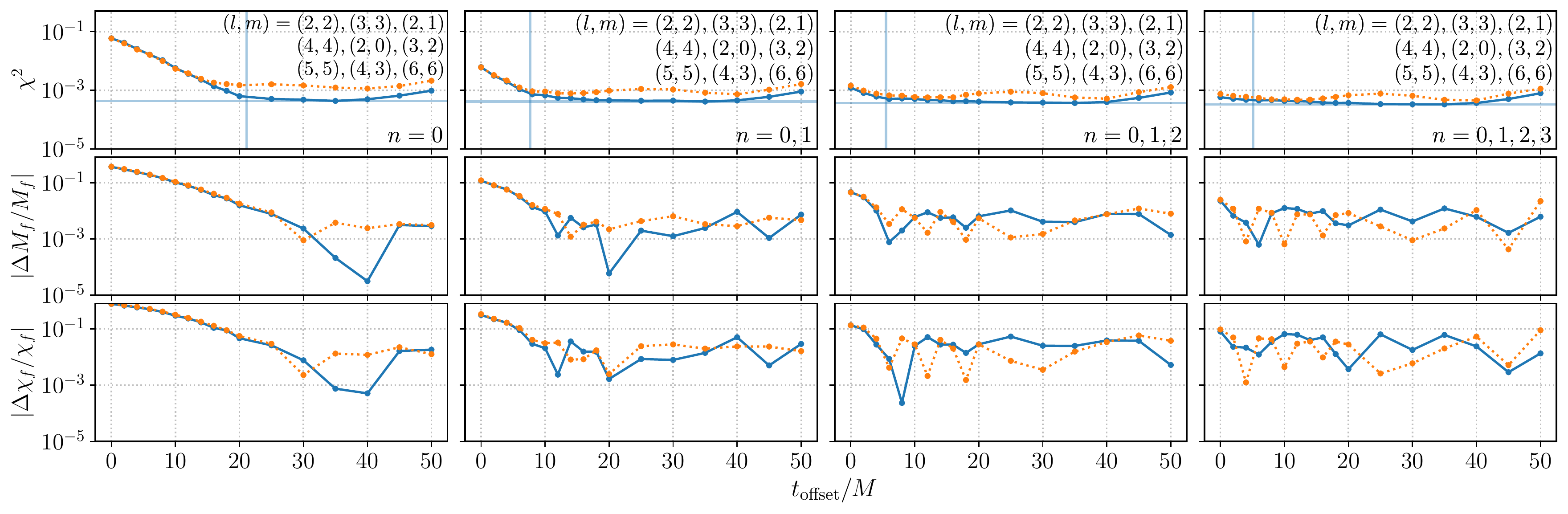}
\caption{\label{fig:N8_ots}Fitting results for binary waveform N8. Plot settings are the same as Fig.~\ref{fig:G0_ots}.}
\end{figure*}

\begin{figure*}
\centering
\includegraphics[width=\textwidth]{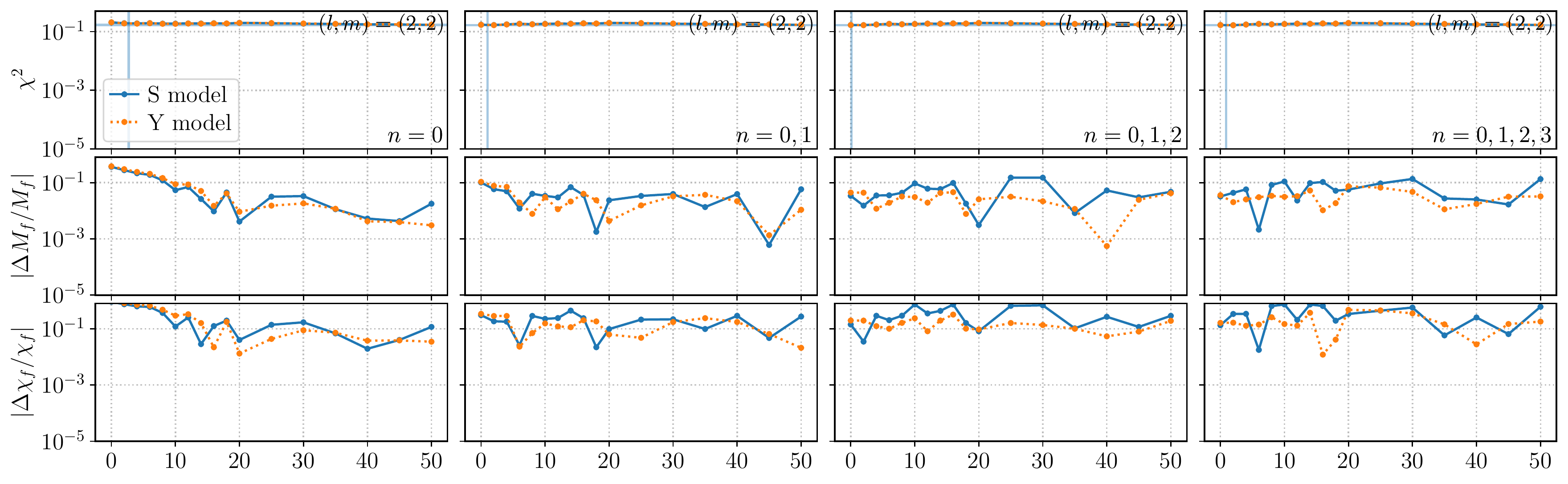}
\includegraphics[width=\textwidth]{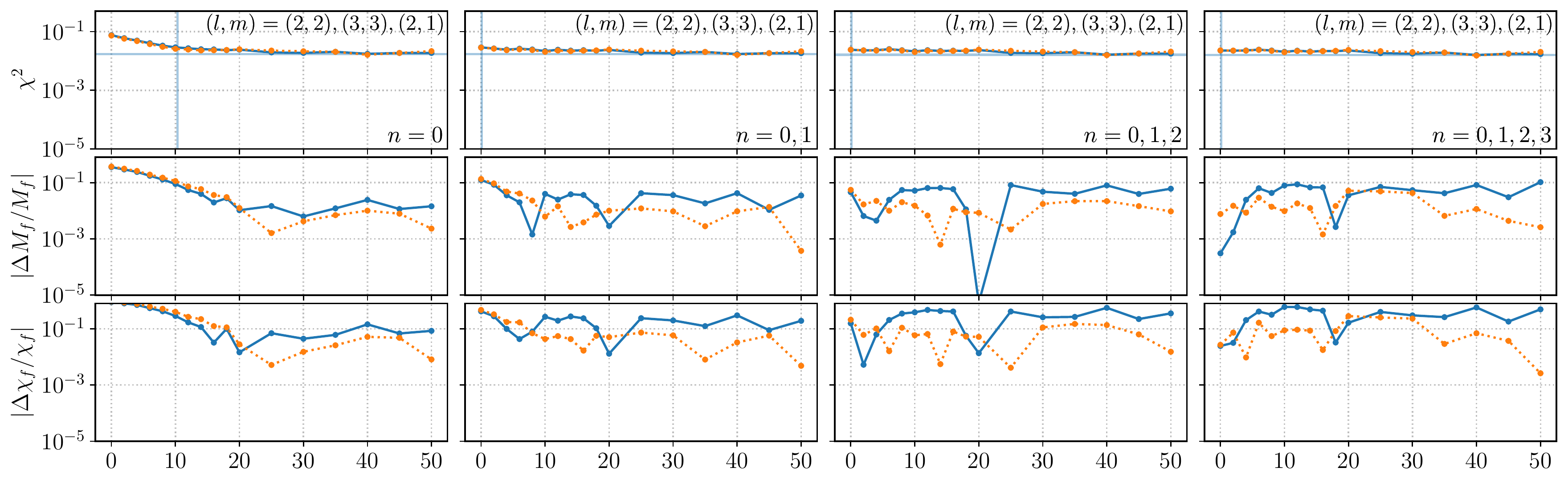}
\includegraphics[width=\textwidth]{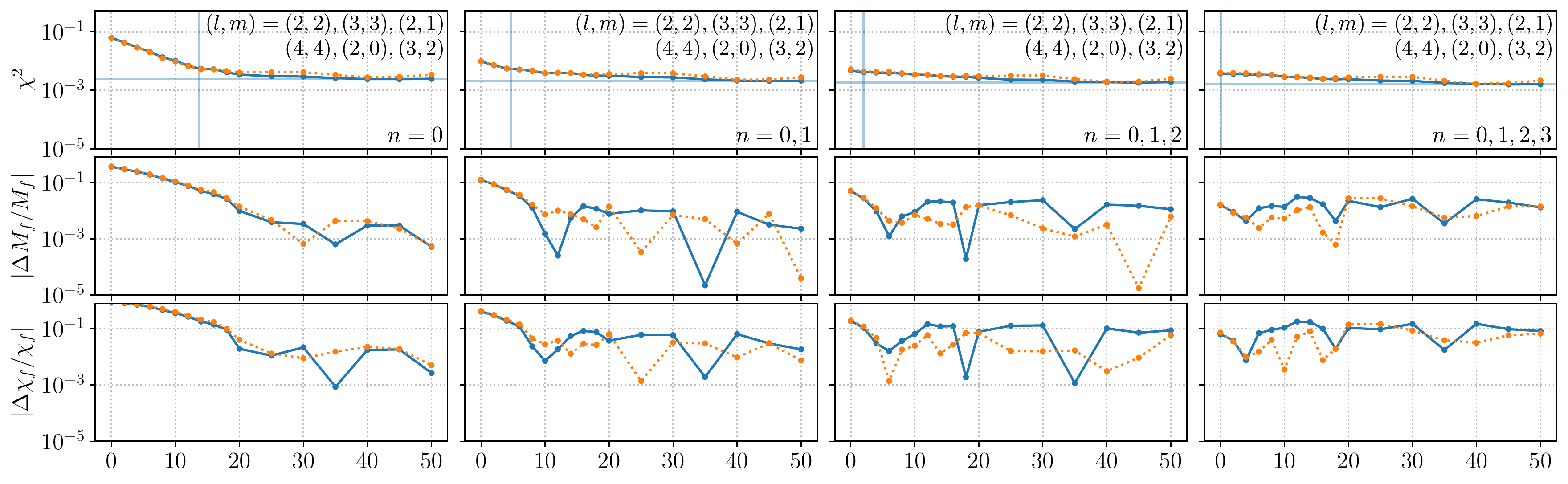}
\includegraphics[width=\textwidth]{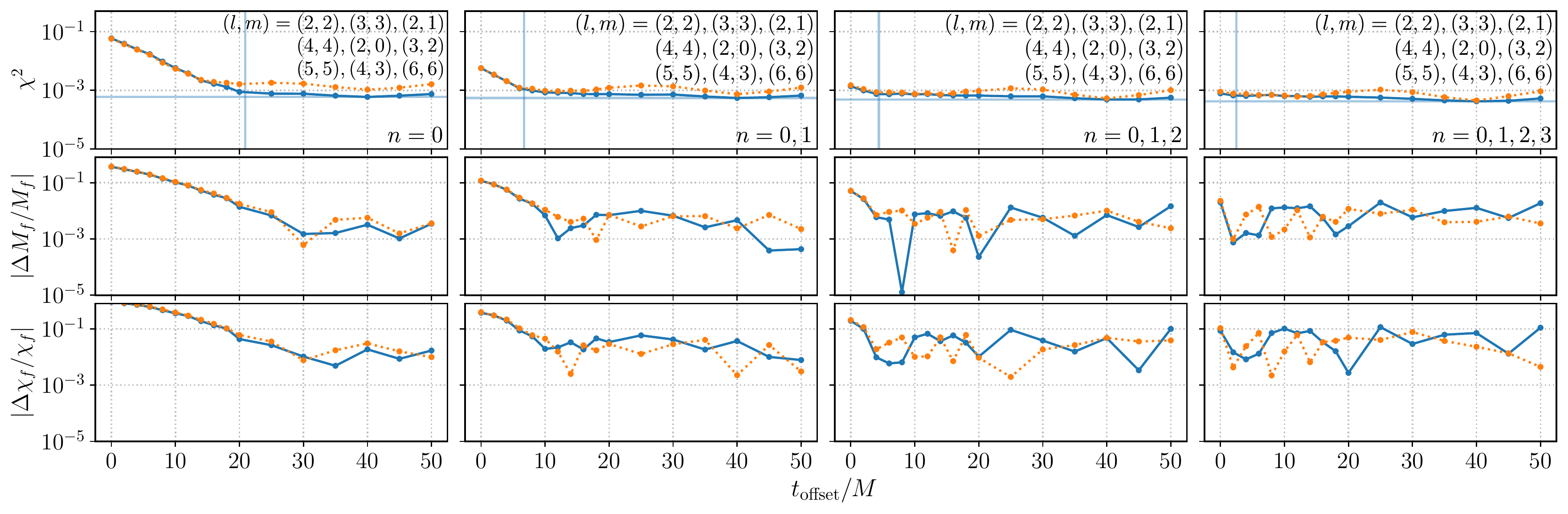}
\caption{\label{fig:N9_ots}Fitting results for binary waveform N9. Plot settings are the same as Fig.~\ref{fig:G0_ots}.}
\end{figure*}

The full fitting results for N1--N9 are shown in Figs.~\ref{fig:N1_ots}--\ref{fig:N9_ots}, plotted in the same way as Fig.~\ref{fig:G0_ots}. 

\section{\label{app:result_levs}Numerical error in SXS waveforms}

\begin{figure*}
\centering
\includegraphics[width=\textwidth]{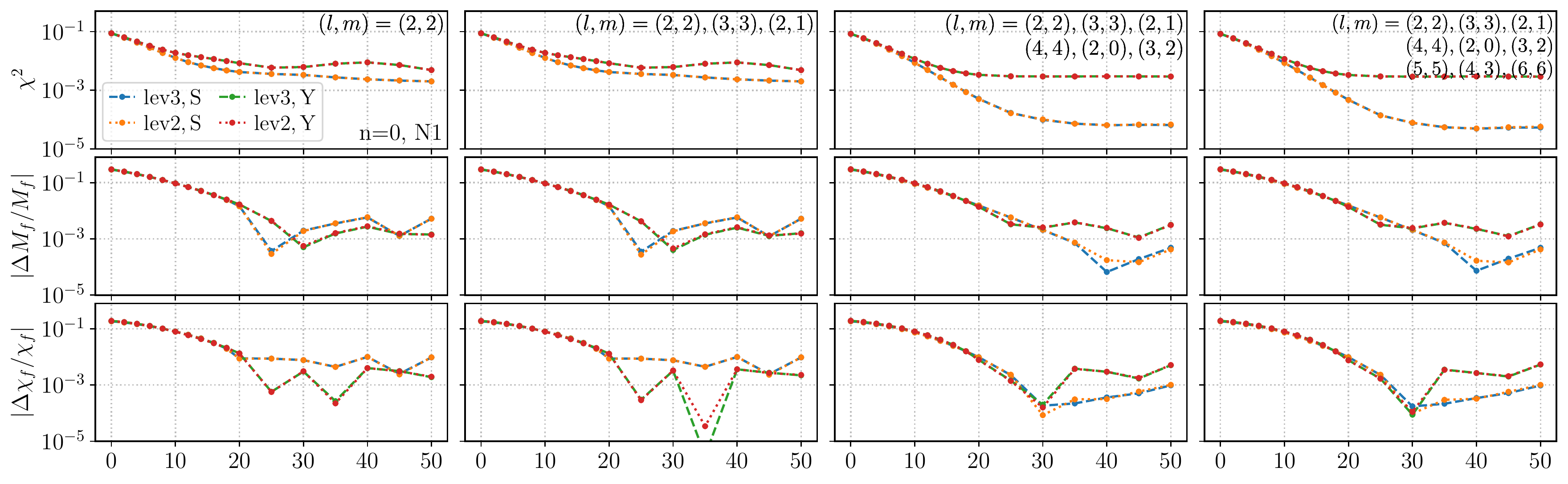}
\includegraphics[width=\textwidth]{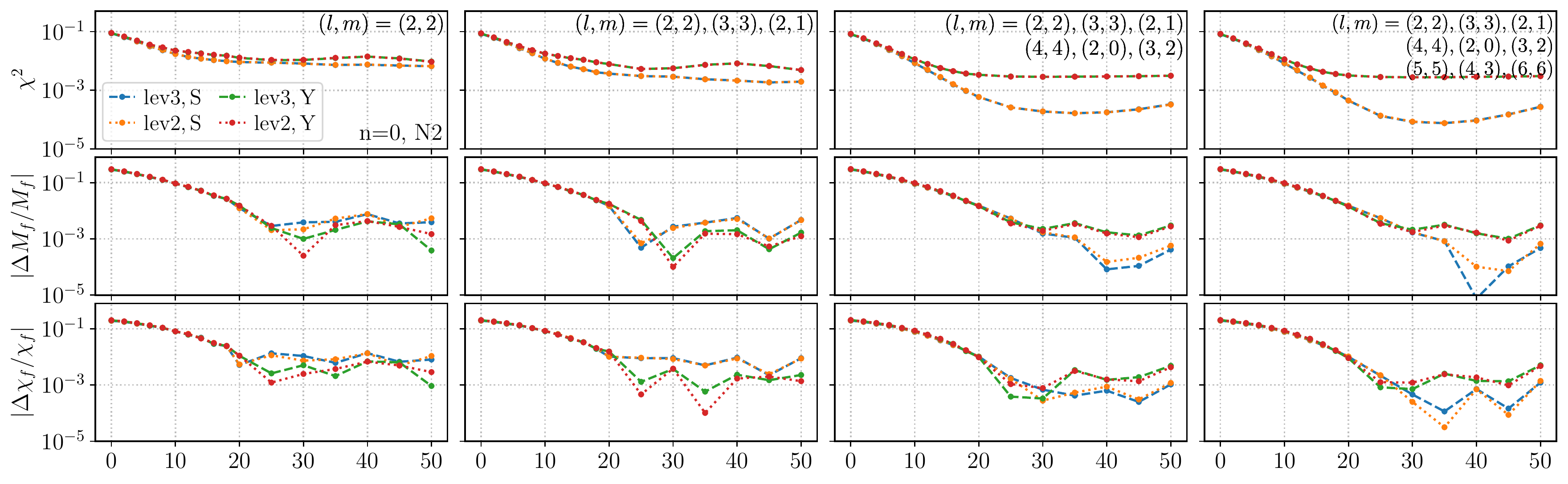}
\includegraphics[width=\textwidth]{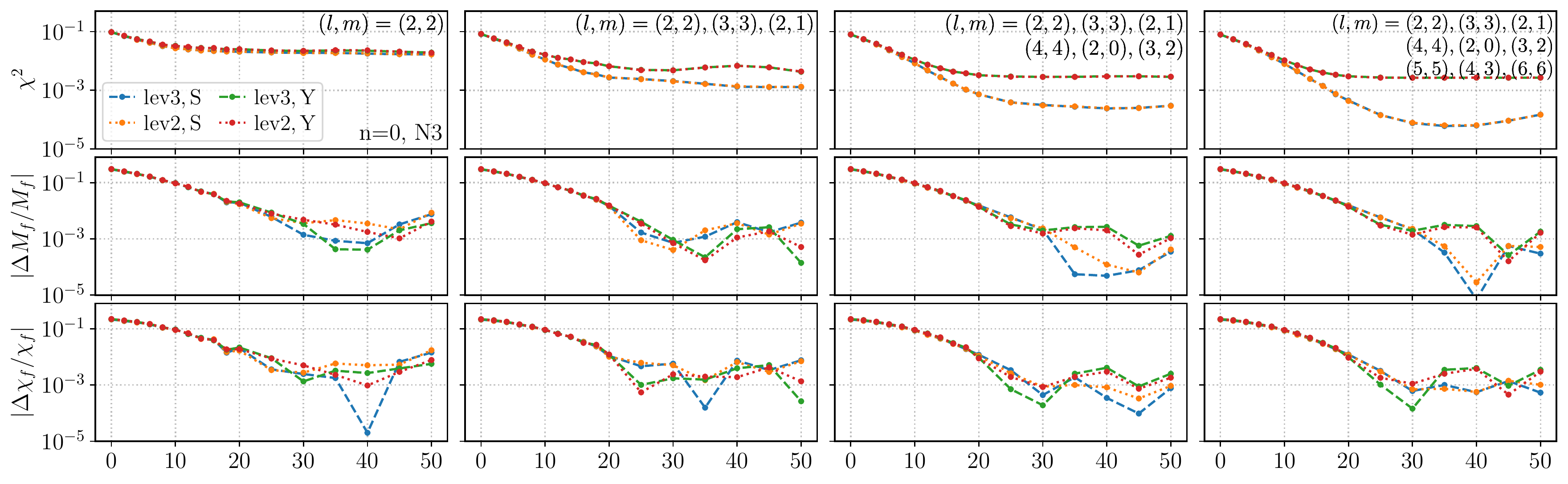}
\includegraphics[width=\textwidth]{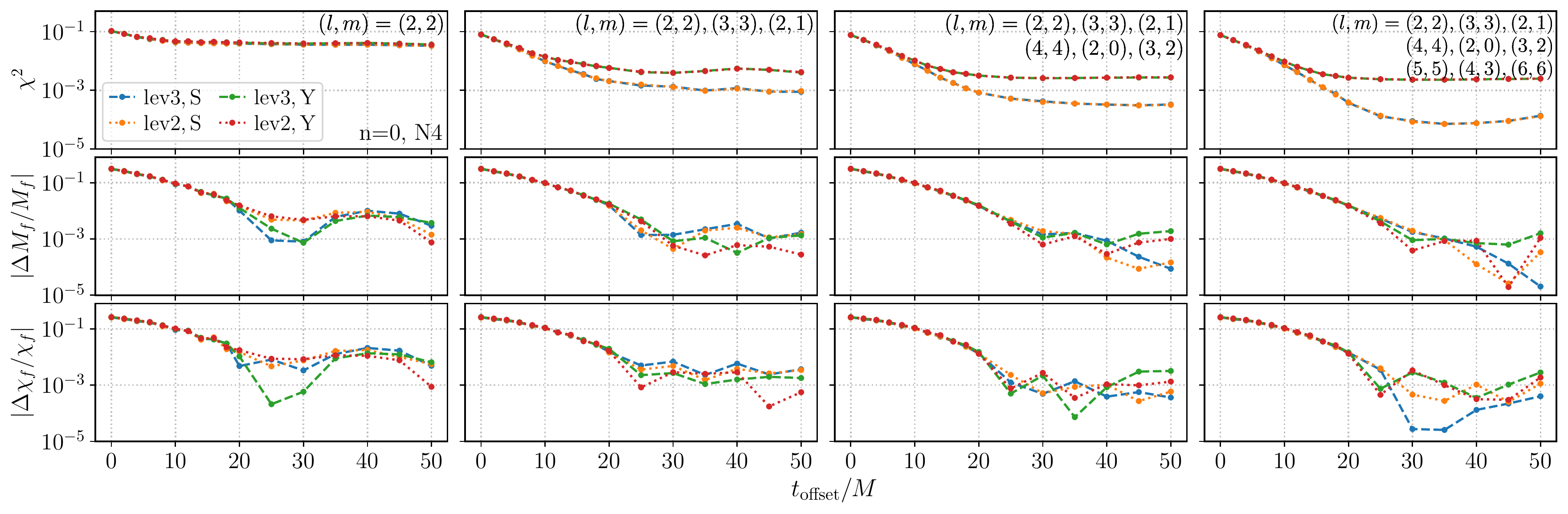}
\caption{\label{fig:N14_ot0_levs}Fitting results for N1--N4 using SXS data with different numerical levels. Plot settings are the same as Fig.~\ref{fig:G0_ot0_levs}.}
\end{figure*}

\begin{figure*}
\centering
\includegraphics[width=\textwidth]{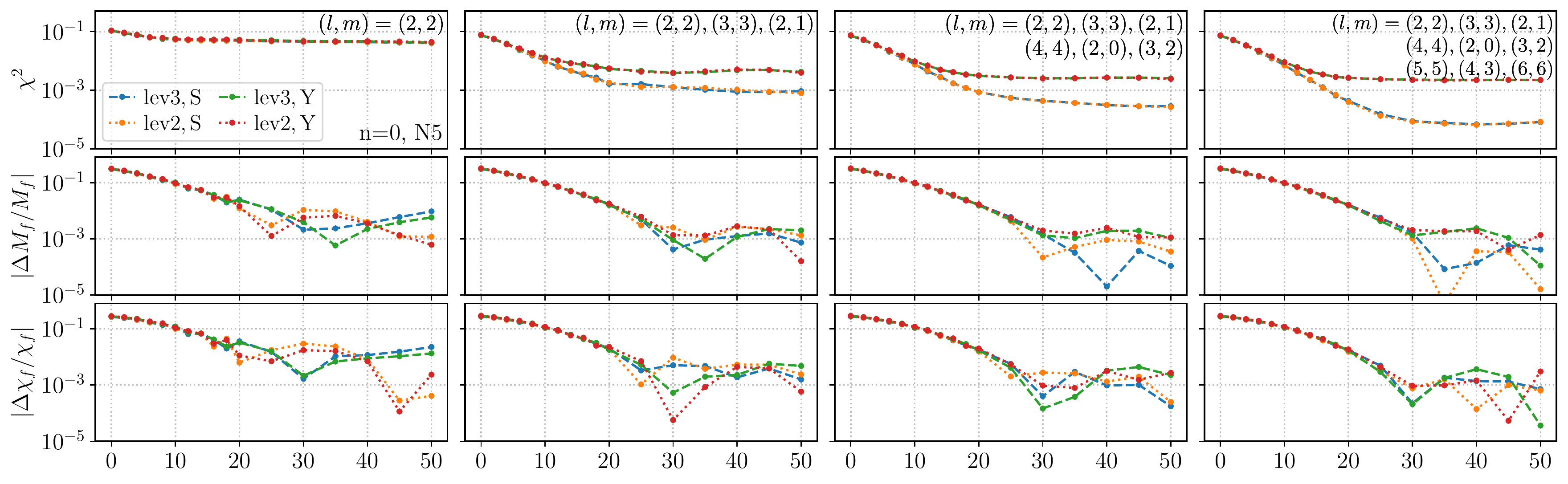}
\includegraphics[width=\textwidth]{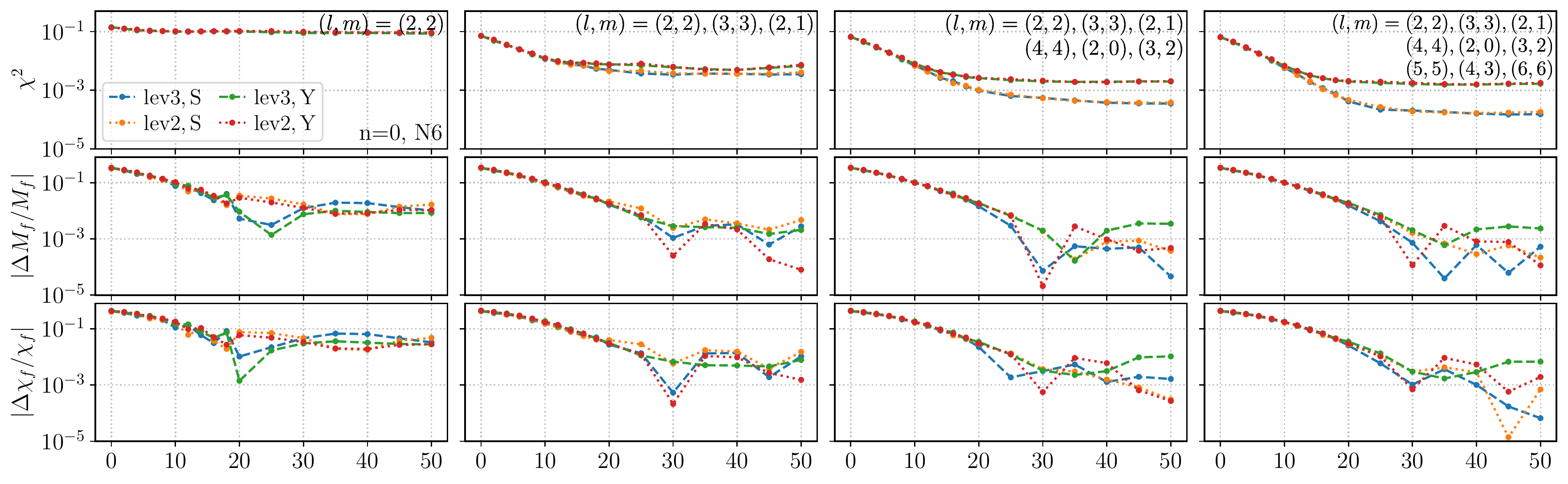}
\includegraphics[width=\textwidth]{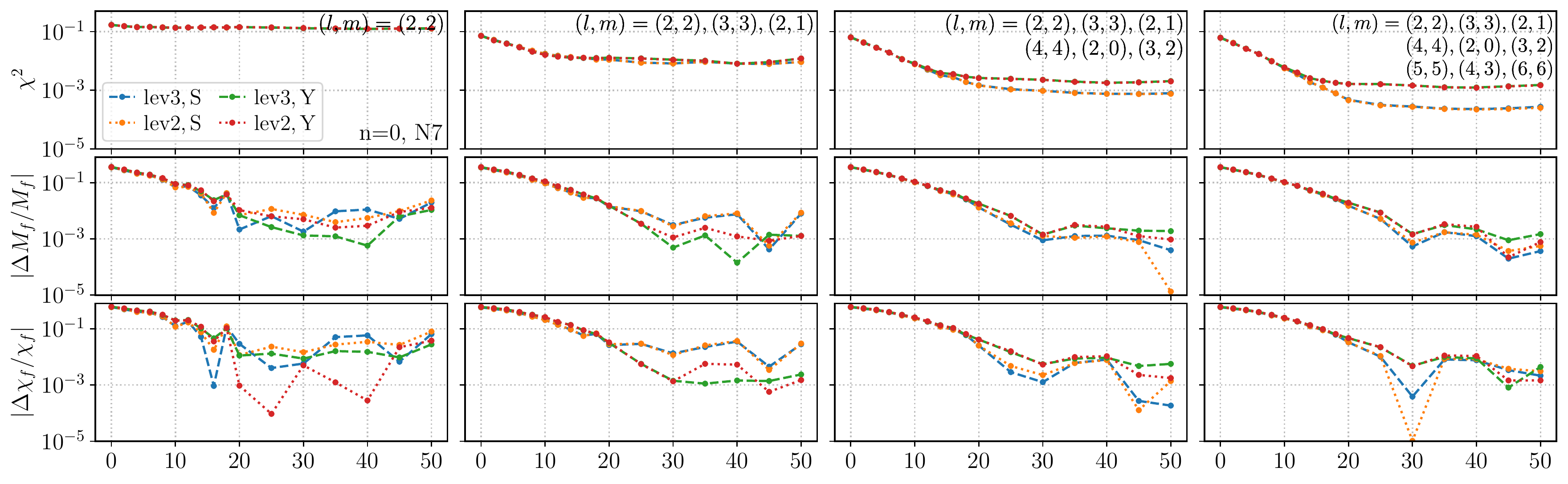}
\includegraphics[width=\textwidth]{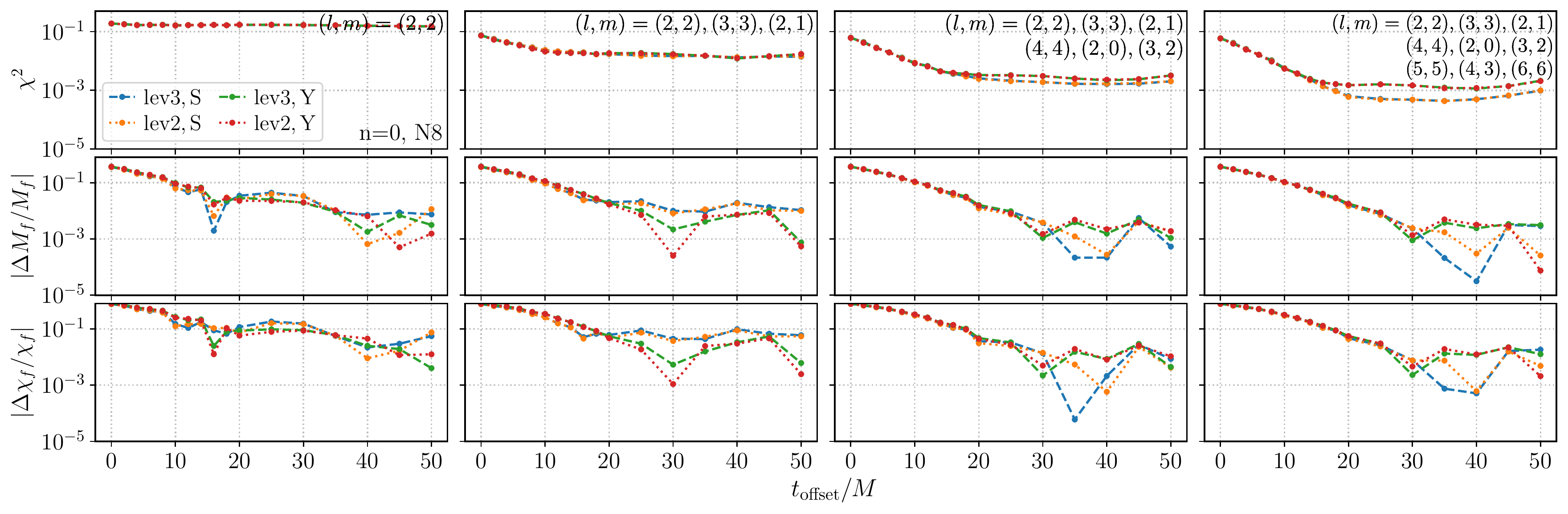}
\caption{\label{fig:N58_ot0_levs}Fitting results for N5--N8 using SXS data with different numerical levels. Plot settings are the same as Fig.~\ref{fig:G0_ot0_levs}.}
\end{figure*}

\begin{figure*}
\centering
\includegraphics[width=\textwidth]{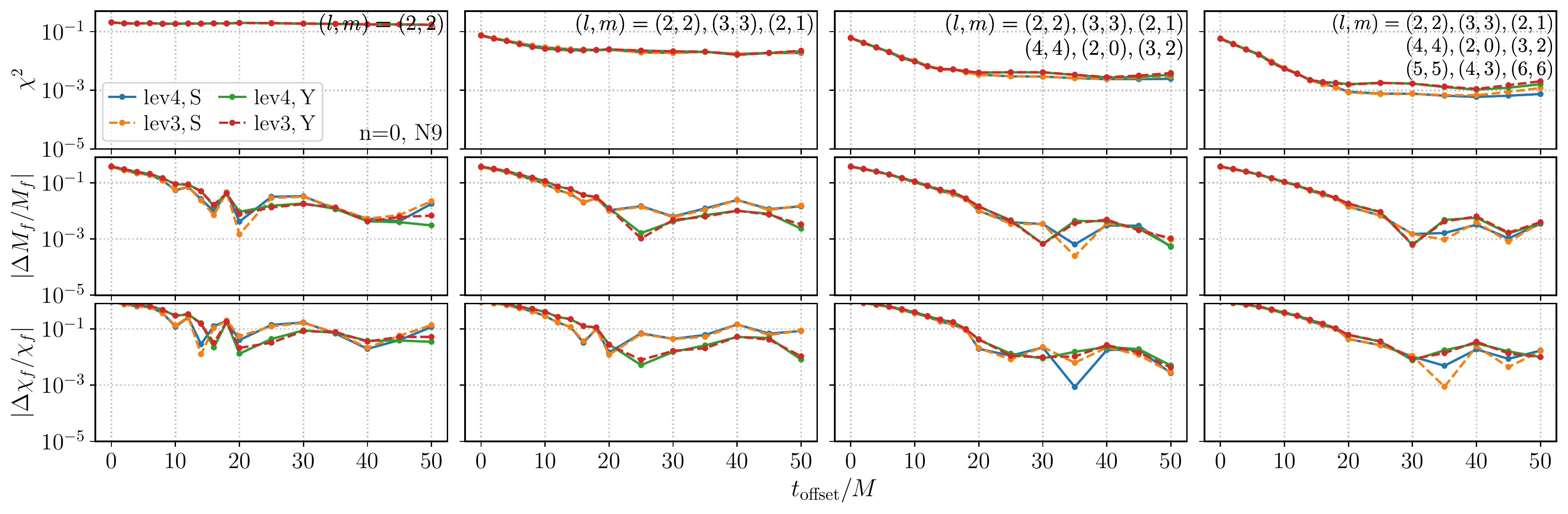}
\caption{\label{fig:N9_ot0_levs}Fitting results for N9 using SXS data with different numerical levels. Plot settings are the same as Fig.~\ref{fig:G0_ot0_levs}.}
\end{figure*}

In the main text, we use the highest numerical resolution levels available for each chosen binary, as listed in Tables~\ref{tab:SXS_N} and \ref{tab:SXS_A}. In this appendix, we show the results of the lower SXS numerical resolution levels for binaries N1--N9. Figs.~\ref{fig:N14_ot0_levs}--\ref{fig:N9_ot0_levs} are plotted in a similar way as Fig.~\ref{fig:G0_ot0_levs} for both the $S$ and $Y$ models. It shows that all the differences from various numerical levels are clearly smaller than the difference from using different fitting models ($S$ versus $Y$, or including different $lmn$ modes). Therefore, our conclusions in Sec.~\ref{sec:nonsp_qs} are not impacted by numerical errors of the waveforms.



\twocolumngrid

\bibliography{gr.bib}
\end{document}